\documentclass[11pt]{article}
\usepackage[letterpaper, left=1in, right=1in, top=1in,bottom=1in]{geometry}
\usepackage{amsfonts}       
\usepackage{nicefrac}       
\usepackage{bbm}
\usepackage[ruled,noend]{algorithm2e}
\usepackage{bm}
\usepackage{tikz}
\usetikzlibrary{positioning,chains,fit,shapes,calc,arrows}
\usetikzlibrary{arrows.meta}
\usepackage{graphicx}
\usepackage{pgf}
\usepackage{mathtools}
\usepackage{mathrsfs}
\usepackage{amsmath,amsthm,amssymb}
\usepackage{comment}
\usepackage{xfrac}
\usepackage{accents}

\usepackage{eurosym}             
\usepackage{caption}
\usepackage[export]{adjustbox}   
\usepackage{booktabs, tabularx} 

\usepackage{auxiliary}
\usepackage{bm}
\usepackage{enumerate}
\usepackage{hyperref} 
\usepackage{cleveref}
\usepackage{amsmath}
\usepackage[ruled]{algorithm2e}
\usepackage{thm-restate}
\usepackage{color}              
\usepackage[suppress]{color-edits}
\usepackage[framemethod=tikz]{mdframed}
\usepackage{float}
\usepackage{natbib} 
\usepackage{url}      
\usepackage{color}              %
\addauthor{df}{blue}
\addauthor{ds}{red}
\addauthor{bb}{orange}

\SetAlFnt{\small}
\SetAlCapFnt{\small}
\SetAlCapNameFnt{\small}
\SetAlCapHSkip{0pt}
\IncMargin{-\parindent}

\def\prob{\mu}
\def\comp{\Gamma}
\def\market{\kappa}
\def\inst{\mathcal I}
\def\inv{\eta}

\def\Halmos{\qed}

\def\sigmabf{\bm{\sigma}}
\def\alphabf{\bm{\alpha}}
\def\betabf{\bm{\beta}}
\def\rhobf{\bm{\rho}}
\def\nubf{\bm{\nu}}
\def\thetabf{\bm{\theta}}
\def\gammabf{\bm{\gamma}}
\def\lambdabf{\bm{\lambda}}
\def\xbf{\mathbf{x}}

\DeclareMathOperator*{\argmin}{arg\,min}
\def\offI{\mathrm{\mathrm{OFF}}\textrm{-}\mathrm{I}}
\def\alg{\mathrm{ALG}}
\def\greedy{\mathrm{GREEDY\textrm{-}D}}
\def\altgreedy{\mathrm{ALT\textrm{-}GREEDY\textrm{-}D}}
\def\optD{\mathrm{OPT\textrm{-}D}}

\def\LP{\mathrm{LP}(T,\prob, \market)}
\def\pbp{\mathrm{PBP}(\inst)}

\def\offIkappa{\offI(\inst,\market)}

\def\supply{S}
\def\undersup{\mathcal U}
\def\oversup{\mathcal O}
\def\classU{\mathcal U^{\mathrm{Cl}}}
\def\classO{\mathcal O^{\mathrm{Cl}}}

\def\Tundersup{T^\undersup}
\def\Toversup{T^\oversup}

\def\DLP{\mathrm{DUAL}(T, \prob, \market)}

\addauthor{bb}{orange}
\addauthor{df}{blue}
\addauthor{ds}{red}

\usepackage{chngcntr}
\counterwithin*{proposition}{section}

\title{Online matching and market imbalance}

\author{
Benjamin Barrientos\thanks{Massachusetts Institute of Technology} \hspace{0.05em},  Daniel Freund\thanks{Massachusetts Institute of Technology} \hspace{0.05em}, Daniela Saban\thanks{Stanford Graduate School of Business}
}

\date{First version: February 2025\\
Current version: February 2025}

\begin{document}

\maketitle

\begin{abstract}
Our work introduces the effect of supply/demand imbalances into the literature on online matching with stochastic rewards in bipartite graphs. We provide a parameterized definition that characterizes instances as over- or undersupplied (or balanced), and show that higher competitive ratios against an offline clairvoyant algorithm are achievable, for both adversarial and stochastic arrivals, when instances are more imbalanced. The competitive ratio guarantees we obtain are the best-possible for the class of \emph{delayed} algorithms we focus on (such algorithms may adapt to the history of arrivals and the algorithm's own decisions, but not to the stochastic realization of each potential match). 

We then explore the real-world implications of our improved competitive ratios. First, we demonstrate analytically that the improved competitive ratios under imbalanced instances is not a one-way street by showing that a platform that conducts effective supply- and demand management should incorporate the effect of imbalance on its matching performance on its supply planning in order to create imbalanced instances. Second, we empirically study the relationship between achieved competitive ratios and imbalance using the data of a volunteer matching platform.
\end{abstract}
\newpage

\section{Introduction}
\label{sec: introduction}

Over the past decades, the emergence of online platforms has brought about a new economy that relies on algorithmic matching procedures for its core operations. These platforms cover various industries such as labor markets, accommodation, online advertising, and transportation, and have caused a revolution in the way we travel, work, and consume goods and services. Platforms like Uber, Airbnb, and Google AdWords are designed to optimize their operations by matching incoming demand with available supply. 
The success of these platforms relies on (i) effectively executing the online matching that connects incoming demand to available supply and (ii) ensuring an appropriate level of supply to handle the demand. 

The complex question of how to match incoming demand with available supply has spurred ample research on online matching algorithms that dates back to the seminal work of \citet{karp1990optimal}. 
In the original problem, a decision-maker is given a fixed set of supply nodes offline and observes the arrival of one new demand node per period over a known time horizon. 
Each time a demand node arrives, the decision-maker needs to irrevocably match the incoming demand after observing which supply nodes the demand node is adjacent to. The decision-maker aims to design an online matching algorithm that maximizes the number of matches made. Since this foundational work, the field has evolved to explore a diverse set of extensions, including the vertex-weighted variant \citep{aggarwal2011online,jaillet2014online}, the edge-weighted variant \citep{feldman2009free,haeupler2011online} and the use of \textit{stochastic rewards} on the edges \citep{mehta2012online,mehta2014online}, among others. In all of these variants, algorithmic performance is measured through the competitive ratio (CR), i.e., the worst-case gap between an algorithm's achieved matching size and that of an offline oracle. Upper bounds on the CR are usually proven through worst-case instances, on which no online algorithm can outperform a given bound; these worst-case instances are usually balanced in the sense that the expected demand precisely matches the available supply.

Real-world instances, however, usually do not  exhibit the kind of balance that is inherent in worst-case instances (see our analysis in \Cref{sec: numerical_exp}). For instance, the demand for Uber rides may vastly outstrip the supply during rush hour whereas Uber drivers may experience a low utilization in-between peak demand periods. 
As a result,  real-world instances are often imbalanced in one direction or the other. Indeed, the market imbalance between supply and demand may even result from the platform's offline planning: firms like Uber and Airbnb leverage their forecasting abilities to guide promotional tactics on both market sides. This process strives to establish an equilibrium where supply is neither excessive nor lacking. The exact tradeoff between the two is captured by the platform's overage and underage costs, i.e., the cost of underutilized resources and that of missed matching opportunities, respectively. Unless the overage and underage cost are approximately equal, it is likely in the firms interest to strive for a somewhat imbalanced marketplace.\footnote{Of course, the planning process of a matching platform is more complicated than a simple newsvendor calculation, but directionally one may still expect overage and underage to guide the desired imbalance of demand and supply.}

In order to identify the appropriate (im)balance of supply and demand, a platform needs to know how its online matching algorithms translate the available supply and demand into a resulting matching size. Specifically, as we describe below, algorithms may obtain significantly higher CRs in imbalanced markets. As a result, we find that a matching platform may find itself in a situation where its (e.g., monthly expected) demand is twice as large as its supply; a natural expectation would then be that by doubling its supply, the platform would be able to double the number of matches it makes between supply and demand. After all, it faces sufficient demand for all of the additional supply to be consumed. However, this reasoning ignores that the matching algorithm's performance deteriorates in a more balanced market. Indeed, we find that, depending on the market dynamics, a doubling in supply may grow the matching size by just 50\%. This gap leads us to answer the following questions in our work:

\begin{enumerate}
    \item \emph{How does the market imbalance of supply and demand affect the performance of matching algorithms? Are higher CRs achievable in imbalanced settings?}
    \item \emph{How does the effect of imbalance on a firm's matching ability affect its underage/overage tradeoff?}
\end{enumerate}

\subsubsection*{Outline of our contributions.}
\label{ssec: outline}
Section \ref{sec: model} sets up our model as an online bipartite matching problem spanning $T$ steps, wherein the supply nodes are known to the platform. Conversely, the arriving demand nodes remain unknown to the platform, and at each timestep, a node arrives, compelling the platform to make an irrevocable decision regarding the match for the incoming demand.
Each specific match carries a success probability that determines whether the demand consumes the matched resource. This is often referred to as the stochastic reward setting. Our study focuses on a class of \textit{delayed algorithms}, which address the complexities of having delayed realizations, e.g., when a customer may not immediately purchase an item but instead delays its purchase to a later point --- this class of algorithms, with the benchmark of our choice (see below) is particularly amenable to analyzing the effect of imbalance. At time $t$, a delayed algorithm has access to information about previous matches up until time $t$ but lacks information about their stochastic realizations. 
Below, we begin by describing how our paper models market imbalance. We then address our first research question by summarizing the impact that  imbalance has on the performance of matching algorithms. Thereafter, we use these results to derive insights on the trade-offs between underage and overage of a matching platform.

\emph{\textbf{Modeling imbalance.}} Our model captures the imbalance of a given instance through the effect of loosening/tightening a particular constraint in a deterministic linear program (DLP) that upper bounds the performance of any algorithm (see  \Cref{sec: model}). 
Specifically, we parameterize instances by a parameter $\market$, where $\market\leq 1$ indicates that matching supply nodes $\market$ times in expectation is sufficient for the DLP to achieve its objective. We thus label these instances as \textit{$\market$-oversupplied}, as the DLP only requires a portion of the supply to effectively meet the same demand. Similarly, we call an instance \textit{$\market$-undersupplied}  when allowing supply nodes to be matched $\market\geq 1$ times increases the objective of the DLP upper bound by a factor of $\market$. We also extend our definition to mixed instances that have both an undersupplied and an oversupplied component to capture the fact that instances may be locally over- and locally undersupplied. Moreover, we use the data of an online matching platform (see Section \ref{sec: numerical_exp}) to evaluate the level of imbalance in real-world matching instances.

\emph{\textbf{The effect of imbalance on the performance of matching algorithms.}} Building on our definition of imbalance, we formally present our findings in \Cref{sec: overview_results}.  
We analyze two different arrival regimes, adversarial and stochastic. In the adversarial arrival setting (Section \ref{sec3}) we show that for any instance characterized by the imbalance parameter~$\market$ (which we refer to as $\market$-imbalanced) there exists a delayed algorithm that achieves a CR of $\max \left\{ \frac{1}{\market+1}, \frac{\market}{\market+1}\right\}$ when compared to the DLP. This result, with a slightly different information structure, also holds for the generalized imbalanced definition that encapsulates graphs that are locally over- and locally undersupplied (\Cref{sec: alternative_def}).  

In the stochastic setting (Section \ref{sec: stochasting_setting}), we show that for any instance characterized by the imbalance parameter~$\market$ there exists a delayed algorithm with a CR of $\max \left\{\frac{1-e^{-\kappa}}{\kappa}, 1-e^{-\kappa} \right\}$.\footnote{We do not extend this result to our generalized imbalance definition because the best-possible CRs in the stochastic setting lack the symmetry in $\market$ that the adversarial setting provides (see \Cref{sec: overview_results}).}
We prove that these bounds are optimal by constructing a sequence of instances in which the CR of the optimal delayed algorithm converges to the specified values (\Cref{section: impossibility}). Hence, balanced instances, characterized by $\market = 1$, give rise to the worst-case CRs whereas imbalanced instances give rise to significantly higher ones (e.g., as $\market\to\infty$, they converge to~$1$). 

\emph{\textbf{Imbalance and underage/overage considerations.}} The previous results may affect a platform's demand and supply planning. To illustrate, consider a platform $A$ that encounters adversarial arrivals and delayed stochastic realizations of matches ({detailed in \Cref{sec: example_introduction}}). Further, assume that our presented lower bounds hold tightly and denote the value of the DLP by \texttt{OPT}. Now, suppose platform $A$ is $2$-undersupplied, i.e., the platform faces enough demand to consume twice as much supply (in expectation) as it has at its disposal. With the assumptions stated above, this implies that platform $A$ matches demand in such a way that $\texttt{OPT} \times \frac{2}{1+2}= \frac{2}{3} \cdot \texttt{OPT}$ supply nodes get consumed (in expectation). Given that platform $A$ faces enough demand to consume twice as much supply (in expectation) as it has at its disposal, it is natural to assume that by doubling its available supply it would be able to double the demand it is able to serve. Indeed, with twice as much supply the objective of the DLP doubles to $2 \cdot \texttt{OPT}$. However, in doing so the platform creates a balanced market ($\market=1$), in which the CR of its matching algorithm is just $1/2$. Therefore, it may occur that the added supply only translates into $2\cdot \texttt{OPT} \times\frac{1}{1+1}=\texttt{OPT}$ supply nodes being consumed, i.e., rather than doubling the number of successful matches, it only increases the latter by 50\%. Of course, these considerations rely on the arrivals being adversarial. A $2$-undersupplied platform $B$ that faces stochastic arrivals would be able to create $(1-e^{-2}) \cdot \texttt{OPT} \approx 0.86 \cdot  \texttt{OPT}$ successful matches in expectation, yet by doubling its supply that number would only increase by about $47\%$ to $2(1-1/e) \cdot \texttt{OPT}\approx 1.26 \cdot \texttt{OPT}$. Thus, in both arrival settings, disappointment might ensue when a doubling of supply does not yield twice as many successful matches. In Section~\ref{sec: overview_results}, we adapt these insights to a setting where decision-makers face a linear cost for the supply on the platform.

\subsubsection*{Related Literature.}
\label{ssec: related_literature}

{
We survey three separate areas: online matching with stochastic rewards under (i) adversarial and (ii) stochastic arrivals, and (iii) the effect of imbalance in other matching problems.

\emph{\textbf{Online bipartite matching with stochastic rewards (stochastic arrivals)}.}  From an algorithmic point of view, our work falls into the literature on online bipartite matching with stochastic rewards. Introduced by \citet{mehta2012online}, papers in this stream generalize the online bipartite matching model \citep{karp1990optimal} by allowing matches to succeed or fail with a known consumption probability. \citet{mehta2012online} demonstrated that the famous RANKING algorithm of \citet{karp1990optimal} achieves a CR of 0.534 when all consumption probabilities are identical and introduced another algorithm with a CR of 0.567 for identical and vanishingly small probabilities; \citet{mehta2014online} show that 0.534 is also achievable for unequal but vanishingly small probabilities. Recently, \citet{huang2020online} designed an algorithm -- based on the randomized online primal dual framework of \citet{devanur2013randomized} -- with a CR of~$0.576$ for the case of identical and vanishingly small probabilities. The CR for the more general setting of unequal but vanishingly small probabilities was improved by \citet{goyal2023online} to $0.596$ and then by \citet{huang2023online} to  $0.611$ (the latter two works compare to a different benchmark).  For delayed algorithms (which they call non-adaptive), \citet{mehta2012online} showed that no algorithm can obtain a CR better than $1/2$, and \citet{mehta2014online} devised an algorithm that matches that CR under general probabilities. {For the more general problem of Online Submodular Welfare Maximization, \citet{udwani2024online} shows that adaptivity to stochastic rewards does not offer improved CRs.} 

Our results for the adversarial setting (Theorem \ref{theorem: adversarial CR LB}, Proposition \ref{prop: adversarial CR UB}) provide new CRs and corresponding upper bounds that are parameterized by $\market$. In particular, we provide a delayed algorithm that achieves a CR of $\max\{1/(\market+1),\market/(\market+1)\}$ on $\market$-imbalanced instances and show that no delayed algorithm can do better.
}

{
\emph{\textbf{Online bipartite matching with stochastic rewards (stochastic arrivals)}.} 
Our stochastic arrival setting aligns with the \textit{known I.I.D.} framework introduced by \citet{feldman2009online} for online bipartite matching. In contrast to the adversarial setting, at each time step, the arriving demand node is drawn independently from previous arrivals from a given known distribution of demand types. 
\citet{feldman2009online}, \citet{manshadi2012online}, and  \citet{jaillet2014online} respectively developed algorithms with CRs of 0.67, 0.702, and 0.706. However, these algorithms do not consider stochastic rewards and, moreover, they are based on comparisons to more intricate benchmarks. 
\citet{haeupler2011online} developed an algorithm for an edge-weighted setting with stochastic arrivals an deterministic consumption; their algorithm achieves a CR of $.667$. Despite not modeling stochastic rewards, their algorithm is non-adaptive in a way that facilitates extending it to a delayed algorithm in the setting with stochastic rewards. Indeed, \citet{brubach2020online} build on \citet{haeupler2011online} to derive a delayed algorithm with CR $(1-1/e)$ for the stochastic rewards setting. Our corresponding bound (\Cref{theorem: stochastic CR LB}) is  $\max \left\{\frac{1-e^{-\kappa}}{\kappa}, 1-e^{-\kappa} \right\}$, which coincides with theirs when $\market = 1$. 
We remark that \citet{brubach2020online} also provide a CR of 0.702 for the stochastic reward setting, but this does not hold for a delayed algorithm and requires additional assumptions.
}

{
\textbf{\emph{Imbalance in other matching problems.}} We know of few papers that analyze the effect of imbalance on matching markets. The celebrated result of \citet{ashlagi2017unbalanced} focused on (a tiny) imbalance in stable matching, identifying that it yields a significant advantage for the short side; related settings have since been studied in the literature \citep{kanoria2021matching,cai2022short}. 
However, the stable matching setting differs significantly from online bipartite matching, both with respect to objective and constraints. More closely related to our work, \citet{ma2020group}  consider a steady-state fairness objective and prove that online matching algorithms may reach asymptotic optimality as the imbalance between supply or demand increases. Our results also characterize how algorithms benefit from imbalance, but we focus on a slightly different setting (standard online stochastic matching problem) and (ii) rather than providing only an asymptotic result, we characterize  explicitly the CR as a function of the level of imbalance. In that regard, our work is closer to \citet{abolhassani2022online} who also provide a parameterized guarantee, though only for over-, not undersupplied, instances and only when the imbalance is an integer-multiple.

\section{Model and Preliminaries}
\label{sec: model}

We consider a central decision maker (platform) facing an online matching problem that evolves over a finite horizon of $T\geq1$ discrete time periods. The platform knows the set of supply vertices $\supply$ in advance. Demand vertices arrive one at a time in an on-line fashion, i.e., in each period $t = 1, \ldots, T$, a vertex $v_t$ arrives and the decision maker observes its type. The type of a demand vertex $v$ is defined through the vector ${(\prob_{u,v})}_{u\in \supply} \in [0,1]^{|\supply|}$,  where $\prob_{u,v}$ indicates the probability that a demand-side vertex $v \in V$ chooses to consume supply vertex $u \in \supply$ upon the decision maker matching $v$ to $u$. To simplify notation, and when clear from context, we usually refer to demand nodes by their arrival time $t$ and use $v_t$ to refer to the type of the $t$-th arrival. Throughout, we denote vectors in bold.

{At the beginning of period $t$, the platform observes the type of vertex $t$ and must choose whether to match vertex $t$ irrevocably to a supply-side vertex $u$ or leave $t$ permanently unmatched.
If the platform matches $t$ to $u$, $t$ chooses to consume $u$, \emph{and}  $u$ is still available (i.e., it has not been consumed before), then we refer to the match of $t$ and $u$ as a \textit{successful match} or a \textit{consumption} and refer to $u$ as a \emph{consumed supply node}. Otherwise, the match is not successful (either because $t$ chooses not to consume or because $u$ has already been consumed before) and no consumption occurs. That is, our model of stochastic rewards distinguishes between the platform matching demand node $t$ to $u$ and that match being successful (i.e., $t$ consuming $u$). }

\textbf{\emph{Algorithms.}}
The decision-maker's objective is to maximize the expected amount of supply that demand nodes consume over $T$ periods. As discussed before, our results aim to characterize the performance of \emph{delayed algorithms},
i.e., ones that only observe whether a match succeeded at the end of the time horizon. 
Formally, an algorithm $\alg$ describes a non-anticipating (possibly randomized) sequence of match decisions $\{
\mathbf{Y}_t
\}^T_{t
=1} $ for every $v_t\in V$, where $\mathbf{Y}_t
=(Y_{u,t})_{u \in \supply}$  and $Y_{u,t}$ is a random indicator variable that is~$1$ if and only if the arrival $v_t$ is matched to $u$ (regardless of whether the match is successful). Moreover, we define the random indicator variable $X_{u,t}$ that is $1$ if and only if the algorithm $\alg$  matches $t$ to $u$ \emph{and} the match is successful; we similarly define $\mathbf{X}_t
=(X_{u,t})_{u \in \supply}$. In other words, the indicator $Y_{u,t}=1$ stands for arrival $t$  having been matched to $u$ whereas $X_{u,t} = 1$ means that the match was successful, i.e., (i) $t$ was matched to $u$, (ii) $t$ chose to consume $u$, and (iii) $u$ has not been previously consumed.\footnote{As any feasible assignment algorithm matches each $t$ to at most one supply vertex $u\in \supply$ we have that, when $Y_{u,t}=1$, then  all remaining variables $X_{u',t}$ and $Y_{u',t}$ for $u' \in \supply\setminus\{u\}$ must be zero.} In our delayed setting, an algorithm is restricted to make decisions that \textit{cannot be based on the realization} of the random variables $(\mathbf{X}_{t'})_{t'=1}^{t-1}$. Rather, decisions can only be based on the previous matching decisions up to time $t$, $(\mathbf{Y}_{t'})_{t'=1}^{t-1}$. Formally, the stochastic process is $(\mathbf{Y}_{t})_{t=1}^{T}$ is $\mathcal{H}_t$-adapted,
where $\{\mathcal{H}_t\}_{t=1}^T$ is the filtration describing the algorithm's history. That is, letting $s$ be the algorithm's random
seed, we have $\mathcal{H}_1=\sigma(s, v_1)$ and  $\mathcal{H}_t=\sigma(\mathcal{H}_{t-1}, \mathbf{Y}_{t-1},  v_t)$ for every $t \in \{2, \ldots, t\}$,  where we use $v_t$ to denote the type of the $t$-th arrival. Throughout, we abuse notation by using $\alg$ to refer to both an algorithm and its value, when the intended meaning is clear from the context.

\subsubsection*{Arrival models,  competitive ratio, and imbalance.}
We derive results for two commonly studied demand arrival models: stochastic and adversarial. In the \textit{stochastic arrival model}, which we sometimes refer to as the stochastic setting, we are given a finite set $V$ of possible demand types and a probability distribution $\{p_{v}\}_{v \in V}$ over these types, from which each $v_t$ is drawn independently.  In the \textit{adversarial arrival model}, we are given a finite set $V$ of possible demand types and the probabilities of consumption are set to be equal, i.e. $\prob_{u,t} \in \{ 0, \prob\}$ for $\prob \in (0,1]$ where $\prob$ can be instance-dependent but we do not make any additional assumptions over the arrivals, other than their types being in a generic set $V$ and that for any demand node $v\in V$ there exists $u\in $ such that $\prob_{u,v}=\prob$. For either arrival model, we call the set of $u$ such that $\prob_{u,v}>0$ the neighborhood of $v$ and denote it by  $N(v) = \{u \in \supply : \prob_{u,v} > 0 \}$. For the adversarial model, we also define the set $E=\{(u,t): u\in \supply, t\in V, \mu_{u,t}=\mu\}$.

Our results rely on the notion of CRs to evaluate the performance of delayed algorithms. To that end, we start by defining an instance. An instance $\inst$ is composed of the set $\supply$ of supply vertices $\supply$, a set of potential demand nodes $V$, and the consumption probabilities $\{ \prob_{u,v}\}_{u \in \supply, v \in V} \subseteq (0,1)^{\supply \times V}$ associated with each type $v \in V$.  In the stochastic arrivals settings, we assume that the instance also contains the probability distribution $\{p_v\}_{v \in V}$ over the demand-side arrival types. We slightly abuse notation and use $\inst$ to denote an instance for either arrival model.  
In the stochastic arrivals setting, the \textit{expected value} of an algorithm $\alg$ given instance $\inst$, $\mathbb{E}[\alg(\inst)]$ refers to the expected value of $\alg$ under stochastic arrivals where the expectation is taken over the sequence of arrivals, the (possibly random) decisions of the algorithm, and the consumption realizations. On the other hand,  in the adversarial setting, we instead define the expected value of an algorithm given an instance and an arrival sequence. That is, the expected value of an algorithm $\alg$ given instance $\inst$ and an arrival sequence  {$\sigmabf=(1, \ldots, T)$} is denoted by $\mathbb{E}[\alg(\inst, \sigmabf)]$, where the expectation is taken over the possibly random decisions of the algorithm and the consumption realizations. 

\begin{figure}[h]
\centering
\tikzset{every picture/.style={line width=0.75pt}} 

\begin{tikzpicture}[x=0.75pt,y=0.75pt,yscale=-.55,xscale=.9]

\draw   (237.14,68.73) .. controls (237.17,77.56) and (230.04,84.76) .. (221.2,84.79) .. controls (212.37,84.83) and (205.17,77.7) .. (205.14,68.86) .. controls (205.1,60.02) and (212.23,52.83) .. (221.07,52.79) .. controls (229.91,52.76) and (237.1,59.89) .. (237.14,68.73) -- cycle ;
\draw   (236.86,32.73) .. controls (236.9,41.56) and (229.76,48.76) .. (220.93,48.79) .. controls (212.09,48.83) and (204.9,41.7) .. (204.86,32.86) .. controls (204.82,24.03) and (211.96,16.83) .. (220.79,16.8) .. controls (229.63,16.76) and (236.82,23.89) .. (236.86,32.73) -- cycle ;
\draw   (108.14,64.73) .. controls (108.17,73.56) and (101.04,80.76) .. (92.2,80.79) .. controls (83.37,80.83) and (76.17,73.7) .. (76.14,64.86) .. controls (76.1,56.02) and (83.23,48.83) .. (92.07,48.79) .. controls (100.91,48.76) and (108.1,55.89) .. (108.14,64.73) -- cycle ;
\draw   (236.86,140.73) .. controls (236.9,149.56) and (229.76,156.76) .. (220.93,156.79) .. controls (212.09,156.83) and (204.9,149.7) .. (204.86,140.86) .. controls (204.82,132.03) and (211.96,124.83) .. (220.79,124.8) .. controls (229.63,124.76) and (236.82,131.89) .. (236.86,140.73) -- cycle ;
\draw   (237.14,104.73) .. controls (237.17,113.56) and (230.04,120.76) .. (221.2,120.79) .. controls (212.37,120.83) and (205.17,113.7) .. (205.14,104.86) .. controls (205.1,96.02) and (212.23,88.83) .. (221.07,88.79) .. controls (229.91,88.76) and (237.1,95.89) .. (237.14,104.73) -- cycle ;
\draw   (108.14,110.73) .. controls (108.17,119.56) and (101.04,126.76) .. (92.2,126.79) .. controls (83.37,126.83) and (76.17,119.7) .. (76.14,110.86) .. controls (76.1,102.02) and (83.23,94.83) .. (92.07,94.79) .. controls (100.91,94.76) and (108.1,101.89) .. (108.14,110.73) -- cycle ;
\draw    (108.14,64.73) -- (204.86,32.86) ;
\draw    (108.14,64.73) -- (205.14,68.86) ;
\draw    (108.14,110.73) -- (205.14,68.86) ;
\draw    (108.14,110.73) -- (205.14,104.86) ;
\draw    (108.14,110.73) -- (204.86,140.86) ;

\draw (213,59) node [anchor=north west][inner sep=0.75pt]   [align=left] {{$v_2$}};
\draw (213,24) node [anchor=north west][inner sep=0.75pt]   [align=left] {{$v_1$}};
\draw (213,132) node [anchor=north west][inner sep=0.75pt]   [align=left] {{$v_4$}};
\draw (213,95) node [anchor=north west][inner sep=0.75pt]   [align=left] {{$v_3$}};

\draw (84,57) node [anchor=north west][inner sep=0.75pt]   [align=left] {{$u_1$}};
\draw (84,102) node [anchor=north west][inner sep=0.75pt]   [align=left] {{$u_2$}};

\end{tikzpicture}
\caption{Depending on $\prob\in\{1/4,.5,1\}$ the instance is $1/2$-oversupplied, balanced, or $2$-undersupplied.}\label{fig:example_imbalance}
\end{figure}
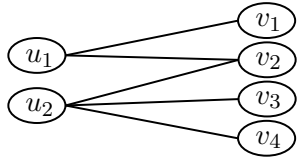

We measure an algorithm's performance through its CR relative to the following LP benchmark (with $\market=1$),\footnote{As it is common in the literature on online matching with stochastic rewards, we  compare the performance of $\alg$ with the LPs to maintain tractability; towards the end of this section we discuss alternative benchmark choices.}
and let $x_{u, v}$ capture the fraction of demand node $v$ that is assigned to supply node~$u$.
\begin{minipage}[t]{0.47\textwidth}
\centering $\offIkappa$ \textbf{(Adversarial)}
\begin{align}
    \max_{\mathbf{x}\geq \mathbf{0}} \quad & \sum_{(u, t) \in E}{ \prob \cdot x_{u, t}} \notag\\
    \textrm{s.t.} \quad & \sum_{t: (u,t) \in E } \prob \cdot  x_{u, t} \leq \market, \quad && \forall u \in \supply \label{adversarial_lp: kappa-constraint}\\
    & \sum_{u: (u,t) \in E} x_{u, t} \leq 1, \quad && \forall t \in [T] \label{adversarial_lp: demand_constraint}
\end{align}
\end{minipage}%
\hspace{.4em}
\begin{minipage}[t]{0.47\textwidth}
\centering $\offIkappa$ \textbf{(Stochastic)}
\begin{align}
     \max_{\mathbf{x}\geq \mathbf{0}} \quad & \sum_{u\in \supply, v\in V}{ \prob_{u,v} \cdot x_{u, v}} \notag\\
    \textrm{s.t.} \quad & \sum_{v \in V} \prob_{u,v} \cdot  x_{u, v} \leq \market, \quad && \forall u \in \supply \label{stochastic_lp: kappa-constraint}\\ 
    & \sum_{u \in \supply} x_{u, v} \leq T p_v, \quad && \forall v \in V \label{stochastic_lp: demand_constraint}
\end{align}
\end{minipage}

Superficially, the parametrization of these LPs by $\market$ resembles the $b$-matching formulations in the literature \citep{kalyanasundaram2000optimal}, though we use them to define imbalance below. Throughout, we denote by $\offI(\inst) = \offI(\inst, 1)$ the optimal value of the respective linear program for either arrival setting (note that the adversarial arrival model requires no dependence on~$\sigmabf$, because the linear program is unaffected by the arrival order of the demand nodes). Each summand in the objective function of $\offI(\inst)$, in either LP, rewards a solution fractionally for assigning $v$ to $u$ with a coefficient that captures the probability that $v$ would consume $u$. With $\market=1$, constraints (\ref{adversarial_lp: kappa-constraint}) and (\ref{stochastic_lp: kappa-constraint}) capture that each supply node $u$ can be consumed at most once. Lastly, constraint (\ref{adversarial_lp: demand_constraint}) stipulates that any demand node $v$ can be matched at most once under adversarial arrivals; similarly, in the stochastic setting, the upper bound in (\ref{stochastic_lp: demand_constraint})  is the expected number of times a node of type $v$ appears in the arrival sequence. In \Cref{proof: offline is upper bound} we prove the following lemma asserting that the LPs are valid upper bounds for any algorithm. 

\begin{lemma}
    \label{lemma: offline is upper bound}
    Let $\inst$ be an instance and $\alg$ be any algorithm with $(X_{u,t})_{u \in \supply, t \in [T]}$ the sequence of matching decisions that where successful. Then, in either setting,
    $\offI(\inst) \geq  \mathbb E\left[\sum_{ (u,t) \in \supply \times [T]} X_{u,t} \right]$.
\end{lemma}

We now define the following notion of CR as our performance metric.

\begin{definition}
Let $\comp \in [0,1]$ and $\mathcal{F}$ be a family of instances for the stochastic arrivals setting; an algorithm $\alg$ is said to be \textit{$\comp$-competitive for family $\mathcal{F}$} if
$        \inf_{\inst \in \mathcal{F}}{\mathbb E[\alg(\inst)] }/{\offI(\inst)} \geq \comp$.
    Similarly, for a family of instances in the adversarial setting, $\mathcal{F}$,  an $\alg$ is $\comp$-competitive for $\mathcal{F}$ if
        $\inf_{\inst \in \mathcal{F}} \inf_{\sigmabf} \frac{\mathbb E[\alg(\inst, \sigmabf)] }{\offI(\inst)} \geq \comp$.
\end{definition}

When the family of instances $\mathcal{F}$ is not specified, we refer to $\mathcal{F}$ as the set of all possible instances. However, we usually parameterize instances based on the demand-supply imbalance as we discuss next.

\textbf{\emph{Oversupplied and Undersupplied Instances.}}
Our findings include sharp characterizations of how supply-demand imbalances affect the performance of online algorithms. The example in Figure \ref{fig:example_imbalance} motivates our characterization of imbalanced instances. The figure displays an instance with 2 supply nodes and 4 demand nodes; with  $\prob_{uv}=\frac{1}{2} \, \forall u\in \supply,v\in V$, it is natural to think of this instance as a \emph{balanced} one. Indeed, there exists an assignment ($v_1,v_2$ to $u_1$ and $v_3,v_4$ to $u_2$) under which the expected demand realizing at each supply node is exactly 1 and there exists no assignment under which the expected demand realizing at each supply node is greater 1. In contrast, with $\prob_{uv}=1 \, \forall u\in \supply,v\in V$, we would view the instance with the same nodes as undersupplied; in particular, with the same assignment as before we would now be assigning 2 (expected) demand to every supply node. In that regard, it is natural to think of the instance with with $\prob_{uv}=1 \, \forall u\in \supply,v\in V$ as 2-undersupplied, which captures that if we had twice as much supply at each supply node the instance would be balanced. Finally, suppose we had  $\prob_{uv}=\frac{1}{4} \, \forall u\in \supply,v\in V$; then, the same assignment as before we would now be assigning just $1/2$ (expected) demand to every supply node. In that regard, it is natural to think of the instance with with $\prob_{uv}=1/4 \, \forall u\in \supply,v\in V$ as $1/2$-oversupplied, which captures that there exists an assignment under which the expected demand at each supply node is $1/2$ and there exists no assignment under which the expected demand at each supply node is larger.

We generalize these intuitions to arbitrary graphs through the parameter $\market$ in the definition of $\offIkappa$. The resulting LPs

connect with our example from Figure \ref{fig:example_imbalance} as follows: in our 2-undersupplied instance we find that $\offIkappa=\kappa \cdot \offI(\inst)$ for $\kappa=2$, i.e., as we increase $\kappa$ to loosen the supply constraint the objective increases, least up to $\kappa=2$, linearly in $\kappa$. In contrast, in our $1/2$-oversupplied instance, as we decrease $\kappa$ to tighten the supply constraint the objective remains the same down to $\kappa=1/2$. To capture this phenomenon we define $\kappa$ under- and oversupplied instances as follows: 
\begin{definition} \label{def: under/oversupplied}
We call an instance $\inst$: 
(1) \textit{$\market$-undersupplied} if $\market = \max\{ \overline{\market} \geq 1 \colon \offI(\inst, \overline{\market}) = \overline{\market} \cdot \offI(\inst)\}$, (2) \textit{$\market$-oversupplied} if $
            \market = \min\{ \overline{\market} \leq 1 \colon \offI(\inst,\overline{\market}) = \offI(\inst)\}$, and (3) 
         \textit{balanced} if it is both $1$-oversupplied and $1$-undersupplied.
    \end{definition}
Observe that $\market=1$ is an element of the sets in the definition, guaranteeing that the sets are non-empty. We make the technical assumption that the value of the maximum for $\market$-undersupplied instances and the minimum for $\market$-oversupplied instances, as defined in Definition \ref{def: under/oversupplied}, is attained at rational numbers. Intuitively, an instance is $\market$-undersupplied if the respective offline problem with capacity $\market$ has exactly $\market$ times the same value as the problem with capacity $1$. An instance is $\market$-oversupplied if the offline problem with capacity $\market$ is equal to the offline problem with capacity~1.

A direct consequence of \Cref{lemma: offline is upper bound} is that $\offIkappa/\market$ provides an upper bound to the expected value of any online algorithm on a $\market$-undersupplied instance. Analogously, if an instance is $\market$-oversupplied for $\kappa < 1$, then $\offIkappa$ provides an upper bound to the expected value of any online algorithm.

\subsection*{Discussion of modeling assumption}

\paragraph{Definition of imbalance.}

The advantage of our imbalance definition (\Cref{def: under/oversupplied}) is its intuitive  interpretability; on undersupplied instances one may achieve a higher objective by increasing the capacity of all the supply nodes and on oversupplied instances one maintains the objective when reducing the capacity of all the supply nodes. However, a practical limitation of our definition is that it may suggest that many instances are balanced even though they seem (intuitively) imbalanced. In particular, this holds true when a graph is locally under- and locally oversupplied: \Cref{def: under/oversupplied} treats such graphs as balanced as they are neither \emph{globally} undersupplied nor oversupplied. To address this limitation, we propose the following alternative definition of imbalance:

\begin{definition}
    \label{def: new_over_undersupplied}
    Let $\inst$ be a given instance. For $\market \geq 1$ we say that the tuple  $(\undersup, \oversup)$ is a $\market$-imbalanced pair if it fulfills the following conditions:
    \begin{enumerate}[(i)]
        \item (Partition) $\undersup$ and $\oversup$ are disjoint and $\undersup \cup \oversup = \supply$.
        \item (Undersupplied set) If the capacity of every node in $\undersup$ gets augmented by $\market$, then the offline solution grows by $(\market-1) | \undersup|$. In other words,
        \begin{align*}
             &\offI(\inst) + (\market-1)|\undersup| = \max_{\mathbf{x}\geq \mathbf{0}} \quad  \sum_{(u, t) \in E}{ \prob \cdot x_{u, t}} \tag{$\offI^{\undersup, \mathcal O}(\inst, \market)$} \label{prob: offlineUndersupplied}\\
    \mathrm{s.t.} \quad & \sum_{t: (u,t) \in E } \prob \cdot  x_{u, t} \leq \market \quad  \forall u \in \undersup, 
    \quad \sum_{t: (u,t) \in E } \prob \cdot  x_{u, t} \leq 1 \quad \forall u \in \oversup,
    \quad \sum_{u: (u,t) \in E} x_{u, t} \leq 1 \quad  \forall t \in [T].
        \end{align*}
        
        \item (Oversupplied set) There exists an optimal solution $\mathbf{x}^*$ to $\offI(\inst)$ that uses at most $1/\market$ capacity at nodes in $\oversup$. In other words, there exists $\mathbf{x}^*$ optimal solution to $\offI(\inst)$ that satisfies $\prob \sum_{t: (u,t)} x_{u,t}^* \leq 1/\market$ for all $u \in \mathcal O$.
    \end{enumerate}
\end{definition}

Under this new definition, we can provide an analogous CR result for adversarial arrivals as long as the algorithm knows which nodes are in $\oversup$ and which nodes are in $\undersup$.

In practice, even under adversarial arrivals, platforms can likely use historical data to determine which nodes belong to the under- or oversupplied sets (\Cref{sec: alternative_def} contains a method to compute these for offline instances). For instance, in spatial settings (ridehailing, volunteer matching), some neighborhoods are consistently over- or undersupplied during certain hours. Similarly, in refugee resettlement some locations offer consistently higher employment probabilities and matching optimization would thus consistently use all of their capacity, but not that at locations with lower probabilities. Though we do not capture the intricacies of these applications, these examples illustrate practical settings in which platforms would reasonably know $\oversup$ and $\undersup$.

\paragraph{Choice of Benchmark.} 
{Most of the recent literature on online matching with stochastic rewards has used tighter benchmarks than the standard matching LP, $\offI$. With these benchmarks, including the Configuration LP by \citet{huang2020online}, the Path Based Program by \citet{goyal2023online} for adversarial arrivals, and the benchmark of \citet{brubach2020online}  for stochastic arrivals, authors have been able to achieve better CRs than possible vis-a-vis $\offI$. However, for our purposes, these benchmarks come with a significant disadvantage: as they add exponentially many constraints to the linear program, it is challenging to compute their objective value even for instances with relatively few nodes and edges (see Appendix \ref{ssec: pbp_infeasible}). This added complexity makes it impractical to compute the tighter upper bound on an instance's objective values, whereas $\offI$ offers a computationally feasible approach. Moreover, due to its compactness, $\offI$ yields a more interpretable benchmark that lends easily lends itself to the study of imbalanced instances. Motivated by these advantages, we choose to follow the more traditional benchmark $\offI$ to evaluate the performance of our algorithms, thereby sacrificing potentially tighter CRs for gains in tractability and interpretability. Nonetheless, some of our numerical results (see Appendix \ref{ssec:imbalanced}) also explore the performance of some of these newer algorithms under imbalanced instances.

\section{Overview of Main Results}
\label{sec: overview_results}

In this section, we provide an overview of how demand-supply imbalances yield improved CRs for matching algorithms and how these improvements in turn affect a platform's optimal demand-supply imbalance. 

We begin with an algorithmically trivial example to illustrate the impact of imbalance:  an instance $\inst$ that is \textit{complete bipartite} with $L$ supply nodes. For some fixed positive integer $n$ we have a time horizon of~$T = n\cdot L$ arrivals with consumption probabilities equalling $\prob_{u,t} = 1/n$ for each arrival $t$ and each supply node~$u$. Note that $\offI(\inst)$ is able to match all the demand and gets a value of $L$. In contrast, consider the delayed algorithm $\alg_D$ that assigns each arrival to the supply node with the fewest demand assigned so far; this ensures in this instance that over the course of the horizon every supply node gets matched to exactly $n$ demand nodes. The expected reward of this algorithm in this particular instance is given by
\begin{equation}
\label{eq: alg_D in overview_results}
\mathbb E[\alg_D(\inst)] = \sum_{u \in \supply} \left( 1 - \left(1-\frac{1}{n}\right)^n \right)= L \left(  1 - \left(1-\frac{1}{n}\right)^n \right).
\end{equation}
As $n \rightarrow \infty$ the CR of this algorithm converges to $(1-1/e)$ in this instance, from above. Furthermore, no delayed algorithm can do better in this instance (see  \Cref{section: impossibility}). 

However, this upper bound on the CR occurs in a balanced instance, as is common in the online matching literature. When instances are imbalanced, the CR of the algorithms may improve. To illustrate, we can generalize the above instance to have $\market \cdot nL$ arrivals\footnote{By our assumption on $\market$, we can express $\market = p/q$ with $p,q \in \mathbb Z_{>0}$. When $n$ is a multiple of $q$, then $\market \cdot nL$ is an integer and the instance is well-defined for all values of $L$. In particular, we take $\market$ as small as $1/n$.} and compare the performance of $\alg_D$ against that of our offline benchmark. 
Intuitively, as the market imbalance ratio~$\market$ increases beyond~$1$, the benchmark remains constant at~$L$. However, the performance of $\alg_D$ improves due to a higher number of arrivals compared to the original instance, which leads to more supply nodes being consumed in expectation. Conversely, when $\market$ is less than~$1$, both the algorithm's and the benchmark's value decrease, but the latter decreases faster which allows the  expected reward of the algorithm to approach that of the {benchmark} (as $\market\to0$). This occurs because each supply node is matched with fewer demand nodes, which leads to the number of supply nodes consumed being approximately equal to $(\kappa nL)\cdot \mu$.\footnote{When $\market < 1$, the probability that a node will be consumed in the imbalanced instance is given by $ 1 - \left(1-\frac{1}{n}\right)^{\market n}$. For $\market = 1/n$, this simplifies to $1/n$, matching the value for a single match in the offline benchmark.} Though we focus on delayed algorithms, similar improvements occur for algorithms that adapt to consumption realizations.\footnote{{As these algorithms have more information, our lower bounds on what is achievable by delayed algorithms naturally extend. At the end of this section, we discuss the regimes in which imbalance may be so large that these lower bounds become higher than the best-known lower bounds for existing adaptive algorithms.}}

Motivated by this special case, we investigate the achievable CRs for delayed algorithms on imbalanced instances. Our main guarantee for the setting with adversarial arrivals is as follows:
\begin{theorem}
    \label{theorem: adversarial CR LB}
    Under the adversarial arrival model, there exists a delayed algorithm with a CR of at least $\max \left \{ \frac{1}{1+\market}, \frac{\market}{1+\market} \right\}$ for the sets of $\market$-undersupplied and $\market$-oversupplied instances.
\end{theorem}

We establish \Cref{theorem: adversarial CR LB} by lower bounding, for any $\market$-oversupplied or $\market$-undersupplied instance $\inst$ and arrival sequence $\sigmabf$, the CR of a greedy algorithm ($\greedy$) that assigns each arriving demand node to an adjacent supply node that has been assigned the fewest demand nodes so far. We lower-bound the ratio $\mathbb E[\greedy(\inst, \sigmabf)]/\offIkappa$ using a factor-revealing linear program; the formulation of this linear program is obtained by leveraging the techniques developed in the analysis of the BALANCE algorithm \cite{mehta2007adwords}  to derive constraints that involve the operational dynamics of both $\greedy$ and $\offI$ on a given instance and arrival sequence. We then characterize a feasible solution to the dual of the factor-revealing linear program, which establishes a lower bound on the CR. Further details on these techniques, as well as the full proof of  the result can be found in Section~\ref{sec3} and its associated appendix. 

For our generalized imbalance definition (\Cref{def: new_over_undersupplied}), we can provide an analogous guarantee that requires the algorithm to have knowledge of the sets $(\undersup, \oversup)$ that form a $\market$-imbalanced pair.\footnote{We omit an analogous guarantee for the stochastic arrivals since the lower bound in \Cref{theorem: stochastic CR LB} is not symmetric in~$\market$. Hence, a similar guarantee would depend on the relative size of the over- and undersupplied components of the instance.} 
This result applies to a much wider class of instances than \Cref{theorem: adversarial CR LB} given that it allows for the sets~$\undersup$ and $\mathcal{O}$ to both be non-empty. In fact, if one of these sets is empty (i.e., $\undersup = \supply$ or $\mathcal{O} = \supply$), the instance is either $\market$-undersupplied or $1/\market$-oversupplied and we recover the guarantee from \Cref{theorem: adversarial CR LB}. 
The proof of this result is deferred to \Cref{sec: alternative_def}.

\begin{theorem}
    \label{theo: CRguarantee_extended_def}
    Let $\inst$ be an instance that admits a $\market$-imbalanced pair $(\undersup, \oversup)$. Then, there exists a delayed algorithm with $(\undersup, \oversup)$ as an input and a CR of at least $\frac{\market}{1+\market}$ in the adversarial arrival model.
\end{theorem}
Our algorithm leverages knowledge of $(\undersup, \oversup)$ to always prioritize matching demand to oversupplied rather than undersupplied nodes. The main technical ingredients of \Cref{sec: alternative_def} show that the DLP solution also does not ever assign a demand node, that is adjacent to an oversupplied node, to an undersupplied node. With that, the CR is a quick corollary of \Cref{theorem: adversarial CR LB}.

Under stochastic arrivals we use the parametrization in \Cref{def: under/oversupplied} to obtain the following: 
\begin{theorem}
\label{theorem: stochastic CR LB}
Under the stochastic arrival model, there exists a delayed algorithm that is at least $\max \left\{\frac{1-e^{-\market}}{\market}, 1-e^{-\market} \right\}$-competitive for the sets of $\market$-undersupplied and $\market$-oversupplied instances.
\end{theorem}

To establish \Cref{theorem: stochastic CR LB}, we adapt the algorithm described by \citet{brubach2020online} to account for the imbalance $\market$. Our algorithm computes an optimal solution to $\offIkappa$ for a given instance; then, when a demand node arrives, it is matched to an adjacent supply node with a probability proportional to this solution. Using the techniques in \citet{brubach2020online} and \Cref{def: under/oversupplied}, we establish a lower bound on the probability that each individual edge results in a successful match, which  allows us to  demonstrate the desired bound on the CR. Further details and the full proof of the result are in Section~\ref{sec: stochasting_setting}.

{We demonstrate the tightness of our CR bounds by presenting impossibility results that establish upper bounds on the performance of any delayed algorithm within the given arrival model.}

\begin{proposition}
    \label{prop: adversarial CR UB}
    In the adversarial arrival model, no delayed algorithm can achieve a competitive ratio strictly greater than $\max \left \{ \frac{1}{1+\market}, \frac{\market}{1+\market} \right\}$  for the sets of $\market$-undersupplied and $\market$-oversupplied instances.
\end{proposition}

We prove \Cref{prop: adversarial CR UB} in \Cref{section: impossibility} via an upper-triangular instance \citep{karp1990optimal}, on which no delayed  algorithm can exceed our bound. Extending upper-triangular instances to our setting requires us to carefully calibrate various parameters to ensure that the instance is well-defined and of the right imbalance. We highlight that similar instances were previously considered by \citet{mehta2012online} to show that no delayed-feedback algorithm can achieve a CR greater than $1/2$, a result that we generalize to imbalanced instances. Under stochastic arrivals we similarly obtain the following upper bound (proof in \Cref{section: impossibility}): 

\begin{proposition}
    \label{prop: stochastic CR UB}
    In stochastic settings, no delayed algorithm can achieve a competitive ratio strictly greater than $\max \left\{\frac{1-e^{-\market}}{\market}, 1-e^{-\market} \right\}$  for the sets of $\market$-undersupplied and $\market$-oversupplied instances.
\end{proposition}

Our proof builds on the instance from the beginning of this section (with $L=1$).

\subsubsection*{Interpretation of results.} In practice, platforms can influence the level of imbalance under which their matching algorithms operate through various strategic levers. Increasing supply typically incurs additional costs but, at the same time, the ability to capture more demand can significantly boost revenue.
Consequently, platforms must carefully balance the potential costs of expanding supply against the revenue benefits derived from serving a larger customer base. As the resulting market imbalance affects the downstream performance of the matching algorithm, it is in the platform's interest to incorporate the expected downstream performance into its upstream inventory level decision.  In the remainder of this section we explore the interplay between the upstream decision and the downstream performance and compare our bounds to existing CRs.

\emph{Overage and underage costs consideration.}
We begin by combining our CR bounds on the performance of a downstream matching algorithm with an inventory decision made upstream. To do so, we assume that two platforms $A$ and $B$ are in a setting with the following features:
\begin{itemize}
    \item In $\offI(\inst)$, supply and demand are perfectly matched, i.e., there exists an optimal solution $\mathbf{x}$ to $\offI(\inst)$ in which  $\forall u\in \supply:\; \sum_{t:(u,t)\in E}\prob_{u,t} x_{u,t}=1$, and $\forall t\in[T]:\;\sum_{u:(u,t)\in E}x_{u,t}=1$;
    \item Denoting by $d=\sum_{t:(u,t)\in E}\prob_{u,t}x_{u,t}$, the expected (realizing) demand, we assume that the inventory decision of each platform has a proportional effect on the imbalance of $\offIkappa$ downstream, i.e., purchasing $d/\market$ units of inventory leads to a $\market$-imbalanced instance. E.g., buying $2d$ units of inventory, which is twice as much as  $\offI(\inst)$ has at its disposal, yields a $1/2$-oversupplied instance; 
    \item The platforms face demand under which the CRs of their matching algorithms hold tightly, i.e., in an $\market$ over- or under-supplied instance, the number of matches that platforms A and B respectively expect is
    \begin{align*}
    & \max \left \{ \frac{1}{1+\market}, \frac{\market}{1+\market} \right\} \cdot \left(d \cdot \min \left\{1, \frac{1}{\market}\right\} \right)=\frac{d}{1+\kappa}\\
    \text{and}\quad &\max \left\{\frac{1-e^{-\market}}{\market}, 1-e^{-\market} \right\} \cdot \left(d \cdot \min \left\{1, \frac{1}{\market}\right\} \right) = \frac{(1-e^{-\market}) d }{\market}.
    \end{align*} 
\item Each platform generates  $r$ revenue for each successful match and incurs a cost of $c<r$ for each unit of inventory.
\end{itemize}

The first two structural assumptions hold true in the family of  worst-case instances (see \Cref{section: impossibility}) we construct and the last holds true asymptotically when taking the limit in that family. For the revenue and cost, we adopt standard assumptions from inventory management. In practice, these assumptions are unlikely to jointly hold exactly true, but nonetheless they capture an abstraction that allows us to cleanly connect our CR results downstream to the upstream decision. 

The platforms need to decide how much inventory to order, relative to the demand.\footnote{We implicitly assume here that the platform only has one global supply lever as opposed to having the ability to add supply in a targeted manner. For volunteer or ridehailing platforms this is a reasonable assumption as platforms need to target potential new volunteers/drivers without access to detailed personal information. }
We let $d\inv$ denote the inventory order; based on the above, $\inv>1$ corresponds to creating a $1/\inv$-oversupplied instance and $\inv<1$ corresponds to creating a $1/\inv$-undersupplied instance. The choice of $\inv$ is based on (i) the cost of inventory $d\inv\cdot c$ and (ii) the revenue that the inventory translates to (via the matching algorithm's performance), i.e., $\frac{dr}{1+1/\inv}=dr\frac{\inv}{\inv+1}$ for platform A and $(1-e^{-1/\inv})\inv \cdot dr$ for platform B. Combining the cost and the revenue, we obtain the two platform's profit functions:
\begin{equation*}
\text{Profit}_{A}(\inv) = d\left(r \cdot \frac{\inv}{1+\inv} - \inv c\right),\quad
    \text{Profit}_{B}(\inv) = d\left((1-e^{-1/\inv})\inv \cdot r - \inv c\right).
\end{equation*}
As both profit functions are concave, we can identify the optimal $\inv$ for each platform by taking the derivative and setting it to 0 in each case. This yields for $A$ and $B$ respectively (see Appendix \ref{appendix: underage-costs}):
\begin{equation*}
        \inv_A^* = \frac{\frac{r-c}{r} - 1 + \sqrt{1 - \frac{r-c}{r}}}{1 - \frac{r-c}{r}} = \sqrt{\left( 1 - \frac{r-c}{r}\right)^{-1}} - 1 \quad \text{and}\quad \inv_B^* =  - \left(W_{-1}\left(-\frac{1}{e} \cdot \frac{r-c}{r}\right)+1 \right)^{-1},
    \end{equation*}
    where $W_{-1
    }$ stands for the negative branch of the Lambert $W$ function.\footnote{$W$ is the multivalued function that satisfies $x = W(x) \exp(W(x))$ for any $x$. For $x\in [-\frac{1}{e},0)$, there are two possible real values for $W(x)$; the branch with $W(x) \leq -1$ is referred to as the negative branch $W_{-1}(x)$.} The quantities reveal that, similar to the newsvendor setting, the optimal stocking decision, relative to the demand $d$, depends only on the ratio $(r-c)/r$.  To explore the behavior of the optimal stocking levels, we plot these levels (Figure \ref{fig:optimal_stocking_levels}) for both adversarial and stochastic settings. 
We first find a pronounced initial peak in stocking levels for stochastic arrivals (platform B); this is due to the exponential term in the profit expression and contrasts with a more gradual increase for adversarial arrivals. Second, as $(r-c)/r$ increases, the optimal stocking levels for platform B transitions to a linear behavior; in particular, the optimal stocking levels for both platforms intersect around a ratio $(r-c)/r$ of around 0.7839. {Lastly, it is worth comparing these decisions to naive upstream supply decisions that do not take into account the effect that imbalance has on the CR: assuming an $\alpha$-approximation, the profit-maximizing quantity would be either $1$ or $0$ with the threshold of $(r-c)/r$ occurring at $1-\alpha$, e.g., with $\alpha=1/2$ (the worst-case upper bound on non-adaptive algorithms in our setting), we would have $\eta_\alpha^*=0$ for $(r-c)/r<1/2$ and $\eta_\alpha^*=1$ for $(r-c)/r\geq 1/2$. In other words, the obtained stocking level could be 0 for $(r-c)/r<1/2$ though it would be profitable to hold inventory; it could be too large for $(r-c)/r\in(.5, .76)$ when $\eta_\alpha^*=1$ though a lower inventory would suffice (given the benefits of imbalance); and it could be too small for $(r-c)/r>.76$ when $\eta_A^*$ and $\eta_B^*$ are both greater 1 because the low cost of supply makes it worthwhile to create greater imbalance, but the non-parameterized CR does not adapt to that regime.
}

\emph{Comparison to adaptive algorithms.} 
    For $(r-c)/r \leq 0.606$ and $(r-c)/r \geq 0.831$, the resulting market imbalance values $\market(\inv)$  fall below 0.736 and exceed 1.358, respectively. In these regimes, our CRs are higher than 0.702 for stochastic instances and 0.576 for adversarial instances (see Figure \ref{fig:adversarial_stochastic_comparison}), which reflects the best-known CRs for (adaptive) algorithms (which require additional assumptions to achieve these CRs). Of course, these algorithms are likely to significantly outperform their CRs on instances that exhibit imbalance; as such, the takeaway of our analysis should not be we provide improved CRs relative to the literature, but rather that imbalance offers itself as a natural parametrization within which stronger CRs are obtainable.

\begin{figure}[h]
    \centering
    \begin{minipage}{0.48\textwidth}
        \centering
        \includegraphics[width=\textwidth]{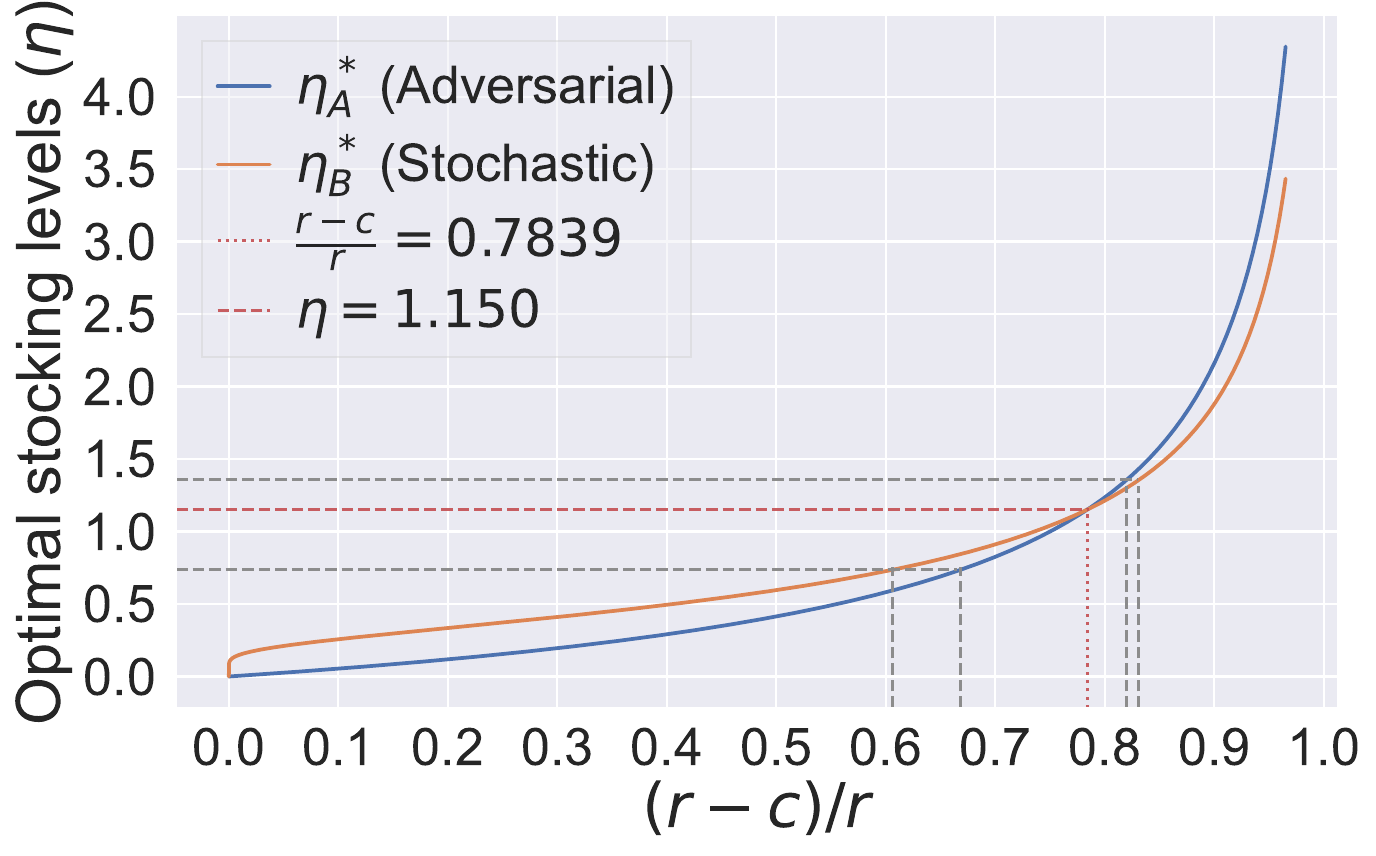}
        \caption{\small{Optimal stocking levels $\inv_A^*$ and $\inv_B^*$ as functions of $(r-c)/r$. The horizontal dotted lines correspond to $\inv\in\{0.736,1.358\}$. The vertical dotted lines correspond to their intersection with the optimal stocking level.}}
        \label{fig:optimal_stocking_levels}
    \end{minipage}\hfill
    \begin{minipage}{0.48\textwidth}
        \centering
        \includegraphics[width=\textwidth]{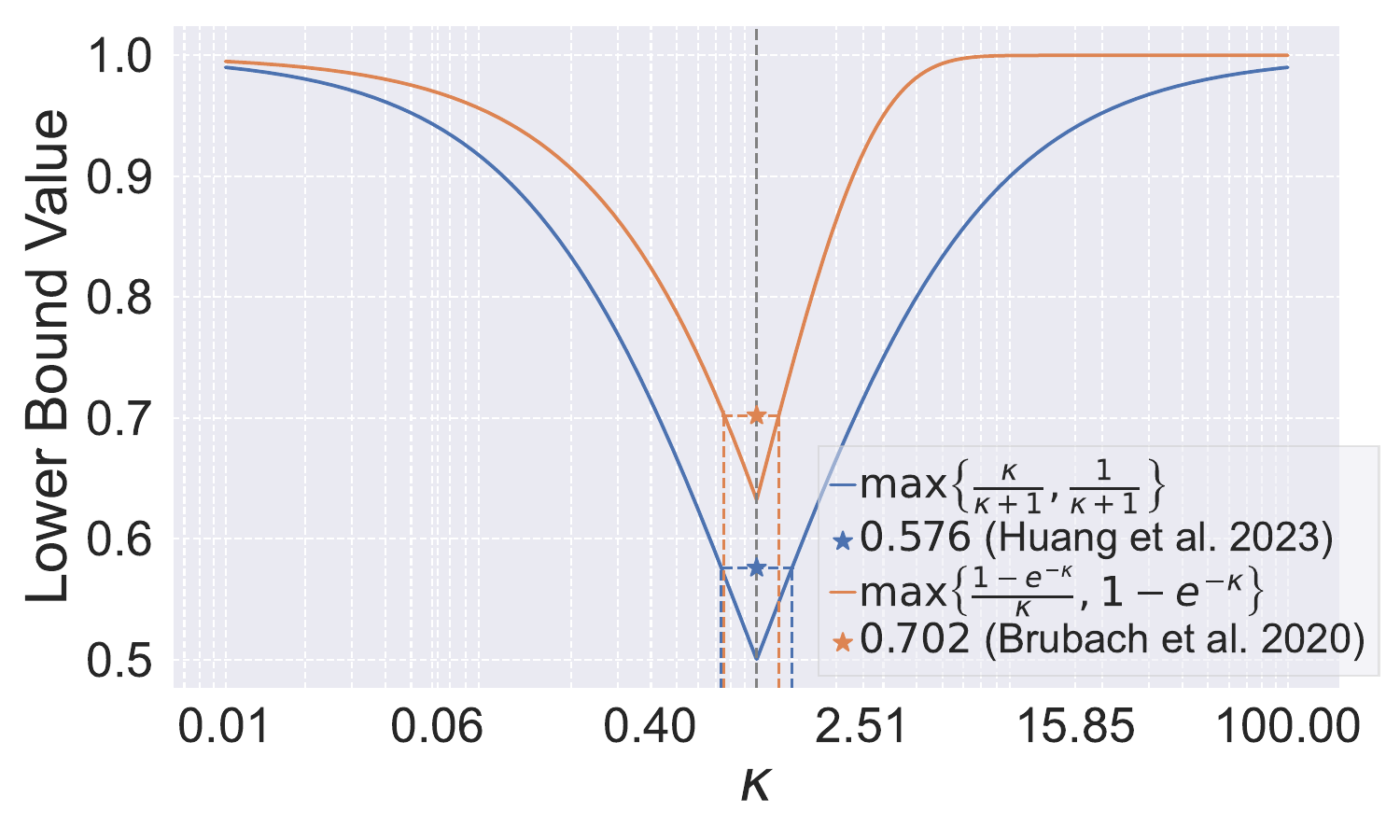}
        \caption{\small{Comparison of our tight CRs for adversarial and stochastic settings across $\market$. The dotted lines intersect with the $x$-axis at $0.736$, $0.754$, $1.210$, and $1.358$ respectively.}}
        \label{fig:adversarial_stochastic_comparison}
        \vfill \hfill
    \end{minipage}
    \vspace{-.15in}
\end{figure}

\section{Adversarial arrivals: lower bound}
\label{sec3}
In this section, we provide a greedy algorithm, $\greedy$  and show that the performance of $\greedy$ can be lower-bounded by the bounds in \Cref{theorem: adversarial CR LB}. The proof consists of four steps. First, following the techniques of \citet{mehta2007adwords, mehta2013online}, we derive inequalities that connect the objectives of $\offIkappa$ and the performance of our greedy algorithm (Lemma~\ref{claim: AdvOfflineOptBound}). Second, we leverage these inequalities to derive a factor-revealing linear program (LP) that bounds for given $\kappa$ and $T$ the worst-case, over all instances $\inst$, ratio between $\offIkappa$ and the performance of the greedy algorithm on $\inst$ (Lemma~    \ref{lemma: DLP lower bounds greedy/off}). By identifying a feasible solution to that LP, we derive a lower bound to the CR (Lemma~\ref{prop: AdvCR}).

We first introduce our greedy algorithm $\greedy$ (Algorithm~\ref{alg: adv_greedy_alg}).  In each period~$t$, $\greedy$ matches the $t$-th arrival to one of the supply nodes in $N(t)$ that have been assigned the fewest demand nodes in the periods up to $t$ (recall that, throughout this section, 
we make the assumption of $\prob_{u,v} \in \{ 0,\prob\}$, for all pairs $(u,v) \in \supply \times V$). As $\greedy$ uses past matching decisions, but not past consumption outcomes, it is a delayed algorithm. 

\RestyleAlgo{ruled}
\begin{algorithm}[ht]
\caption{$\greedy$ .}\label{alg: adv_greedy_alg}
\label{algo: greedy_algorithm_adversarial}
\textbf{Input:} An instance $\inst$, a sequence of arrivals $\sigmabf $. 

\textbf{Output:} Sequence of decisions for the online problem.

For all $u \in \supply$, initialize $n_u^g = 0$.

\For{$t= 1$ \KwTo $T$}{
Observe the arrival $t$, and choose $u^* \in  \argmin_{u \in N(t)} n^g_u$, breaking ties arbitrarily.

Assign arrival $t$ to node $u^*$.

Update the number of matches to $u^*$: $n^g_{u^*} \leftarrow n^g_{u^*} + 1$.
}
\end{algorithm}

With $\greedy$ formally introduced, we now state the following proposition, which implies \Cref{theorem: adversarial CR LB}:

\begin{proposition}
\label{prop: fst_part_adversarial}
Let $\inst$ be any instance and $\sigmabf$ any sequence of arrivals. Then,
    \begin{equation*}
        \frac{\mathbb E[\greedy(\inst, \sigmabf)]}{\offI(\inst)} \quad \geq \quad  \begin{cases}
        \frac{1}{1 + \market} & \text{ if } \inst \text{ is } \market\text{-oversupplied},\\
\frac{\market}{1 + \market} & \text{ if } \inst \text{ is } \market\text{-undersupplied}.
\end{cases}
    \end{equation*}
\end{proposition}

To prove the result we first couple how $\greedy$ acts relative to $\offIkappa$ {for any $\market$}, on any execution. Let $\inst$ be any instance and $\sigmabf = (1, \ldots, T)$ be an arrival sequence. 
Consider an execution of $\greedy$ on $\inst$.\footnote{Without loss of generality, we assume that $\greedy$ breaks ties deterministically which makes $n_u^g$ a deterministic quantity. This can be done, for example, by imposing an order $u_1, \ldots, u_{|\supply|}$ over the supply nodes and breaking ties by selecting the highest ranked supply node according to this order.} Let $n_u^g$ denote the total number of demand nodes that $\greedy$ assigns to $u$ at the end of its execution. For any  $j \in \{0\}\cup[T]$, define the set $A_j = \left\{ u \in \supply:  n_u^g = j \right\}$; that is, $A_j$ is the set of supply nodes to which  $\greedy$ assigns exactly $j$ demand nodes under that run of the algorithm. Define $\rho_j = |A_j|$. {Note that these definitions, with appropriate modifications, mirror those in the slab-type proofs of \citet{mehta2007adwords, mehta2013online} and are similarly employed in \citet{manshadi2022online}.} 
We use these definitions to write the expected objective of $\greedy$ as
\begin{equation*}
    \mathbb E[\greedy(\inst, \sigmabf)] = \sum_{u \in \supply} \left(1-(1-\prob)^{n_u^g} \right) = \sum_{j=0}^T \rho_j \left( 1-(1-\prob)^{j} \right).
\end{equation*}
To connect the greedy algorithm with $\offIkappa$, we denote by $\xbf^*$ an optimal solution to $\offIkappa$, and by $x_u^* = \sum_{t \in [T]: (u, t) \in E} x^*_{u, t}$, the fractional mass of demand that $\xbf^*$ assigns to a given supply node $u \in \supply$. 
\begin{equation*}
    \text{Then, we can re-write}\quad\offIkappa ~ = ~  \prob \sum_{ (u,t) \in E} x^*_{u,t} ~ = ~  \prob \sum_{t=0}^T \sum_{ (u,t) \in E} x^*_{u,t} ~ = ~ \prob \sum_{j=0}^T \sum_{u \in A_j} x_u^*.
\end{equation*}
\begin{equation*}\text{Combining the two expressions, we obtain}\quad
         \frac{\mathbb E[\greedy(\inst, \sigmabf)]}{\offIkappa} ~ = ~ \frac{\sum_{j=0}^T \rho_j \left( 1-(1-\prob)^{j} \right)}{\prob \sum_{j=0}^T \sum_{u \in A_j} x_u^*}.
\end{equation*}
The following lemma (proof in Section \ref{ssec:AdvOfflineOptBound}) uses the feasibility of $\xbf^* \in \offIkappa$ and the behaviour of $\greedy$ to characterize a set of constraints that $\rhobf$ and $\xbf^*$ must jointly satisfy {on a given instance}.

\begin{restatable}{lemma}{AdvOfflineOptBound}
\label{claim: AdvOfflineOptBound}
 Fix any instance $\inst$ and arrival sequence $\sigmabf$.  Let $\xbf^*$ be an optimal solution to $\offIkappa$. Then, the following inequalities hold
  \begin{enumerate}
      \item \textit{Feasibility}. For all $k \in \{0\} \cup [T]$, we have that
      $\sum_{u \in A_k} x_u^* \leq  \frac{\market}{\prob}\cdot \rho_k.
      $
      \item \textit{Greedy bounds}. For all $s \in \{0\}\cup[T-1]$, we have that
      \begin{equation*}
    \sum_{k=0}^s \sum_{u \in A_k} x_u^* \leq \sum_{j=0}^s j \cdot \rho_j +  (s+1)  \sum_{j=s+1}^{T} \rho_j
\quad \text{and}\quad
    \sum_{k=0}^T \sum_{u \in A_k} x_u^* \leq \sum_{j=0}^T j \cdot \rho_j.
\end{equation*}
\end{enumerate}
\end{restatable}

The feasibility inequality is based on \eqref{adversarial_lp: kappa-constraint} and $\mathbf{x}^*$ being feasible for $\offIkappa$. For the greedy bounds we use how $\greedy$ assigns each demand node and the constraints \eqref{adversarial_lp: demand_constraint}, which explains why these  inequalities have no direct dependence on $\market$. 

Leveraging the inequalities in Lemma~\ref{claim: AdvOfflineOptBound}, we construct a series of optimization problems to provide a lower-bound 
on ${\mathbb E[\greedy(\inst, \sigmabf)]}/{\offIkappa}$.
Specifically, given an instance $\inst$ with consumption probabilities~$\prob$, and $|V|=|\sigmabf|=T$, we define the following linear maximization problem $\DLP$ as:

\begin{align*}
\DLP = \max_{\alphabf, \betabf, \delta} \quad & \delta\\
\textrm{s.t.} \quad & \sum_{s=j}^{T} \alpha_s + \beta_j + \delta \leq 0, & \forall j \in \{ 0\} \cup [T]  \tag{D-1} \label{cons: D1}\\
\quad & 0 \leq \frac{1- (1-\prob)^j}{\prob} + \frac{\market}{\prob} \beta_j + j \sum_{s=j}^{T} \alpha_s + \sum_{s=0}^{j-1} (s+1) \alpha_s, & \forall j \in [T] \tag{D-2} \label{cons: D2}\\
& 0 \leq \frac{\market}{\prob} \beta_0, \tag{D-3} \label{cons: D3}\\
& \alphabf \leq \mathbf{0}, \tag{D-4} \label{cons: D4}\\
& \betabf \leq \mathbf{0}. \tag{D-5} \label{cons: D5}
\end{align*}
In \Cref{proof: DLP lower bounds greedy/off} we show that $\DLP$ lower bounds $\mathbb E[\greedy(\inst, \sigmabf)]/\offIkappa$:

\begin{restatable}{lemma}{AdvDLPLBGreedy}
    \label{lemma: DLP lower bounds greedy/off}
    Let $\inst$ be an instance  with consumption probabilities equal to $\mu$, and $|\sigmabf|=T$. Then, for any $\market>0$, we have that  $\mathbb E[\greedy(\inst, \sigmabf)]/\offIkappa$ is lower-bounded by the value of $\DLP$.
\end{restatable}

Our next result provides, for any $T \in \mathbb N$ and $\prob, \market > 0$, an explicit lower bound on $\DLP$. 

\begin{restatable}{lemma}{AdvCR}
    \label{prop: AdvCR}
     Let $T \in \mathbb N$ and $\prob, \market > 0$.
     \begin{enumerate}
         \item Suppose that $T \geq \lceil \frac{\market}{\prob} \rceil + 1$. Then $\DLP$ is lower bounded by
         \begin{equation*}
         \frac{1}{1 + \market} + \frac{\market}{1 + \market} \left( \frac{\market(1-\prob)}{\market +\prob} \right)^{T - \lceil \frac{\market}{\prob}\rceil - 1}- \frac{(1-\prob)^T + (1-\prob)^{T - \lceil \frac{\market}{\prob}\rceil - 1} (\prob ( \lceil \frac{\market}{\prob} \rceil+1) -1)}{ \prob \lceil \frac{\market}{\prob}\rceil} \left(\frac{\market}{\market + \prob} \right)^{T - \lceil \frac{\market}{\prob}\rceil}.
     \end{equation*}
        \item Suppose that $T <\lceil \frac{\market}{\prob} \rceil + 1$. Then $\DLP$ is lower bounded by $
            \left({1- (1-\prob)^T}\right)/{T \prob}$.
     \end{enumerate}
\end{restatable}

We prove the lemma in \Cref{proof: AdvCR} by constructing a feasible solution to $\DLP$. 
We require one additional auxiliary result (proof in \Cref{proofs: AdvTheoremLastIneq}) to prove the main result of this section.
\begin{restatable}{lemma}{AdvTheoremLastIneq}
    \label{lemma: AdvTheoremLastIneq}
    For any $\prob \in (0,1]$ and $\market > 0$ we have $
            1 - \frac{\prob \lceil \frac{\market}{\prob}\rceil}{1+\market} \geq (1-\prob)^{\lceil \frac{\market}{\prob} \rceil}$.
\end{restatable}

\noindent{\textbf{Proof of \Cref{prop: fst_part_adversarial}.}
Let $\inst$ be an instance with consumption probability equal to $\mu$ and with $|\sigmabf| =T$. Suppose that $\inst$ is $\market$-undersupplied.
Then, we have that:  
\begin{equation*}
    \frac{\mathbb E[\greedy(\inst, \sigmabf)]}{\offI(\inst)} ~=~ \market \cdot \frac{\mathbb E[\greedy(\inst, \sigmabf)]}{\offIkappa} ~\geq~ \market \cdot \DLP,
\end{equation*}
where the equality follows from the definition of $\market$-undersupplied the inequality follows from the fact that $\DLP$ lower bounds $\mathbb E[\greedy(\inst, \sigmabf)]/\offIkappa$ (\Cref{lemma: DLP lower bounds greedy/off}).  

Observe that, if $\inst$ is  $\market$-oversupplied instances, then we have that: 
\begin{equation*}
    \frac{\mathbb E[\greedy(\inst, \sigmabf)]}{\offI(\inst)} ~=~ \frac{\mathbb E[\greedy(\inst, \sigmabf)]}{\offIkappa} ~\geq~ \DLP,
\end{equation*}
where the equality follows from the definition of $\market$-oversupplied the inequality follows from \Cref{lemma: DLP lower bounds greedy/off}.

Therefore, to prove \Cref{prop: fst_part_adversarial} for both under- and oversupplied instances, it suffices to show that $\DLP \geq \frac{1}{1+\market}$ for any values of $T$, $\prob$ and $\market$. We will establish this by considering two cases. 

Suppose that $T < \lceil \frac{\market}{\prob} \rceil + 1$. Then, using  \Cref{prop: AdvCR} and the fact that that $(1-(1-\prob)^T)/T$ is decreasing in $T$, we obtain
\begin{equation*}
     \DLP ~\geq~ \frac{1- (1-\prob)^T}{T \prob} ~ \geq ~ \inf_{T \leq \lceil \frac{\market}{\prob} \rceil} \frac{1- (1-\prob)^T}{T \prob} ~= ~ \frac{1- (1-\prob)^{\lceil \frac{\market}{\prob} \rceil}}{\lceil \frac{\market}{\prob} \rceil \prob} ~\geq~  \frac{1}{\market + 1},
\end{equation*}
where the last inequality follows from rearranging terms in the inequality stated in \Cref{lemma: AdvTheoremLastIneq}. 

It remains to prove that $\DLP \geq \frac{1}{\market +1 }$ for $T \geq \lceil \frac{\market}{\prob} \rceil + 1$. Again, by \Cref{prop: AdvCR}, we have that: 
\small
    \begin{align*}
        && \frac{1}{1 + \market} + \frac{\market}{1 + \market} \left( \frac{\market(1-\prob)}{\market +\prob} \right)^{T - \lceil \frac{\market}{\prob}\rceil - 1} - \frac{(1-\prob)^T + (1-\prob)^{T - \lceil \frac{\market}{\prob}\rceil - 1} (\prob ( \lceil \frac{\market}{\prob} \rceil+1) -1)}{ \prob \lceil \frac{\market}{\prob}\rceil} \left(\frac{\market}{\market + \prob} \right)^{T - \lceil \frac{\market}{\prob}\rceil} \geq \frac{1}{1+\market}\\
        \Leftrightarrow && \frac{\market}{1 + \market} \left( \frac{\market(1-\prob)}{\market +\prob} \right)^{T - \lceil \frac{\market}{\prob}\rceil - 1} \geq \frac{(1-\prob)^T + (1-\prob)^{T - \lceil \frac{\market}{\prob}\rceil - 1} (\prob ( \lceil \frac{\market}{\prob} \rceil+1) -1)}{ \prob \lceil \frac{\market}{\prob}\rceil} \left(\frac{\market}{\market + \prob} \right)^{T - \lceil \frac{\market}{\prob}\rceil}.
    \end{align*}
    \normalsize
    Dividing through by $(\frac{\market}{\market + \prob})^{T - \lceil \frac{\market}{\prob} \rceil}$ and multiplying through by $\prob \lceil \frac{\market}{\prob}\rceil$ we can reduce this inequality to
    \begin{align*}
        & & \frac{\prob \lceil \frac{\market}{\prob}\rceil}{1+\market} (1-\prob)^{T - \lceil \frac{\market}{\prob}\rceil - 1} (\market +\prob) &\geq (1-\prob)^T + (1-\prob)^{T - \lceil \frac{\market}{\prob}\rceil - 1} (\prob ( \lceil \frac{\market}{\prob} \rceil+1) -1)\\
        &\Leftrightarrow & \frac{ \prob \lceil \frac{\market}{\prob}\rceil}{1+\market} (\market +\prob) &\geq (1-\prob)^{\lceil \frac{\market}{\prob} \rceil + 1} + \prob ( \lceil \frac{\market}{\prob} \rceil+1) -1\\
        &\Leftrightarrow & \frac{\prob \lceil \frac{\market}{\prob}\rceil (\prob - 1)}{1+\market} + 1-\prob &\geq  (1-\prob)^{\lceil \frac{\market}{\prob} \rceil + 1}\\
        &\Leftrightarrow & 1 - \frac{\prob \lceil \frac{\market}{\prob}\rceil}{1+\market} &\geq (1-\prob)^{\lceil \frac{\market}{\prob} \rceil},
    \end{align*}
    which is equivalent to the inequality stated in \Cref{lemma: AdvTheoremLastIneq}. Therefore, the result follows.  
\hfill\qed

\section{Stochastic arrivals: lower bound}
\label{sec: stochasting_setting}

In this section, we prove \Cref{theorem: stochastic CR LB} via a derivation of the stochastic matching ($\mathrm{SM}$) algorithm proposed by \citet{brubach2016new}. In their approach, after solving the LP benchmark and obtaining the optimal solution $\xbf$, they match an arriving demand node $v$ to a node $u \in N(v)$ with probability~$\frac{x_{u,v}}{T p_v}$. We extend their algorithm and proof to accommodate imbalanced instances.
\BlankLine

\RestyleAlgo{ruled}
\begin{algorithm}[H]
\label{alg: sm}
\SetKwInOut{Input}{input}\SetKwInOut{Output}{output}
\SetKwData{Left}{left}\SetKwData{This}{this}\SetKwData{Up}{up}
\SetKwFunction{Union}{Union}\SetKwFunction{FindCompress}{FindCompress}
\SetKwInOut{Input}{Input}\SetKwInOut{Output}{Output}
\Input{An instance $\inst$ and $\market > 0$.}
\Output{A set of offline decisions for the delayed problem.}
\BlankLine

\caption{$\mathrm{SM}$}

  Compute an optimal solution $\xbf[\market]$ to $\offIkappa$.

  When a demand of type $v$ arrives, assign $v$ to any supply node $u \in N(v)$ with probability $\frac{x_{u,v}[\market]}{T p_v}$.

\end{algorithm}
\BlankLine

\begin{proposition}
\label{prop: fst_part_stochastic}
Given an instance $\inst$ and $\market>0$, denote by $\mathrm{SM}(\inst, \market)$ the reward of Algorithm~\ref{alg: sm} in a single execution. Then, we have that
    \begin{equation*}
        \frac{\mathbb E[\mathrm{SM}(\inst, \market)]}{\offI(\inst)} ~ \geq ~ \begin{cases}
            \frac{1-e^{-\market}}{\market}& \text{ if $\inst$ is $\kappa$-oversupplied,}\\
            1-e^{-\market} & \text{ if $\inst$ is $\kappa$-undersupplied.}
        \end{cases}
    \end{equation*}
\end{proposition}

\noindent{\textbf{Proof of \Cref{prop: fst_part_stochastic}.}
    Consider any instance $\inst$ and $\market > 0$ as inputs to $\mathrm{SM}(\inst, \market)$. We define the following:
    \begin{itemize}
        \item[-] $A(u,t)$ is the event that demand arrival $t$ consumes supply node $u$, i.e., that $\mathrm{SM}(\inst, \market)$ matches arrival~$t$ to $u$, that $u$ has not been consumed in periods $1,\ldots,t-1$, and that $t$ chooses to consume $u$;
        \item[-] $B(u,t)$ is the event that $u$ has not been consumed before the $t$-th arrival, i.e., that arrivals $1,\ldots,t-1$ do not consume node $u$;
        \item[-] $C(u,t)$ is the event that  $\mathrm{SM}(\inst, \market)$ matches arrival $t$ to $u$ and $t$ chooses to consume $u$ (regardless of whether that is possible given that $u$ may have been consumed previously).
    \end{itemize}
    With these events defined, we observe that $A(u,t) \subseteq C(u,t)$. Then, from the feasibility of $\xbf^*[\market]$, we derive 
    \begin{equation*}
        \mathbb P(A(u,t)) ~\leq ~ \mathbb P(C(u,t)) ~ = ~ \sum_{v \in V} p_v \cdot \frac{x^*_{u,v}[\kappa]}{T p_v} \cdot \prob_{u,v}  ~\leq ~ \frac{\kappa}{T}.
    \end{equation*}
    Denoting for any event $X$ the complement of $X$ as $\overline{X}$, we obtain that $\mathbb P( \overline{A(u,t)}) \geq \mathbb P( \overline{C(u,t)}) \geq (1-\frac{\kappa}{T})$. 
    \begin{align*}
    \text{Hence,}\quad
        \mathbb P(B(u,t)) ~ = ~\mathbb P(\cap_{i=1}^{t-1}  \overline{A(u,i)}) ~ \geq ~ \mathbb P(\cap_{i=1}^{t-1}  \overline{C(u,i)}) ~ = ~ \prod_{i=1}^{t-1} \mathbb P(\overline{C(u,i)}) ~ \geq ~ \left(1- \frac{\kappa}{T} \right)^{t-1}.
    \end{align*}
    Lastly, for any $u\in \supply$ and $v\in V$, consider the event $D_{uv}$ that a node of type $v$ consumes the node $u$ in any period. We can bound the probability of $D_{uv}$ as
    \begin{align*}
        \mathbb P[D_{uv}] &= \sum_{t=1}^T \mathbb P[\text{arrival $t$ is of type $v$}, B(u,t), \text{$\mathrm{SM}(\inst, \market)$ matches $t$ to $u$, $t$ chooses to consume $u$}]\\
        &\geq \sum_{t=1}^T p_v \cdot \left(1- \frac{\kappa}{T} \right)^{t-1} \cdot \frac{x^*_{u,v}[\kappa]}{T p_v} \cdot \prob_{u,v}
         = \frac{1}{T} \cdot \frac{1 - (1-\kappa/T)^T}{\kappa/T} \cdot x^*_{u,v}[\kappa] \prob_{u,v}\\
        & \geq (1-e^{-\kappa}) \cdot \frac{x^*_{u,v}[\kappa ] \prob_{u,v}}{\kappa}.
    \end{align*}
    Summing over all $D_{uv}$ we conclude
    \begin{align*}
        \mathbb E[\mathrm{SM}(\inst, \market)]=\sum_{u\in \supply, v\in V} \mathbb P[D_{uv}] &\geq \frac{1-e^{-\market}}{\market} \cdot \offIkappa\\
        &= \begin{cases}
            (1-e^{-\market}) \cdot \offI(\inst) & \text{if $\inst$ is $\kappa$-undersupplied,}\\
            \frac{1-e^{-\market}}{\market} \cdot \offI(\inst) & \text{if $\inst$ is $\kappa$-oversupplied.}\quad\Halmos
        \end{cases}
    \end{align*}

\section{Numerical experiments}
\label{sec: numerical_exp}
In this part of our analysis, we investigate the imbalance of real-world matching markets and how it may affect the CR of $\greedy$. We rely on data from a volunteer platform that matches individual providers with individual demand requests (to ensure anonymity, we omit a description of the nature of the requests); the nature of the platform is such that (i) the platform centrally matches supply and demand, (ii) upon arrival of a node t, the edges between t and the supply nodes present are known to the platform (demand comes with a latitude/longitude and each supply node has a declared radius within which they are willing to fulfill demand; moreover, each demand/supply may have particular restrictions which are known to the platform), and (iii) when a demand node is matched to a supply node, the platform does not immediately learn whether or not the supplier actually fulfills the desired request (we model this through the delayed realization). 

In practice, supply renews on our platform with heterogeneous periodicities as some suppliers volunteer on a monthly, a bi-weekly, a weekly or a one-off basis. For the sake of our analysis here, we ignore this replenishment and instead focus on a 2-week horizon in 2023, wherein we consider a static set of supply nodes that we assume are each available to fulfill at most one demand request. We obtain the set of supply nodes by including each supplier that was available to fulfill at least one demand request during the time interval of interest. For simplicity, we model each supplier to have a radius of 5mi and ignore the additional constraints that may arise. 

With respect to demand, we include the requests that came onto the platform during the 2-week horizon; thus, our analysis does not include unfulfilled demand requests that may have remained on the platform from previous weeks. For the arrival order of the demand requests, we rely on the same ordering in which requests occurred historically. 

In total, we use our data to create 69 graphs (many of which are not connected), on which we then compute 6,900 distinct values of $\offI(\inst)$, $\greedy(\inst)$, and $\market$ by varying $\mu$ to take on 100 different values from 1/1000 to 1, i.e., $\mu\in\{ 10^{-3} + i \cdot \frac{1-10^{-3}}{100} \, : \, i \in \{0, \ldots, 100\}\}$.  We discuss in \Cref{ssec: computation_kappa_instance} how we compute $\market$ for a given instance (based on \Cref{def: new_over_undersupplied}). In \Cref{fig:side_by_side_scatter} we show two scatter plots that display the relationship between the empirically observed  ratio $\greedy(\inst)/\offI(\inst)$ and the imbalance as well as the analytically proven bound on the CR as a function of the imbalance; one scatter plot shows all 6,900 values of $\offI(\inst)$, $\greedy(\inst)$ whereas the other only shows those for one particular graph (\Cref{fig: map_graph_instance}). Our analysis gives rise to the following observations.

\begin{figure}[h!]
    \centering
    \begin{minipage}{0.48\textwidth}
        \centering
        \includegraphics[width=\linewidth]{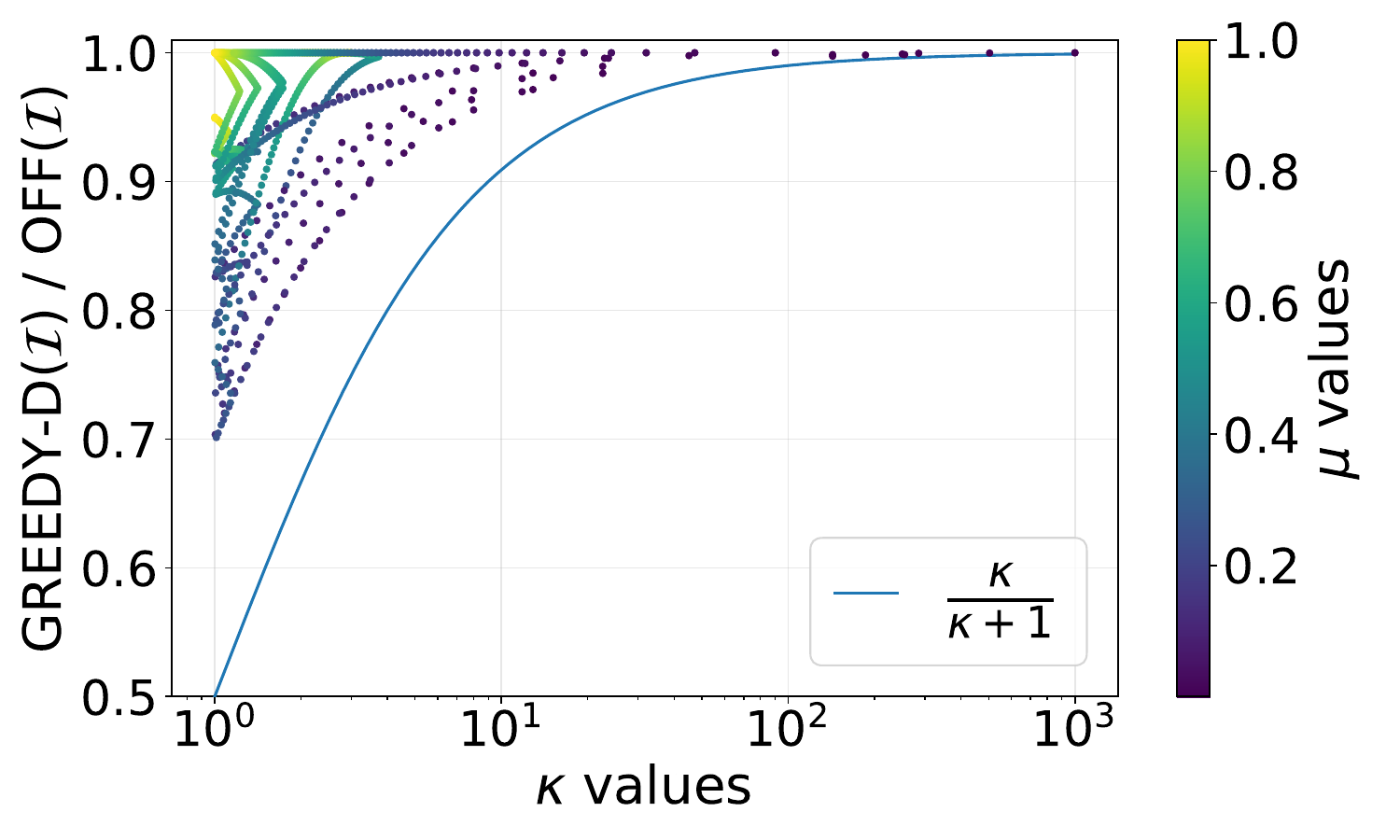}
    \end{minipage}
    \hfill
    \begin{minipage}{0.48\textwidth}
        \centering
        \includegraphics[width=\linewidth]{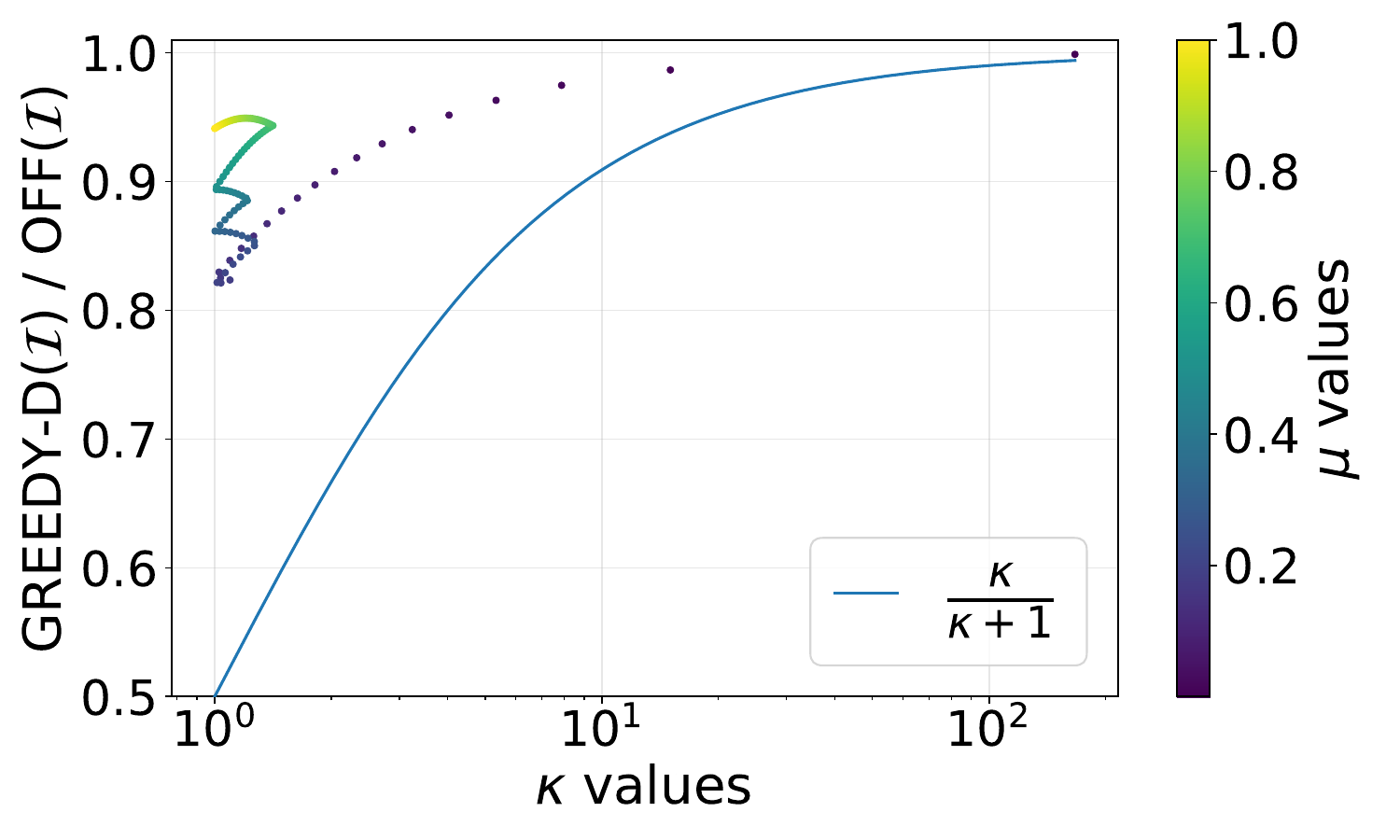}

    \end{minipage}
    \caption{\small{Scatter plots for the CR of $\greedy$: aggregated instances (left) and a particular instance (right).}}
    \label{fig:side_by_side_scatter}
\end{figure}

\noindent\textbf{Greater imbalance yields better CRs.} In line with our theoretical results, we find that the ratio $\greedy(\inst)/\offI(\inst)$ generally improves when instances are more imbalanced. For example, for balanced instances $(\market=1$), we see the ratio go down to $.7$, but for $\market\approx10$, the lowest ratios lie above $.95$. Especially in the range of $\market\leq10$, we see that the improvement in the ``worst instances'' seems to track the trajectory of the  worst-case lower bound fairly well. Of course, as all of these instances are based on real-world data, and not on adversarial upper-triangular instances, we would not expect the lower bound to ever be tight.

\noindent\textbf{Preponderance of balanced instances when} $\prob = 1$. Though our analysis shows a wide range of values of $\market$ that are achieved, we find that most (though not all) instances are at $\market=1$ when~$\prob=1$. This occurs because, as we alluded to before, many of our instances are disconnected and contain sparse subgraphs with very few nodes. For example, the graph in \Cref{fig: map_graph_instance} contains a connected component in which a demand and a supply node have an edge between each other and no other edges. These two nodes by themselves render the instance balanced according to our definition. Of course, in reality, the CR would be higher than $\frac{1}{\market+1}=1/2$: the CR across different connected components is a convex combination of the CR on each individual component with the weight dictated by the value of $\offI(\inst)$ restricted to the component. Thus the pair would be discounted in a way that is not captured in our imbalance definitions (though, in the specific example of a connected component consisting of just one demand node and one supply node, the greedy algorithm makes the optimal decision anyway).

\noindent\textbf{Oscillating values of} $\market$ \textbf{for increasing $\prob$}.  Focusing on a particular choice among our 69 graphs (see \Cref{fig: map_graph_instance}), we find the trajectory of $\market$ to display an oscillating behavior  as $\prob$ increases (see right panel of \Cref{fig:side_by_side_scatter}). This holds due to the following dynamics. For simplicity denote by $(\undersup_\prob, \oversup_\prob)$ the imbalanced pair for a given $\prob$. When $\prob$ is close to $0$, all supply nodes are in $\oversup_\prob$, and increasing $\prob$ yields a decrease in imbalance. As $\prob$ grows larger, the imbalance~$\market$ decreases and hits~1 exactly when the first supply node becomes part of $\undersup_\prob$. Consider the nodes that are now in $\undersup_\prob$: as we increase $\prob$, these nodes become more undersupplied and consequently $\market$ increases again. This continues until some node in $\oversup_\prob$ becomes less oversupplied than nodes in $\undersup_\prob$ are undersupplied; at that point,  increasing $\prob$ lowers $\market$ once again, and this process repeats itself up to the point where $\prob=1$.

\begin{figure}[h]
    \centering
    \begin{minipage}[b]{0.48\textwidth}
        \centering
        \includegraphics[width=\linewidth]{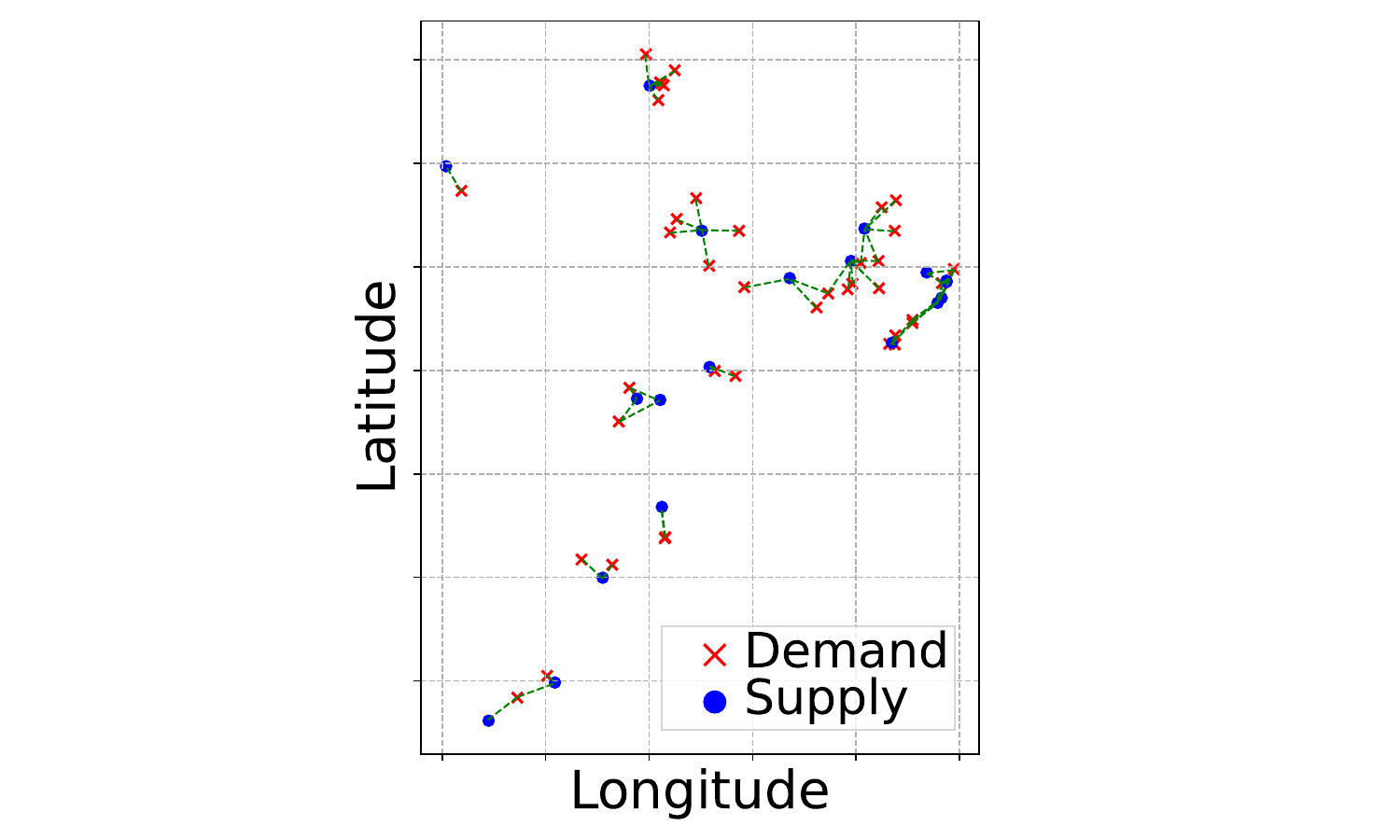}
    \end{minipage}
    \hfill
    \begin{minipage}[b]{0.48\textwidth}
        \centering
        \includegraphics[width=\linewidth]{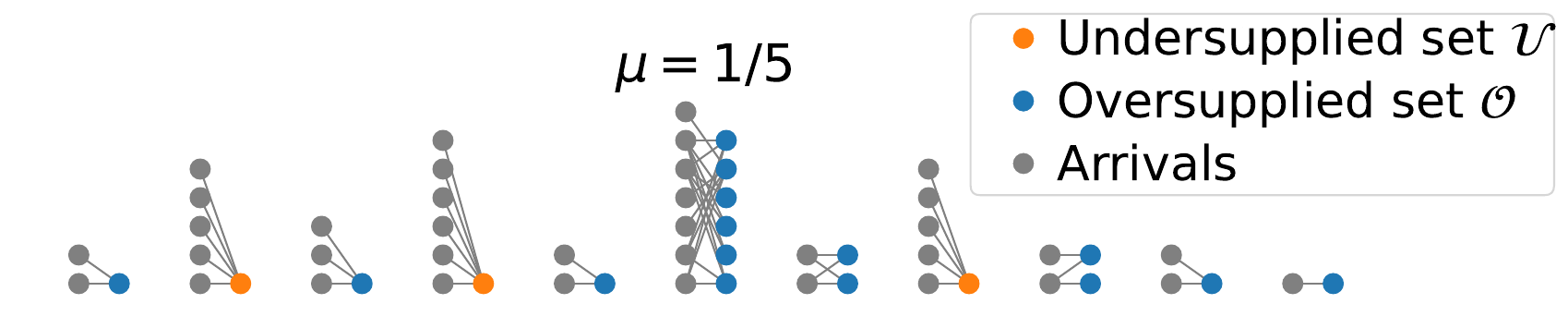}
        \includegraphics[width=\linewidth]{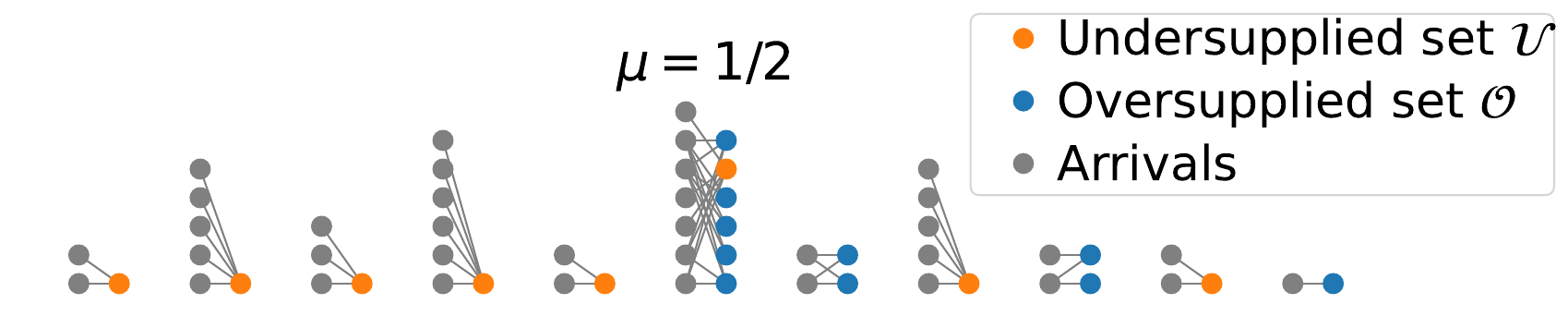}
        \includegraphics[width=\linewidth]{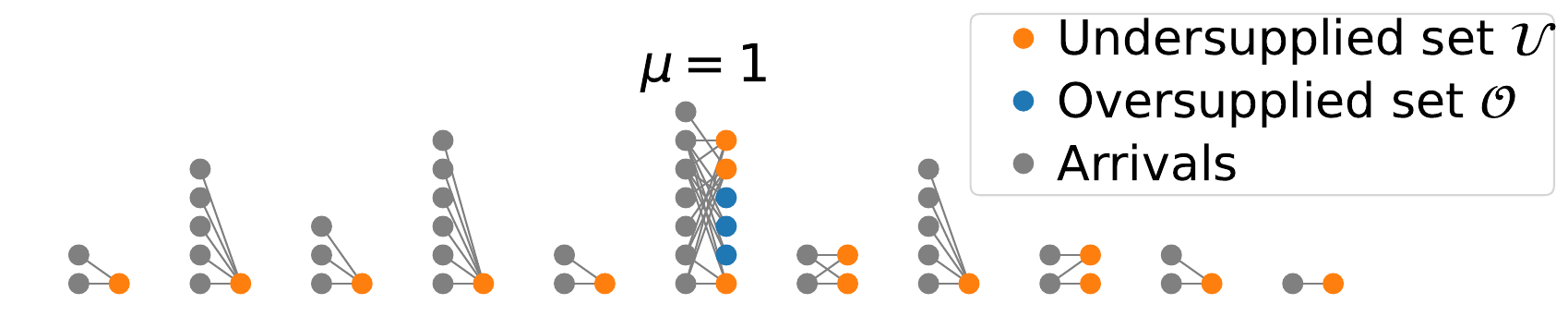}
    \end{minipage}

    \caption{\small{Map/graph of the particular instance (left) and the instance's connected components for different values of $\mu$ (right). 
{Coordinates were randomly perturbed to preserve the privacy of the platform users.}} }
    \label{fig: map_graph_instance}
\end{figure}

\section{Concluding Remarks}
\label{section: conclusion}

Our paper explores the impact of supply and demand imbalances on online bipartite matching with stochastic rewards. We introduce a novel parametrization to measure imbalance and prove that better CRs attainable on imbalanced instances. We further reveal that when contemplating decisions related to the expansion of a platform's demand or supply, the dependence of the CR on the resulting imbalance ought to be taken into account. To the best of our knowledge, ours is the first study to (i) derive CRs that are parameterized by the supply-demand imbalance, (ii) illustrate how these ought to affect an upstream supply planning question.


\bibliographystyle{plainnat}
\bibliography{references}

\appendix

\newpage
\appendix
\section{Doubling supply does not double successful matches}
\label{sec: example_introduction}

This section complements 
\Cref{sec: introduction} and illustrates that there is an $2$-undersupplied instance in which doubling the available supply does not double the number successful matches.
Let $\prob = 1/n$ for a large $n \in \mathbb{N}$. Consider the instance $\inst$ constructed as follows (see \Cref{fig:2-undersupplied_instance_1}): there is one supply node $u_1$ connected to $2/\prob$ demand nodes $v_1, \ldots, v_{2/\prob}$, where each edge exists independently with probability $\prob$. Since there are $2/\prob$ demand nodes $\offI(\inst) = 1$ and $\offI(\inst, 2) = 2 \cdot \offI(\inst)$, i.e., $\inst$ is $2$-undersupplied. Clearly, the optimal algorithm achieves $\text{ALG}(\inst) = 1 - (1-\prob)^{2/\prob} \approx 1 - e^{-2}$.

Now, consider the instance $\inst'$ (see \Cref{fig:2-undersupplied_instance_2}) that is achieved by duplicating the available supply; we consider a second supply node $u_2$ connected to every demand node. By concavity of $1-(1-\prob)^x$, the optimal algorithm in $\inst'$ allocates $1/\prob$ demand nodes to each supply node, which results in $$\text{ALG}(\inst') = 2 \cdot \left(1 - (1-\prob)^{1/\prob}\right) \approx 2 \cdot (1 - e^{-1}) = 2 \cdot (1 - e^{-1}) \cdot \offI(\inst).$$

\begin{figure}[h]
    \centering
    \tikzset{every picture/.style={line width=0.75pt}} 
    \begin{minipage}[b]{0.45\textwidth}
        \centering
        
\begin{tikzpicture}[x=0.75pt,y=0.75pt,yscale=-0.8,xscale=.9]
\draw   (237.14,68.73) .. controls (237.17,77.56) and (230.04,84.76) .. (221.2,84.79) .. controls (212.37,84.83) and (205.17,77.7) .. (205.14,68.86) .. controls (205.1,60.02) and (212.23,52.83) .. (221.07,52.79) .. controls (229.91,52.76) and (237.1,59.89) .. (237.14,68.73) -- cycle ;
\draw   (236.86,32.73) .. controls (236.9,41.56) and (229.76,48.76) .. (220.93,48.79) .. controls (212.09,48.83) and (204.9,41.7) .. (204.86,32.86) .. controls (204.82,24.03) and (211.96,16.83) .. (220.79,16.8) .. controls (229.63,16.76) and (236.82,23.89) .. (236.86,32.73) -- cycle ;
\draw   (109.14,106.73) .. controls (109.17,115.56) and (102.04,122.76) .. (93.2,122.79) .. controls (84.37,122.83) and (77.17,115.7) .. (77.14,106.86) .. controls (77.1,98.02) and (84.23,90.83) .. (93.07,90.79) .. controls (101.91,90.76) and (109.1,97.89) .. (109.14,106.73) -- cycle ;
\draw    (109.14,106.73) -- (204.86,32.86) ;
\draw    (109.14,106.73) -- (205.14,68.86) ;
\draw   (237.86,106.73) .. controls (237.9,115.56) and (230.76,122.76) .. (221.93,122.79) .. controls (213.09,122.83) and (205.9,115.7) .. (205.86,106.86) .. controls (205.82,98.03) and (212.96,90.83) .. (221.79,90.8) .. controls (230.63,90.76) and (237.82,97.89) .. (237.86,106.73) -- cycle ;
\draw   (236.86,177.73) .. controls (236.9,186.56) and (229.76,193.76) .. (220.93,193.79) .. controls (212.09,193.83) and (204.9,186.7) .. (204.86,177.86) .. controls (204.82,169.03) and (211.96,161.83) .. (220.79,161.8) .. controls (229.63,161.76) and (236.82,168.89) .. (236.86,177.73) -- cycle ;
\draw    (109.14,106.73) -- (205.86,106.86) ;
\draw    (109.14,106.73) -- (204.86,177.86) ;

\draw (213,63) node [anchor=north west][inner sep=0.75pt]   [align=left] {{$v_2$}};
\draw (213,26) node [anchor=north west][inner sep=0.75pt]   [align=left] {{$v_1$}};
\draw (85,100) node [anchor=north west][inner sep=0.75pt]   [align=left] {{$u_1$}};
\draw (225, 125) node [anchor=north west][inner sep=0.75pt]  [rotate=-90.1]  {$.\ .\ .$};
\draw (213,100) node [anchor=north west][inner sep=0.75pt]   [align=left] {{$v_3$}};
\draw (205,172) node [anchor=north west][inner sep=0.75pt]   [align=left] {{$v_{2/\prob}$}};

\end{tikzpicture}
        \caption{The described instance $\inst$. Each edge has a probability of success $\prob = 1/n$.}
        \label{fig:2-undersupplied_instance_1}
    \end{minipage}
    \hfill
    \begin{minipage}[b]{0.45\textwidth}
        \centering
        \begin{tikzpicture}[x=0.75pt,y=0.75pt,yscale=-.8,xscale=.9]

\draw   (237.14,68.73) .. controls (237.17,77.56) and (230.04,84.76) .. (221.2,84.79) .. controls (212.37,84.83) and (205.17,77.7) .. (205.14,68.86) .. controls (205.1,60.02) and (212.23,52.83) .. (221.07,52.79) .. controls (229.91,52.76) and (237.1,59.89) .. (237.14,68.73) -- cycle ;
\draw   (236.86,32.73) .. controls (236.9,41.56) and (229.76,48.76) .. (220.93,48.79) .. controls (212.09,48.83) and (204.9,41.7) .. (204.86,32.86) .. controls (204.82,24.03) and (211.96,16.83) .. (220.79,16.8) .. controls (229.63,16.76) and (236.82,23.89) .. (236.86,32.73) -- cycle ;
\draw   (106.14,67.73) .. controls (106.17,76.56) and (99.04,83.76) .. (90.2,83.79) .. controls (81.37,83.83) and (74.17,76.7) .. (74.14,67.86) .. controls (74.1,59.02) and (81.23,51.83) .. (90.07,51.79) .. controls (98.91,51.76) and (106.1,58.89) .. (106.14,67.73) -- cycle ;
\draw    (106.14,67.73) -- (204.86,32.86) ;
\draw    (106.14,67.73) -- (205.14,68.86) ;
\draw   (237.86,106.73) .. controls (237.9,115.56) and (230.76,122.76) .. (221.93,122.79) .. controls (213.09,122.83) and (205.9,115.7) .. (205.86,106.86) .. controls (205.82,98.03) and (212.96,90.83) .. (221.79,90.8) .. controls (230.63,90.76) and (237.82,97.89) .. (237.86,106.73) -- cycle ;
\draw   (236.86,177.73) .. controls (236.9,186.56) and (229.76,193.76) .. (220.93,193.79) .. controls (212.09,193.83) and (204.9,186.7) .. (204.86,177.86) .. controls (204.82,169.03) and (211.96,161.83) .. (220.79,161.8) .. controls (229.63,161.76) and (236.82,168.89) .. (236.86,177.73) -- cycle ;
\draw    (106.14,67.73) -- (205.86,106.86) ;
\draw    (106.14,67.73) -- (204.86,177.86) ;
\draw   (106.14,136.73) .. controls (106.17,145.56) and (99.04,152.76) .. (90.2,152.79) .. controls (81.37,152.83) and (74.17,145.7) .. (74.14,136.86) .. controls (74.1,128.02) and (81.23,120.83) .. (90.07,120.79) .. controls (98.91,120.76) and (106.1,127.89) .. (106.14,136.73) -- cycle ;
\draw    (105.14,138.73) -- (204.86,177.86) ;
\draw    (105.14,138.73) -- (205.86,106.86) ;
\draw    (105.14,138.73) -- (205.14,68.86) ;
\draw    (105.14,138.73) -- (204.86,32.86) ;

\draw (213,63) node [anchor=north west][inner sep=0.75pt]   [align=left] {{$v_2$}};
\draw (213,26) node [anchor=north west][inner sep=0.75pt]   [align=left] {{$v_1$}};
\draw (81,62) node [anchor=north west][inner sep=0.75pt]   [align=left] {{$u_1$}};
\draw (225, 125) node [anchor=north west][inner sep=0.75pt]  [rotate=-90.1]  {$.\ .\ .$};
\draw (213,100) node [anchor=north west][inner sep=0.75pt]   [align=left] {{$v_3$}};
\draw (205,172) node [anchor=north west][inner sep=0.75pt]   [align=left] {{$v_{2/\prob}$}};
\draw (81,132) node [anchor=north west][inner sep=0.75pt]   [align=left] {{$u_2$}};

\end{tikzpicture}
        \caption{The described instance $\inst'$. Each edge has a probability of success of $\prob = 1/n$.}
        \label{fig:2-undersupplied_instance_2}
    \end{minipage}
\end{figure}

\section{Proof of \Cref{lemma: offline is upper bound} (Section \ref{sec: model})}
\label{proof: offline is upper bound}
    Fix any instance $\inst$ and consider $\alg$ any optimal clairvoyant algorithm and $v_1, \ldots, v_T$ a sequence of demand arrivals.  Let $X_{u,t}$ the random indicator variable that indicates if the $t$-th arrival is matched to $u$ and the match is successful, following the decisions from $\alg$. Similarly, 
   let $Y_{u,t}$ the random indicator variable that indicates if the $t$-th arrival is matched to $u$. Note that, for adversarial arrivals, if $(u,t) \in E$ then $\mathbb E[X_{u,t}] = \prob \cdot \mathbb E[Y_{u,t}]$. Similarly, for stochastic arrivals, if $v_t = v$ then $\mathbb E[X_{u,t}] = \prob_{u,v} \cdot \mathbb E[Y_{u,t}]$.
   We will prove that $(\mathbb E[Y_{u,t}])_{u \in \supply, t \in [T]}$ can be transformed into a feasible solution to the respective LP, $\offI(\inst, 1)$ with objective value equal to $\mathbb E[\sum_{(u,t) \in \supply \times [T]}  X_{u,t}]$, depending on the arrival model.

   \begin{enumerate}
       \item \textit{Adversarial arrivals.}  On any given realization $\sum_{t: (u,t) \in E}  X_{u,t} \leq 1$, for all $u \in \supply$ since any supply vertex $u$ can be consumed at most once. By taking expectations, we arrive at $\sum_{t: (u,t) \in E}  \prob \cdot \mathbb E[Y_{u,t}] \leq 1$ and constraint \eqref{adversarial_lp: kappa-constraint} follows. Furthermore, since any arriving demand node $t$ can be matched at most once we have $\sum_{u: (u,t) \in E} Y_{u,t} \leq 1$ for all $t \in [T]$ and hence constraint \eqref{adversarial_lp: demand_constraint} follows by taking expectations. The expected reward of $\alg$ in this arrival model is $\sum_{(u,t) \in \supply \times [T]}  \mathbb E[X_{u,t}] = \mathbb E[\sum_{(u,t) \in E}  X_{u,t}]$ and can be re-written as $\sum_{(u,t) \in E} \prob \cdot \mathbb E[Y_{u,t}]$. Hence, $(\mathbb E[Y_{u,t}])_{(u,t) \in E}$ is a feasible solution to $\offI(\inst)$ with objective $\sum_{(u,t) \in E} \prob \cdot \mathbb E[Y_{u,t}]$.

       \item \textit{Stochastic arrivals.} For all $(u,v) \in \supply \times V$, define $X_{u,v} = \sum_{t \in [T]: v_t = v} X_{u,t}$ the random variable that indicates if a given demand type $v$ consumed a supply node $u$. Similarly, define $Y_{u,v} = \sum_{t \in [T]: v_t = v} Y_{u,t}$. On any given realization we have $\sum_{v \in V}  X_{u,v} \leq 1$, for all $u \in \supply$ since any supply node can be consumed at most once. By taking expectations, we arrive at $\sum_{v \in V}  \prob_{u,v} \cdot \mathbb E[Y_{u,t}] \leq 1$ and constraint \eqref{stochastic_lp: kappa-constraint} follows. Furthermore, any type $v \in V$ can be matched at most $N_v = |\{ v_{t} : v_t = v\}|$ times and hence $\sum_{u \in \supply} Y_{u,v} \leq N_v$ for all $v \in V$. Constraint \eqref{stochastic_lp: demand_constraint} follows by taking expectations and noting that $\mathbb E[N_v] = T p_v$. The expected reward of $\alg$ is $\sum_{(u,t) \in \supply \times [T]}  \mathbb E[X_{u,t}] = \mathbb E[\sum_{(u,v) \in \supply \times V}  X_{u,v}]$ and can be re-written as $\sum_{(u,v) \in \supply \times V} \prob_{u,v} \cdot \mathbb E[Y_{u,v}]$.
   \end{enumerate}
\hfill \Halmos

\section{Results on the underage and overage costs (Section \ref{sec: overview_results})}
\label{appendix: underage-costs}
This section is dedicated to a further explanation of the results exposed at the end of \Cref{sec: overview_results}; in particular, we show that the profit functions $\text{Profit}_{i}(\inv)$ are concave for $i \in \{A,B\}$ and how to derive the optimal stocking decisions.

We start by showing that the profit functions are concave by showing that the second derivative of these functions is non-positive. For the adversarial setting,
\begin{equation*}
    \frac{\partial}{\partial \inv} \text{Profit}_{A}(\inv) = d \cdot \frac{\partial}{\partial \inv} \left(r \cdot \frac{\inv}{1+\inv} - \inv c\right) = d \cdot \left( \frac{r}{(1+\inv)^2} - c \right) \quad \Rightarrow \quad \frac{\partial^2}{\partial \inv^2}\text{Profit}_{A}(\inv) = \frac{-2dr}{(1+\inv)^3}.
\end{equation*}
Hence, the second derivative of $\text{Profit}_{A}(\inv)$ is non-positive for any value of $\inv \geq 0$, since $d, r \geq 0$. Analogously, we compute the second derivative for the stochastic setting:
\begin{align*}
    \frac{\partial}{\partial \inv}\text{Profit}_{B}(\inv) = d \cdot \frac{\partial}{\partial \inv} \left((1-e^{-1/\inv})\inv \cdot r - \inv c\right) = d \cdot \left( r \left(1 - \frac{e^{-1/\inv}(\inv+1)}{\inv} \right) - c \right)\\
    \quad \Rightarrow \quad \frac{\partial^2}{\partial \inv^2}\text{Profit}_{B}(\inv) = -dr \cdot \frac{e^{-1/\inv}}{\inv^3}.
\end{align*}
This implies that the second derivative of $\text{Profit}_{B}(\inv)$ is non-positive for any value of $\inv \geq 0$ by the positivity of the exponential function and $d,r \geq 0$.

Now we turn to provide the derivations of the optimal stocking decision for each arrival model.
 \begin{enumerate}
    \item \textit{Firm $A$ (Adversarial arrivals)}. The first-order condition exhibits that
    \begin{align*}
        \frac{\partial}{\partial \inv} \text{Profit}_{A}(\inv^*_A) = 0 \qquad &\Longleftrightarrow \qquad d \cdot \left( \frac{r}{(1+\inv^*_A)^2} - c \right) = 0\\
        &\Longleftrightarrow \qquad \frac{c}{r} = \frac{1}{(1+\inv^*_A)^2}\\
        &\Longleftrightarrow \qquad \frac{r-c}{r} = 1-\frac{1}{(1+\inv^*_A)^2},
    \end{align*}
    where the first equivalence comes from the definition of the profit function, the second by diving through $d$ and re-arranging, and the last one is just algebraic manipulation. For simplicity, set $y = 1 - \frac{r-c}{r}$. We can re-write the above condition as
    \begin{equation*}
        (1+\inv_A^*)^2 = \frac{1}{y} \quad \Longleftrightarrow \quad (\inv_A^*)^2 + 2 \inv_A^* + (1 - 1/y) = 0,
    \end{equation*}
    which is a quadratic equation in $\inv_A^*$, which can be solved for the positive root
    \begin{equation*}
        \inv_A^*=\frac{-2 + \sqrt{4-4(1-1/y)}}{2} = -1 + \sqrt{1/y} = -1 + \sqrt{1/y} = \frac{-y + \sqrt{y}}{y} = \frac{\frac{r-c}{r} - 1 + \sqrt{1 - \frac{r-c}{r}}}{1 - \frac{r-c}{r}}.
    \end{equation*}
    \item \textit{Firm $B$ (Stochastic arrivals)}. We proceed in the same way as in the previous case. We have
    \begin{align*}
        \frac{\partial}{\partial \inv} \text{Profit}_{B}(\inv^*_B) = 0 \qquad &\Longleftrightarrow \qquad d \cdot \left( r\left( 1 - e^{-1/\inv^*_B}\left(\frac{1+\inv^*_B}{\inv^*_B}\right) \right) - c \right) = 0\\
        &\Longleftrightarrow \qquad \frac{c}{r} = 1 - e^{-1/\inv^*_B}\left(\frac{1+\inv^*_B}{\inv^*_B}\right)\\
        &\Longleftrightarrow \qquad \frac{r-c}{r} = e^{-1/\inv^*_B}\left(\frac{1+\inv^*_B}{\inv^*_B}\right),
    \end{align*}
    where the first equivalence follows by the definition of the profit function, the second follows by dividing by $d$ and re-arranging, and the last is again algebraic manipulation.
    We can solve for $\inv^*_B$ by algebraically manipulating the above equation
    \begin{align*}
    & & \frac{r-c}{r} &= e^{-1/\inv^*_B}\left(\frac{1+\inv^*_B}{\inv^*_B}\right) \\
    &\Longleftrightarrow & \frac{1}{e} \left(\frac{r-c}{r}\right) &= e^{-1/\inv^*_B - 1}\left(1 + \frac{1}{\inv^*_B}\right) \\
    &\Longleftrightarrow & -\frac{1}{e} \left(\frac{r-c}{r}\right) &= e^{-1/\inv^*_B - 1}\left(-\frac{1}{\inv^*_B} - 1\right) \\
    &\Longleftrightarrow & -\frac{1}{\inv^*_B} - 1 &= W_{-1}\left( - \frac{1}{e} \left( \frac{r-c}{r} \right) \right) \\
    &\Longleftrightarrow & \inv^*_B &= \frac{-1}{W_{-1}(-\frac{1}{e} \left( \frac{r-c}{r} \right))+1}
\end{align*}
where the first equivalence follows by dividing both sides by $e$, the next one by multiplying both sides by $(-1)$. The third equivalence follows by definition of the $W_{-1}$ function\footnote{In general one can use the positive and negative branches of the Lambert $W$ function to solve equations for real numbers. In this particular setting, we use the negative branch as that is the one that makes $\inv^*_B \geq 0$.}; $z = y \exp(y)$ can be solved for $y$ with $y = W_{-1}(z)$, and in this particular example $y = -\frac{1}{\inv^*_B} - 1$ and $z = -\frac{1}{e} \left( \frac{r-c}{r} \right)$. The last equivalence follows by re-arranging terms.
\end{enumerate}

\section{Remaining proofs from Section \ref{sec3}}

This section is structured as follows: in Sections \ref{ssec:AdvOfflineOptBound}, \ref{proof: DLP lower bounds greedy/off}, and \ref{proofs: AdvTheoremLastIneq} we prove Lemmas \ref{claim: AdvOfflineOptBound}, \ref{lemma: DLP lower bounds greedy/off}, and \ref{lemma: AdvTheoremLastIneq}. Then, in  Section \ref{proof: AdvCR} we prove Lemma \ref{prop: AdvCR}; the latter is a lengthy proof, involving various auxiliary results which we prove in subsequent subsections.

\subsection{Proof of \Cref{claim: AdvOfflineOptBound}}
\label{ssec:AdvOfflineOptBound}
Let $\xbf^*$ be an optimal solution to $\offIkappa$. As $\xbf^*$ is feasible, then for any $k \in [T] \cup \{0\}$ we must have that,
    \begin{equation*}
        \sum_{u \in A_k} x_u^* = \sum_{u \in A_k} \sum_{t \in [T]: (u,t) \in E} x_{u,t}^* \leq \sum_{u \in A_k} \frac{\market}{\mu} = \frac{\market}{\mu} \rho_k,\end{equation*}
    where the inequality follows from $\xbf^*$ satisfying constraint \eqref{adversarial_lp: kappa-constraint} {for each $u \in \supply$. Therefore, the feasibility constraints are satisfied.} 

    To show the {greedy bounds in} the lemma we introduce a new quantity to allow us to count the number of demand nodes assigned to a supply node $u$ at time $t$. For every pair $(u,t)$ we define $x_{u,t}^g$ as the indicator variable that equals $1$ if and only if $\greedy$ matches $t$ to $u$. Further, we let $n_{u,t}$ denote the number of demand nodes that $\greedy$ assigns to $u$ up to and including period $t$, i.e., 
    \begin{equation*}
        n_{u,t} = \sum_{t'=1}^t x_{u,{t'}}^g.
    \end{equation*}
With this notation, we have $n_{u, T}=n_u^g$ denoting the {total} number of demand nodes that $\greedy$ assigns to~$u$ {throughout the $T$ periods}.
    
Now, fix any $s \in [T]\cup \{0\}$, and note from the definition of $x_u^*$ that 
    \begin{align}
        \sum_{k=0}^s \sum_{u \in A_k} x_u^* &= \sum_{k=0}^s  \sum_{u \in A_k} \sum_{t \in [T]: (u,t) \in E} x_{u,t}^* \notag\\
        &= \sum_{t \in [T]} \sum_{u \in \bigcup_{k=0}^s A_k} x_{u, t}^* \cdot \bm{1}_{\{(u,t) \in E \}}\notag\\
        &= \sum_{t \in [T]} \sum_{u \in \bigcup_{k=0}^s A_k} x_{u, t}^* \cdot \left(\sum_{j=0}^T \sum_{w \in A_j} x_{w,t}^g\right) \cdot \bm{1}_{\{(u,t) \in E \}}\notag\\
        &= \sum_{j=0}^T \sum_{w \in A_j} \left \{ \sum_{t \in [T]} \sum_{u \in \bigcup_{k=0}^s A_k} x_{u, t}^* \cdot x_{w,t}^g \cdot \bm{1}_{\{(u,t) \in E\}} \right\}
        \label{eq:lem2_rewriting} 
    \end{align}
    where each equality holds since $\{A_j\}_{j=0}^T$ is a partition of $\supply$ and we used $\sum_{j=0}^T \sum_{w \in A_j} x_{w,t}^g = 1$ for all $t \in [T]$ {because greedy matches each demand node exactly once as, by assumption, $|N(t)| > 0$ for all $t \in [T]$.} Keeping $s \in [T]\cup \{0\}$ fixed, we now derive two bounds on the last summand in curly braces that depend on the value of $j$ such that $w\in A_j$. \begin{enumerate}
        \item \textit{Worst-case bound}. Let $w \in A_j$. 
        Then
        \begin{equation}
         \label{proof: adv worst_case_bound}
             \sum_{t \in [T]} \sum_{u \in \bigcup_{k=0}^s A_k} x_{u, t}^* \cdot x_{w,t}^g \cdot \bm{1}_{\{(u,t) \in E\}}  
            = \sum_{t \in [T]} x_{w,t}^g \cdot \sum_{u \in \bigcup_{k=0}^s A_k: \, (u,t) \in E} x_{u, t}^* 
            \leq  \sum_{t \in [T]} x_{w,t}^g
            =\, j,
        \end{equation}
        where the first equality comes from re-arranging the summations, the next inequality holds since feasibility of {$\xbf^*$} for $\offIkappa$ requires $\sum_{u \in \bigcup_{k=0}^s A_k: (u, t) \in E} x_{u, t}^* \leq 1$ for all $t \in [T]$, and the final equality {follows by the fact that $w \in A_j$ together with the definition of $A_j$.}

        \item \textit{Conditional bound}. Fix $j > 1$, let $w \in A_j$ and suppose  $s<j$. We will prove {that}
        \begin{equation}
        \label{proof: adv conditional_case_bound}
             \sum_{t \in [T]} \sum_{u \in \bigcup_{k=0}^s A_k} x_{u, t}^* \cdot x_{w,t}^g \cdot \bm{1}_{\{ (u,t) \in E \}}  \leq s+1.
        \end{equation}

        \dscomment{Can you explain why this is true?}\dfcomment{Daniela, can you verify if the explanation is sufficient for you? Either way, I think this is OK for WINE.}
        Suppose that the inequality does not hold and let 
        \begin{equation*}
            t(w,s) = \max \left\{ t \in [T] \, : \, x_{w,t}^g = 1, \, \exists u' \in \bigcup_{k=0}^s A_k, \text{ s.t. } x_{u',t}^* > 0 \text{ and } (u',t) \in E\right\}
        \end{equation*}
        denote the last period in which $\greedy$ assigns an arrival to $w$ whereas $\xbf^*$ assigns positive mass to some node in $\cup_{k=0}^s A_k$.
        By the contradiction hypothesis this quantity is well-defined \bbedit{as there must exist some term $x_{u, t}^* \cdot x_{w,t}^g \cdot \bm{1}_{\{ (u,t) \in E \}} > 0$ for some $u \in \bigcup_{k=0}^s A_k, t \in [T]$ and $(u,t) \in E$. Hence, the above set is non-empty and achieves a maximum.}
        Let $u'$ be any supply node in $\bigcup_{k=0}^s A_k$ with $(u', {t(w,s)}) \in E$. Then, assuming that \eqref{proof: adv conditional_case_bound} does not hold we derive the following contradiction:
        \begin{align*}
            s+1 &<\sum_{t \in [T]} \sum_{u \in \bigcup_{k=0}^s A_k} x_{u, t}^* \cdot x_{w,t}^g \cdot \bm{1}_{\{ (u,t) \in E \}}\\
            &=\sum_{t = 1}^{t(w,s)} \sum_{u \in \bigcup_{k=0}^s A_k} x_{u, t}^* \cdot x_{w,t}^g \cdot \bm{1}_{\{ (u,t) \in E \}}\\
            &\leq \sum_{t=1}^{t(w,s)} x_{w,t}^g\\
            &= n_{w, t(w,s)}\\
            &\leq n_{u', t(w,s)} + 1\\
            &\leq n_{u', T} + 1 \\
            &\leq s+1.
        \end{align*}
        Here, the first equality holds by definition of $t(w,s)$ (as $x_{u, t}^* \cdot x_{w, t}^g=0$ for $t>t(w,s)$), the next inequality holds because $\xbf^*$ is feasible for $\offIkappa$ {and therefore $\sum_{u \in \bigcup_{k=0}^s A_k} x_{u, t}^* \cdot \bm{1}_{\{ (u,t) \in E \}} \leq 1$ for each $t$}, and the equality thereafter uses the definition of $n_{w, t(w,s)}$.  The next inequality uses that $(u',{t(w,s)}) \in E$ and that $\greedy$ decides to assign $t(w,s)$ to $w$ rather than $u'$, implying that at the end of the period $w$ can have at most one more demand node assigned to it than $u'$.  The penultimate inequality is due to  $n_{u,t}$ being increasing in $t$ for a fixed $u$, and the last inequality is due to $u' \in \cup_{k=0}^s A_k$ implying that $\greedy$ assigns at most $s$ demand nodes to $u'$ over the entire time horizon. This concludes the proof of  \eqref{proof: adv conditional_case_bound}.
    \end{enumerate}
    
        We now  combine the conditional and the worst-case {bounds as follows. Fix} $s \in \{0\} \cup [T-1]$. {By } \eqref{eq:lem2_rewriting},{we have that}
    \begin{align*}
        \sum_{k=0}^s \sum_{u \in A_k} x_u^* &=
         \sum_{j=0}^T \sum_{w \in A_j} \left \{ \sum_{t \in [T]} \sum_{u \in \bigcup_{k=0}^s A_k} x_{u, t}^* \cdot x_{w,t}^g \cdot \bm{1}_{\{(u,t) \in E\}} \right\}  \\
        &= \sum_{j=0}^s \sum_{w \in A_j} \left \{ \sum_{t \in [T]} \sum_{u \in \bigcup_{k=0}^s A_k} x_{u, t}^* \cdot x_{w,t}^g \cdot \bm{1}_{\{(u,t) \in E\}} \right\} \\ &\quad + \sum_{j=s+1}^T \sum_{w \in A_j} \left \{ \sum_{t \in [T]} \sum_{u \in \bigcup_{k=0}^s A_k} x_{u, t}^* \cdot x_{w,t}^g \cdot \bm{1}_{\{(u,t) \in E\}} \right\} \\
        &\leq \sum_{j=0}^s \sum_{w \in A_j} j + \sum_{j=s+1}^{T} \sum_{w \in A_j} (s+1)\\
        &= \sum_{j=0}^s j \cdot \rho_j + (s+1)\sum_{j=s+1}^T \rho_j.
    \end{align*}
    Here, the inequality uses the worst case bound \eqref{proof: adv worst_case_bound} on the first term and the conditional bound \eqref{proof: adv conditional_case_bound} on the second term; in the final equality we use the definition of $\rho_j$.
    If $s = T$ then using \eqref{eq:lem2_rewriting}, the worst case bound, and the definition of $\rho_j$ we obtain the last inequality needed to conclude the proof of the lemma
    \begin{align*}
        \sum_{k=0}^s \sum_{u \in A_k} x_u^* &=
         \sum_{j=0}^T \sum_{w \in A_j} \left \{ \sum_{t \in [T]} \sum_{u \in \bigcup_{k=0}^s A_k} x_{u, t}^* \cdot x_{w,t}^g \cdot \bm{1}_{\{(u,t) \in E\}} \right\} \leq \sum_{j=0}^T \sum_{w \in A_j} j = \sum_{j=0}^T j \cdot \rho_j.\hfill\Halmos
    \end{align*}

\subsection{Proof of \Cref{lemma: DLP lower bounds greedy/off}}
\label{proof: DLP lower bounds greedy/off}

We prove the result for fixed $|\sigmabf|=T$ and $\market>0$ and a fixed $\inst$ with consumption probabilities equal to~$\prob$. We begin by formulating a linear program $\LP$ in minimization form that uses the bounds shown in \Cref{claim: AdvOfflineOptBound}. Then, we construct a feasible solution to $\LP$ and show that this particular solution achieves an objective value of $\mathbb E[\greedy(\inst, \sigmabf)]/\offIkappa$.  Since $\LP$ is a minimization problem, this allows us to lower bound the CR for any arrivals $\sigmabf$ with $|\sigmabf|=T$, and instance~$\inst$ with $\market>0$.  Finally, we conclude the proof by noting that $\DLP$ is the dual of~$\LP$, so weak duality implies that the value of $\DLP$ is a lower bound to $\LP$ and thus also a lower bound to $\mathbb E[\greedy(\inst, \sigmabf)]/\offIkappa$.

We start by stating the linear program:
\begin{equation*}
\begin{aligned}
\LP = \min_{\mathbf{w}, \mathbf{y} \in \mathbb{R}^{T+1}} \quad &  \sum_{j=0}^T y_j  \cdot \frac{1-(1-\prob)^{j}}{\prob}\\
\text{s.t.} \quad & \sum_{k=0}^s w_k \leq \sum_{j=0}^s j \cdot y_j + (s+1)\sum_{j=s+1}^T y_j, & \forall s \in \{0\} \cup [T-1]. \quad & (\alpha_s)_{s=0}^{T-1}\\
\quad & \sum_{k=0}^T w_k \leq \sum_{j=0}^T j \cdot y_j, & & (\alpha_T)\\
\quad & w_j \leq \frac{\market}{\prob} \cdot y_j, & \forall j \in \{ 0 \} \cup [T]. \quad & (\beta_j)_{j=0}^{T}\\
\quad & \sum_{k=0}^T w_k = 1, & & (\delta)\\
\quad & \mathbf{w}, \mathbf{y} \geq \mathbf{0}.
\end{aligned}
\end{equation*} 

Next, we construct a feasible solution $(\mathbf{\hat{w}}, \mathbf{\hat{y}})$ to the above problem that achieves the objective value of $\mathbb E[\greedy(\inst, \sigmabf)]/\offIkappa$.
Consider an execution of $\greedy$ on $\inst$ with the arrival order~$\sigmabf$. Let $\rhobf$ the vector of cardinalities of the sets $A_i = \left\{ u \in \supply:  n_u^g = i \right\}$, i.e., $\rho_i=|A_i|$, and $\xbf^*$ be an an optimal solution to $\offIkappa$ with $|\xbf^*|_1>0$. We define~$\nubf$ as $\nu_i = \sum_{u \in A_i} x_{u}^*$ for $i = 0, \ldots, T$. As the sets $\{A_i\}_{i=0}^T$ partition $\supply$ we have $\sum_{i=0}^T \rho_i = |\supply|$. We first observe that $\rhobf$ and~$\nubf$ are non-negative with at least one positive coordinate: for $\rhobf$ this holds because  it represents the cardinalities of the sets $A_i$ where $\sum_i |A_i|=|\supply|$; for $\nubf$ it holds because it is a non-negative sum in every coordinate and at least one $x_u^*$ is positive. Now, define 
    \begin{equation*}
         \forall k = 0, \ldots, T: \quad  \hat{w}_k = \frac{\nu_k}{\sum_{i=0}^T \nu_i},\quad \hat{y}_k = \frac{\rho_k}{\sum_{i=0}^T \nu_i}.
    \end{equation*}
    We prove as follows that $(\mathbf{\hat{w}}, \mathbf{\hat{y}})$ is feasible in $\LP$. 
    \begin{enumerate}
      \item We first rewrite the greedy bounds in \Cref{claim: AdvOfflineOptBound}, for all $s \in \{0\}\cup[T-1]$, by substituting $\nu_k$:
      \begin{align*}
    & \sum_{k=0}^s \nu_k \leq \sum_{j=0}^s j \cdot \rho_j +  (s+1)  \sum_{j=s+1}^{T} \rho_j
    \quad \text{and}\quad
    \sum_{k=0}^T \nu_k \leq \sum_{j=0}^T j \cdot \rho_j.
    \end{align*}
    Dividing on both sides by $\sum_{i=0}^T \nu_i$ we find that these are equivalent to
\begin{align*}
    \sum_{k=0}^s \hat{w}_k = \sum_{k=0}^s \nu_k / \sum_{i=0}^T \nu_i  \leq \left(\sum_{j=0}^s j \cdot \rho_j +  (s+1)  \sum_{j=s+1}^{T} \rho_j\right)/\sum_{i=0}^T \nu_i =\sum_{j=0}^s j \cdot \hat{y}_j +  (s+1)  \sum_{j=s+1}^{T} \hat{y}_j
\end{align*}
\begin{align*}    
    \text{and}\quad
    \sum_{k=0}^T \hat{w}_k = \sum_{k=0}^T \nu_k / \sum_{i=0}^T \nu_i \leq \left( \sum_{j=0}^T j \cdot \rho_j\right)/ \sum_{i=0}^T \nu_i = \sum_{j=0}^T j \cdot \hat{y}_j.
\end{align*}
This guarantees that our solution $(\mathbf{\hat{w}}, \mathbf{\hat{y}})$ satisfies the first two sets of constraints. 
\item We similarly rewrite the feasibility bounds in \Cref{claim: AdvOfflineOptBound} for all $j \in \{0\} \cup [T]$ as $\nu_j\leq  \frac{\market}{\prob}\cdot \rho_j$. Dividing both sides by $\sum_{i=0}^T \nu_i$, we find that this is equivalent to $\hat{w}_j=\nu_j/\sum_{i=0}^T \nu_i \leq \frac{\market}{\prob}\cdot \rho_j/\sum_{i=0}^T \nu_i= \frac{\market}{\prob} \cdot \hat{y}_j$. This guarantees that our solution $(\mathbf{\hat{w}}, \mathbf{\hat{y}})$ satisfies the third set of constraints.
\item The last constraint follows because by definition of $\mathbf{\hat{w}}$, we have $\sum_{k=0}^T \hat{w}_k = \sum_{k=0}^T  \frac{\nu_k}{\sum_{i=0}^T \nu_i} = ~1$.
\end{enumerate}
Since $\mathbf{\hat{w}}, \mathbf{\hat{y}}$ are non-negative vectors, this proves that  $(\mathbf{\hat{w}}, \mathbf{\hat{y}})$ is a feasible solution to $\LP$. Now, note that this solution produces an objective of
\begin{equation*}
     \sum_{j=0}^T \hat{y}_j \cdot \frac{1-(1-\prob)^{j}}{\prob} = \sum_{j=0}^T \frac{\rho_j}{\sum_{i=0}^T \nu_i} \cdot \frac{1-(1-\prob)^{j}}{\prob} = \frac{\sum_{j=0}^T \rho_j \left( 1-(1-\prob)^{j} \right)}{\prob \sum_{i=0}^T \nu_i} = \frac{\mathbb E[\greedy(\inst, \sigmabf)]}{\offIkappa},
\end{equation*}
where the last equality follows from the definition of $\rhobf$ and $\nubf$. Hence, the value of $\LP$ is at most $\mathbb E[\greedy(\inst, \sigmabf)]/\offIkappa$. Noticing that the dual of $\LP$ corresponds to $\DLP$ (see derivation below) with dual variables $(\alphabf, \betabf, \delta)$, weak duality guarantees that
\begin{equation*}
     \frac{\mathbb E[\greedy(\inst, \sigmabf)]}{\offIkappa} \geq \LP \geq \DLP. \hfill\Halmos
\end{equation*}

\subsubsection*{Dual derivation}
We now provide a detailed derivation of the dual $\DLP$ of $\LP$, which is used above. We recall the formulation of the linear program $\LP$, in which we rewrite the inequalities as follows, with the use of an indicator function.
\begin{equation*}
\begin{aligned}
 \min_{\mathbf{w}, \mathbf{y} \in \mathbb{R}^{T+1}} \quad &  \sum_{j=0}^T y_j  \cdot \frac{1-(1-\prob)^{j}}{\prob}\\
\text{s.t.} \quad & \sum_{k=0}^T w_k \cdot \bm{1}_{\{ k \leq s\}} - \sum_{j=0}^T (j \cdot \bm{1}_{\{ j \leq s\}} - (s+1) \cdot \bm{1}_{\{j > s\}}) \cdot y_j \leq 0, & \forall s \in \{0\} \cup [T-1]. \quad & (\alpha_s)_{s=0}^{T-1}\\
\quad & \sum_{k=0}^T w_k - \sum_{j=0}^T j \cdot y_j  \leq 0, & & (\alpha_T)\\
\quad & w_j - \frac{\market}{\prob} \cdot y_j \leq 0, & \forall j \in \{ 0 \} \cup [T]. \quad & (\beta_j)_{j=0}^{T}\\
\quad & \sum_{k=0}^T w_k = 1, & & (\delta)\\
\quad & \mathbf{w}, \mathbf{y} \geq \mathbf{0}.
\end{aligned}
\end{equation*}
Now we will proceed constructing the dual problem using the variables $(\alphabf, \betabf, \delta)$. Since $\LP$ is a minimization problem, we have that $\alphabf, \betabf \leq 0$ and $\delta$ is free. For the constraints, we proceed by constructing them as follows.
\begin{enumerate}
    \item \textit{$y_j$ constraints.} Fix any variable $y_j$. The coefficient associated with this variable in the objective is $(1-(1-\prob)^{j})/\prob$, and since this variable is non-negative the corresponding induced inequality by this variable is
    \begin{align*}
        & \quad - \sum_{s=0}^T (j \cdot \bm{1}_{\{ j \leq s\}} - (s+1) \cdot \bm{1}_{\{j > s\}}) \alpha_s - \frac{\market}{\prob} \beta_j \leq \frac{1-(1-\prob)^{j}}{\prob}\\
        \Longleftrightarrow & \quad j \cdot \sum_{s=j}^T \alpha_s + \sum_{s < j} (s+1) \alpha_s + \frac{\market}{\prob} \beta_j + \frac{1-(1-\prob)^{j}}{\prob} \geq 0,
    \end{align*}
    which corresponds to \ref{cons: D2} for $j \in [T]$ and \ref{cons: D3} for $j = 0$.
    \item \textit{$w_j$ constraints.} Fix any variable $w_j$. The coefficient associated with this variable in the objective is $0$, and since this variable is non-negative the corresponding induced inequality by this variable is
    \begin{align*}
            & \quad\sum_{s=0}^T \bm{1}_{\{ j \leq s\}} \cdot \alpha_s + \beta_j + \delta \leq 0\\
            \Longleftrightarrow & \quad \sum_{s=j}^T \alpha_s + \beta_j + \delta \leq 0.
    \end{align*}
    which corresponds to \ref{cons: D1}.
\end{enumerate}
Finally, the objective function is simply $\delta$ as is the only constraint that has a non-zero constant on the right hand side. The derivation of $\DLP$ follows by the above.

\subsection{Proof of Lemma \ref{lemma: AdvTheoremLastIneq}}
\label{proofs: AdvTheoremLastIneq}
    If $\prob = 1$ the inequality becomes $1 \geq \frac{1}{1+\market}$ which is true for any $\market > 0$. We  prove the result by analyzing how $\lceil\frac{\market}{\prob}\rceil$ behaves in each interval of the form $(j, j+1]$. 
    Suppose $\frac{\market}{\prob} \in (j, j+1]$ for some $j \in \mathbb Z_{\geq 0}$. This is equivalent to $\prob \in [\frac{\market}{j+1}, \frac{\market}{j})$. Then the left hand side of the inequality of the lemma is equal to $1 - \frac{\prob (j+1)}{1+\market}$ and the right hand side is equal to $(1-\prob)^{j+1}$. Define
\begin{align*}
        h(\prob) = 1 - \frac{\prob (j+1)}{1+ \market}\quad\text{and}\quad
        l(\prob) = (1-\prob)^{j+1}.
\end{align*}
We ultimately want to prove that $h(\prob) \geq l(\prob)$ for any $\prob \in [\frac{\market}{j+1}, \frac{\market}{j})$.  First, observe that
\begin{align*}
    h\left( \frac{\market}{j} \right) = 1- \frac{\market}{\market+1} \frac{j+1}{j} = \frac{j-\market}{j(\market+1)}\quad\text{and}\quad
    h\left( \frac{\market}{j+1} \right) &= 1 - \frac{\market}{\market+1} = \frac{1}{\market + 1},
\end{align*}
and similarly
\begin{align*}
    l\left(\frac{\market}{j} \right)=\left( 1 - \frac{\market}{j} \right)^{j+1} = \left(\frac{j-\market}{j}\right)^{j+1}\quad\text{and}\quad    l\left(\frac{\market}{j+1} \right)&=\left( 1 - \frac{\market}{j+1} \right)^{j+1}.
\end{align*}
Now, observe the following equivalences (the second follows by dividing both sides by $(j-\market)/j$)
\begin{align*}
    h\left( \frac{\market}{j} \right) \geq l\left( \frac{\market}{j} \right) \quad \Longleftrightarrow \quad \frac{j-\market}{j(\market+1)} \geq \left(\frac{j-\market}{j}\right)^{j+1} \quad \Longleftrightarrow \quad \frac{1}{\market +1} \geq \left(1-\frac{\market}{j}\right)^{j}. \quad 
\end{align*}
Raising both sides to the $1/\market$-th power, we arrive at $\left(\frac{1}{\market +1}\right)^{1/\market} \geq \left(1-\frac{\market}{j}\right)^{j/\market}$, which holds true because the left hand side is always greater than $1/e$, whilst the right hand side is always less than $1/e$. Next,
\begin{align*}
    h\left( \frac{\market}{j+1} \right) \geq l\left( \frac{\market}{j+1} \right) \quad \Longleftrightarrow \quad \frac{1}{\market +1} \geq \left( 1 - \frac{\market}{j+1} \right)^{j+1}.
\end{align*}
The latter holds true as raising both sides to the $1/\market$-th power gives $\left(\frac{1}{\market +1}\right)^{1/\market} \geq \left(1-\frac{\market}{j+1}\right)^{(j+1)/\market}$. Again, the left hand side is greater than $1/e$ and the left hand side is less than $1/e$. Finally, we claim that showing that $h\left( \frac{\market}{j} \right) \geq l\left( \frac{\market}{j} \right)$ and $h\left( \frac{\market}{j+1} \right) \geq l\left( \frac{\market}{j+1} \right)$  suffices to prove $h(\prob) \geq l(\prob)$ for any $\prob \in [\frac{\market}{j+1}, \frac{\market}{j})$ since $h$ is a linear  and $l$ is a convex\footnote{Its second derivative is $j(j+1)(1-\prob)^{j-1} \geq 0$.} function of $\prob$. Thus,  the  inequality holds for all $\prob \in [\frac{\market}{j+1}, \frac{\market}{j})$. 
\hfill\Halmos 

\subsection{Proof of \Cref{prop: AdvCR}.}
\label{proof: AdvCR}
We distinguish between the cases $T < \lceil \frac{\market}{\prob} \rceil + 1$ and $T \geq \lceil \frac{\market}{\prob} \rceil + 1$. In each case we identify a dual solution, prove that it is feasible, and bound its objective.

\emph{Case 1: $T < \lceil \frac{\market}{\prob} \rceil + 1$. } We construct a solution as follows: for every $s < T$ we set $\alpha_s = 0$. We let
    \begin{align*}
        \alpha_T &= - \frac{1-(1-\prob)^T}{T\prob}
    \end{align*}
    and define the values of $\betabf$ in such a way that the constraints of type \ref{cons: D2} are binding for all $j \in [T]$. In particular, with $\alpha_s=0$ for $s<T$, this requires \begin{equation*}
        0 = \frac{1- (1-\prob)^j}{\prob} + \frac{\market}{\prob} \beta_j + j \alpha_T, \quad \forall j \in [T].
    \end{equation*}
    After substituting our value of $\alpha_T$ and re-arranging this yields
    \begin{equation*}
        \beta_j =  \frac{\frac{j}{T} \cdot (1-(1-\prob)^T)-(1-(1-\prob)^j)}{\market}, \quad \forall j \in [T].
    \end{equation*}
    Similarly, we define $\beta_0=0$ which fulfills the same equality. Finally, we let $\delta = -\alpha_T$.

    We now show that this solution is feasible. First off, it follows immediately from the definition of $\alphabf$ that $\alphabf \leq \mathbf{0}$, i.e., \ref{cons: D4} holds true. Further, by definition of $\betabf$, the constraints of type \ref{cons: D2} and \ref{cons: D3} are all satisfied. Therefore, \ref{cons: D1} and \ref{cons: D5} remain to be shown.

    For \ref{cons: D5} we need to show that $\betabf\leq \mathbf{0}$. We do so by observing that $\beta_0=0=\beta_T$ and proving that $\beta_j$ is a convex function of $j$. As a result, every value of $\beta_j$ with $j = 0, \ldots, T$ lies beneath the line connecting $\beta_0 = 0$ and $\beta_T = 0$ and thus all of these values are non-positive. We show that $\beta_j$ is a convex function of $j$ by bounding the second derivative
     \begin{align*}
         \frac{d^2}{dj^2} \beta_j &= \frac{1}{\market} \frac{d^2}{dj^2} \left(  \frac{j}{T} \cdot (1-(1-\prob)^T)-(1-(1-\prob)^j) \right)\\
         &= \frac{1}{\market} \frac{d}{dj} \left( \frac{1 - (1-\prob)^T + T (1-\prob)^j \log(1-\prob)}{T} \right)\\
         &= \frac{1}{\market} (1-\prob)^j \log^2(1-\prob)\geq 0.
     \end{align*}
    It follows that $\betabf\leq \mathbf{0}$. Finally, we prove that our solution fulfills  \ref{cons: D1}. Rewriting these constraints for $j\in\{0,T\}$  and substituting our values of $\alpha_s$ as well as $\beta_0=\beta_T=0$  yields 
$$
\delta \leq -\beta_j - \sum_{s=j}^T \alpha_s = 0 - \alpha_T=-\alpha_T.
$$
Moreover, for $j\in\{1,\ldots,T-1\}$ we have $-\beta_j - \sum_{s=j}^T \alpha_s > -\beta_0 - \sum_{s=0}^T \alpha_s$, and as a result the constraints of type \ref{cons: D1} hold true for such values of $j$ as well. We have thus shown that our solution satisfies constraints \ref{cons: D1}--\ref{cons: D5} and is therefore feasible. As a last step, we observe that the objective $\delta = -\alpha_T=\alpha_T = \frac{1-(1-\prob)^T}{T\prob}$ matches the stated bound in the lemma which completes the analysis of the first case.

    
\emph{Case 2: $T \geq \lceil \frac{\market}{\prob} \rceil + 1$}. We define $t^* = T - \lceil \frac{\market}{\prob} \rceil$ and $(\alphabf, \betabf, \delta)$ as follows:
        \begin{align*}
        \alpha_T &=  \frac{ (1-\prob)^T - (1-\prob)^{t^*} }{\prob \lceil \frac{\market}{\prob}\rceil}, \tag{1a} \label{1a}\\
        \alpha_s &= 0,  &\forall s = t^*, \ldots, T-1, \tag{1b} \label{1b}\\ 
        \alpha_{t^*-1} &=  \frac{(1-\prob)^{t^*-1} \left(1 - (\lceil \frac{\market}{\prob}\rceil+1)\prob \right) - (1-\prob)^T}{\lceil \frac{\market}{\prob} \rceil \cdot (\market+\prob)}, & \tag{1c} \label{1c}\\
        \alpha_s &= \left(\frac{\market}{\market+\prob}\right)^{t^* - 1 - s} \alpha_{t^* -1} - \frac{\prob (1-\prob)^{s}}{ 1+ \market} \left( 1 - \left( \frac{\market(1-\prob)}{\market +\prob} \right)^{t^* - 1 - s} \right), & \forall s = 0, \ldots, t^* - 2. \tag{1d} \label{1d}\\
        \beta_{0} &= 0, \tag{2a} \label{2a}\\
        \beta_{j} &= \sum_{s=0}^{j-1} \alpha_s, & \forall j = 1, \ldots, {t^*-1} \tag{2b} \label{2b}\\
        \beta_{j} &=  \frac{(1-\prob)^j - (1-\prob)^T + (T-j) \cdot \mu \alpha_T}{\market} + \sum_{s=0}^{T-1} \alpha_s, & \forall j = t^*, \ldots, T-1 \tag{2c} \label{2c}\\
        \beta_{T} &= \sum_{s=0}^{T-1} \alpha_s. \tag{2d} \label{2d}
    \end{align*}
    Lastly, we let
    \begin{align*}
        \delta &= -\sum_{s=0}^T \alpha_s. \tag{3a} \label{4a}
    \end{align*}
    
    We divide the proof in three steps: We first state a series of structural lemmas, which we then use to prove that this solution is in fact feasible. 
    Finally, we compute the objective of this solution and conclude the bounds by weak duality.\\

    \emph{Structure of the solution.}
    We begin by stating two lemmas.

\begin{restatable}{lemma}{AdvTheoremDConstraints}
    \label{lemma: AdvTheoremDConstraints}
    Let $(\alphabf, \betabf)$ as in (\ref{1a})-(\ref{1d}) and (\ref{2a})-(\ref{2d}).
    Then, for all $j = 1, \ldots, T$ we have
    \begin{equation*}
        \frac{1- (1-\prob)^j}{\prob} + \frac{\market}{\prob} \beta_j + j \sum_{s=j}^{T} \alpha_s + \sum_{s=0}^{j-1} (s+1) \alpha_s = \frac{1- (1-\prob)^T}{\prob} + \frac{\market}{\prob} \beta_T + T \alpha_T + \sum_{s=0}^{T-1} (s+1) \alpha_s.
    \end{equation*}
    In other words, the right-hand side of \ref{cons: D2} is the same for $j=1,\ldots,T$.
\end{restatable}


    \begin{restatable}{lemma}{AdvTheoremProofSum}
    \label{lemma: AdvTheoremProofSum}
        Let $\alphabf$ as in (\ref{1a})-(\ref{1d}). Then, we have the following results:
        \begin{enumerate}
            \item \begin{equation*}
            \sum_{s=0}^{t^*-2} \left(s+1+\frac{\market}{\prob} \right) \alpha_s= \alpha_{t^*-1} \cdot \frac{\market(t^*-1)}{\prob} - \frac{1}{\prob} (1 - (1-\prob)^{t^*} - t^* \prob (1-\prob)^{t^*-1})
        \end{equation*}
        \item \begin{equation*}
            \sum_{s=0}^{t^*-2} \alpha_s = -\frac{1}{1 + \market} - \frac{\market}{1 + \market} \left( \frac{\market(1-\prob)}{\market +\prob} \right)^{t^* - 1} +  \alpha_{t^*-1} \left( \frac{\market - (\market + \prob)(\frac{\market}{\market + \prob})^{t^*}}{\prob} \right) + (1-\prob)^{t^*-1}
        \end{equation*}
        \item \begin{align*}
        & -  \alpha_T - \alpha_{t^*-1} \left( 1 + \frac{\market - (\market + \prob)(\frac{\market}{\market + \prob})^{t^*}}{\prob} \right) - (1-\prob)^{t^*-1}\\
        &= - \frac{(1-\prob)^T + (1-\prob)^{T - \lceil \frac{\market}{\prob}\rceil - 1} (\prob ( \lceil \frac{\market}{\prob} \rceil+1) -1)}{ \prob \lceil \frac{\market}{\prob}\rceil} \left(\frac{\market}{\market + \prob} \right)^{T - \lceil \frac{\market}{\prob}\rceil}.
    \end{align*}
\end{enumerate}
    \end{restatable}
We include the proofs of the lemmas in \Cref{proof: AdvTheoremDConstraints} and \Cref{proof: AdvTheoremProofSum} respectively; we use both to prove that our solution fulfills \ref{cons: D2} and also use the latter to compute the objective value of the solution. Lastly, we repeatedly use the following lemma, which we prove in \Cref{proof: AdvTheoremNegativeBeta}.
    
    \begin{restatable}{lemma}{AdvTheoremNegativeBeta}
        \label{lemma: AdvTheoremNegativeBeta}
        For any $\prob \in (0,1]$, 
        \begin{equation*}
            (1-\prob)^j - (1-\prob)^T + (T-j) \cdot \mu \alpha_T \leq 0, \quad \forall j \in \{t^*, \ldots, T-1\}
        \end{equation*}
\end{restatable}

    \emph{Feasibility.} 
    We show the feasibility of our proposed solution by verifying that it fulfills the following:
    \begin{enumerate}
        \item \ref{cons: D3}  ($0\leq \market/\prob \cdot \beta_0$)
        \item \ref{cons: D4} ($\alpha_j\leq 0$ for all $j$)
        \item \ref{cons: D5}, which is $\beta_j\leq 0$ for
        \begin{enumerate}
            \item  $j=0$
            \item  $j$ between $1$ and {$t^*-1$}
            \item  $j$ between $t^*$ and $T-1$
            \item  $j=T$
        \end{enumerate}
        \item \ref{cons: D1}
        \item \ref{cons: D2}
    \end{enumerate} 
    With these lemmas in hand, we first observe that constraints 1) and 3(a) are automatically fulfilled since we set $\beta_0=0$ in equation \eqref{2a}. We show below that 2) also holds true. As a result, it is immediate that 3(c) and 3(d) hold true (recall that we defined, in equations \eqref{2b} and \eqref{2d}, $\beta_j=\sum_{s=0}^{j-1}\alpha_s$ for $j=1,\ldots, {t^*-1}$ and $j=T$). Thus, it remains to show 2), 3(c), 4), and 5).
    

\paragraph{Proving $\alphabf \leq 0$ (\ref{cons: D4}).}
In  equation \eqref{1b} we set $\alpha_s = 0$ for $s = t^*, \ldots, T-1$; thus, we only have to check the cases $s = T$ and $s\leq t^*-1$. 

When $s=T$, we recall from equation~\eqref{1a} that $\alpha_T =  \frac{ (1-\prob)^T - (1-\prob)^{t^*} }{\prob \lceil \frac{\market}{\prob}\rceil}$. This is negative if $(1-\prob)^T \leq (1-\prob)^{T - \lceil \frac{\market}{\prob} \rceil}$, which holds true for $(1-\mu)\in(0,1)$.
    
    We next consider $s=t^*-1$, and find 
    \begin{align*}
        \alpha_{t^*-1} \leq 0 \quad &\Longleftrightarrow \quad (1-\prob)^{t^*-1} \left(1 - \left(\lceil \frac{\market}{\prob}\rceil+1\right)\prob \right) \leq  (1-\prob)^T\\
        &\Longleftrightarrow \quad 1-\prob - \prob\lceil \frac{\market}{\prob} \rceil \leq (1-\prob)^{\lceil \frac{\market}{\prob} \rceil +1}\\
        &\Longleftrightarrow \quad (1-\prob) (1- (1-\prob)^{\lceil \frac{\market}{\prob} \rceil}) \leq \prob\lceil \frac{\market}{\prob} \rceil\\
        &\Longleftarrow \quad (1- (1-\prob)^{\lceil \frac{\market}{\prob} \rceil}) \leq \prob\lceil \frac{\market}{\prob} \rceil.
    \end{align*}
    The first equivalence holds by definition (see \eqref{1c}), the second divides both sides by $(1-\prob)^{t^*-1} = (1-\prob)^{T - \lceil \frac{\market}{\prob} \rceil -1}$, the third follows by re-arranging terms and the last holds $1-\prob \in [0, 1)$.  To prove the final inequality, with $n=\lceil \frac{\market}{\prob} \rceil$ and $x=\mu$,  it suffices to show that $1-(1-x)^n \leq xn$ for all $x \in (0,1]$ and $n \geq 1$. Fix $x \in (0,1]$. For $n = 1$ the inequality is true. Assume by induction that $1-xn\leq (1-x)^n$ and note that
    \begin{equation*}
        1-x(n+1) \leq (1-x)^n - x \leq (1-x)^{n+1}
    \end{equation*}
    where the first inequality follows by the induction hypothesis, and the last follows by $(1-x)^n - (1-x)^{n+1} = (1-x)^n (1 - (1-x)) \leq x$, using that $x \in (0,1]$. Hence we have proven the inductive step and it follows that $\alpha_{t^*-1} \leq 0$.
    
    Lastly, for $s \leq t^*-1$, we find in equation \eqref{1d} that $\alpha_{s} \leq 0$ if and only if 
    $$
        \alpha_{s} \leq 0 \quad  \Longleftrightarrow  \quad \left(\frac{\market}{\market+\prob}\right)^{t^* - 1 - s} \alpha_{t^* -1} \leq \frac{\prob (1-\prob)^{s}}{ 1+ \market} \left( 1 - \left( \frac{\market(1-\prob)}{\market +\prob} \right)^{t^* - 1 - s} \right).
    $$
    As the the left hand side is non-positive and the right hand side is non-negative, this inequality holds true as well. 
    \paragraph{Proving $\betabf \leq 0$ (\ref{cons: D5}).} As argued at the beginning that $\beta_j\leq 0$ for $j\leq t^*$ and $j=T$, we only need to verify $\beta_j\leq 0$ for $j=t^*,\ldots, T-1$. In equation \eqref{2c} we defined
    \begin{equation*}
        \beta_j = \frac{(1-\prob)^j - (1-\prob)^T + (T-j) \cdot \mu \alpha_T}{\market} + \sum_{s=0}^{T-1} \alpha_s \quad \forall j = t^*, \ldots, T-1.
    \end{equation*}
    Since $\alphabf \leq \mathbf{0}$, we have $\sum_{s=0}^{T-1} \alpha_s\leq 0$  and to prove $\beta_j\leq 0$, it suffices to prove $(1-\prob)^j - (1-\prob)^T + (T-j) \cdot \mu \alpha_T \leq 0$, for $j \in \{t^*, \ldots, T-1\}$, which is the statement of Lemma \ref{lemma: AdvTheoremNegativeBeta}.

    \paragraph{Verifying the \ref{cons: D1} constraints.} We now verify that $
    \sum_{s=j}^T \alpha_s + \beta_j + \delta \leq 0$ for $j = 0, \ldots, T$.
    We first substitute, from equation \eqref{4a}, the value of $\delta = - \sum_{s=0}^T \alpha_s$ to obtain the equivalent inequality
    \begin{equation}
    \label{delta-ineq}
        \beta_j - \sum_{s=0}^{j-1} \alpha_s \leq 0, \quad \forall j = 0, \ldots, T.
    \end{equation}
    For $j=0, \ldots, {t^*-1}$ and $j=T$, equations \eqref{2b} and \eqref{2d} define $\beta_j = \sum_{s=0}^{j-1} \alpha_s$ and thus the left-hand side evaluates to 0 for these $j$ and the inequality is true. For $j \in \{t^*, \ldots, T-1\}$ we recall from equation \eqref{2c} that 
    \begin{equation*}
        \beta_{j} = \frac{(1-\prob)^j - (1-\prob)^T + (T-j) \cdot \mu \alpha_T}{\market} + \sum_{s=0}^{T-1} \alpha_s.
    \end{equation*}
    Substituting this expression into \eqref{delta-ineq} we can rewrite the equation, for $j = t^*, \ldots, T-1$,  equivalently as
    \begin{equation*}
        \frac{(1-\prob)^j - (1-\prob)^T + (T-j) \cdot \mu \alpha_T}{\market} + \sum_{s=j}^{T-1} \alpha_s \leq 0, \quad j = t^*, \ldots, T-1.
    \end{equation*}
    Since $\alphabf \leq \mathbf{0}$, it suffices to show that $(1-\prob)^j - (1-\prob)^T + (T-j) \cdot \mu \alpha_T \leq 0$ which holds for such~$j$ by Lemma~\ref{lemma: AdvTheoremNegativeBeta}. \paragraph{Verifying the \ref{cons: D2} constraints.} 
    
    {We next use Lemma~\ref{lemma: AdvTheoremDConstraints} along with equation \eqref{1b} to prove that the \ref{cons: D2} constraints are binding, i.e.,}
    \begin{align*}
        && 0 &= \frac{1- (1-\prob)^T}{\prob} + \frac{\market}{\prob} \beta_T + T \alpha_T + \sum_{s=0}^{t^*-1} (s+1) \alpha_s\\
        \Longleftrightarrow && 0 &= \frac{1- (1-\prob)^T}{\prob} + T\alpha_T + \sum_{s=0}^{t^*-1} \left(s+1+\frac{\market}{\prob} \right) \alpha_s\\
        \Longleftrightarrow && 0 &= \frac{1- (1-\prob)^T}{\prob} + T\alpha_T + \left(t^* + \frac{\market}{\prob} \right) \alpha_{t^*-1} + \sum_{s=0}^{t^*-2} \left(s+1+\frac{\market}{\prob} \right) \alpha_s. \tag{D-2'} \label{eqn: AdvTheoremD4Feasibility}
    \end{align*}
    Above, the intermediate step applies $\beta_T=\sum_{s=0}^{T-1} \alpha_s=\sum_{s=0}^{t^*-1}\alpha_s+\sum_{s=t^*}^{T-1}\alpha_s=\sum_{s=0}^{t^*-1}\alpha_s$ (by equation \eqref{2d} and \eqref{1b}). Thus, we may equivalently prove that \ref{eqn: AdvTheoremD4Feasibility} is true.
    We rely on the following equality from \Cref{lemma: AdvTheoremProofSum} (a):
    \begin{equation*}
            \sum_{s=0}^{t^*-2} \left(s+1+\frac{\market}{\prob} \right) \alpha_s= \alpha_{t^*-1} \cdot \frac{\market(t^*-1)}{\prob} - \frac{1}{\prob} (1 - (1-\prob)^{t^*} - t^* \prob (1-\prob)^{t^*-1}).
    \end{equation*}

    The equality allows us to replace the right-hand side of the last equality in \ref{eqn: AdvTheoremD4Feasibility} as follows
    \begin{align*}
        & \frac{1- (1-\prob)^T}{\prob} + T\alpha_T + \left(t^* + \frac{\market}{\prob} \right) \alpha_{t^*-1} + \sum_{s=0}^{t^*-2} \left(s+1+\frac{\market}{\prob} \right) \alpha_s\\
         = &  \frac{1- (1-\prob)^T}{\prob} + T\alpha_T + \left(t^* + \frac{\market}{\prob} \right) \alpha_{t^*-1} + \alpha_{t^*-1} \cdot \frac{\market(t^*-1)}{\prob} - \frac{1}{\prob} (1 - (1-\prob)^{t^*} - t^* \prob (1-\prob)^{t^*-1})\\
        = & \frac{1- (1-\prob)^T}{\prob} + T\alpha_T + t^* \left(1 + \frac{\market}{\prob} \right) \alpha_{t^*-1}  - \frac{1}{\prob} (1 - (1-\prob)^{t^*} - t^* \prob (1-\prob)^{t^*-1}).
    \end{align*}
    Finally, substituting the values of $\alpha_T$ and $\alpha_{t^*-1}$ from \eqref{1a} and \eqref{1c} we get
    \begin{align*}
        &\quad \frac{1- (1-\prob)^T}{\prob} + T \cdot \frac{(1-\prob)^T - (1-\prob)^{t^*}}{\prob \lceil \frac{\market}{\prob} \rceil} + t^* \left( 1 + \frac{\market}{\prob} \right) \frac{(1-\prob)^{t^*-1} (1 - (\lceil \frac{\market}{\prob} \rceil+1)\prob) - (1-\prob)^T }{\lceil \frac{\market}{\prob} \rceil (\market +\prob)} \\
        &\quad - \frac{1}{\prob} (1 - (1-\prob)^{t^*} - t^* \prob (1-\prob)^{t^*-1})\\
        &= \frac{(1-\prob)^{t^*} - (1-\prob)^T}{\prob} +  \frac{T \cdot((1-\prob)^T - (1-\prob)^{t^*})}{\prob \lceil \frac{\market}{\prob} \rceil} + t^* \frac{ (1-\prob)^{t^*} - (1-\prob)^T - \prob \lceil \frac{\market}{\prob} \rceil (1-\prob)^{t^*-1} }{\prob \lceil \frac{\market}{\prob} \rceil}\\
        &+ t^* (1-\prob)^{t^*-1}\\
        &= \frac{(1-\prob)^{t^*} - (1-\prob)^T}{\prob} +  \frac{T \cdot((1-\prob)^T - (1-\prob)^{t^*})}{\prob \lceil \frac{\market}{\prob} \rceil} + (T - \lceil \frac{\market}{\prob} \rceil) \frac{ (1-\prob)^{t^*} - (1-\prob)^T - \prob \lceil \frac{\market}{\prob} \rceil (1-\prob)^{t^*-1}}{\prob \lceil \frac{\market}{\prob} \rceil} \\
        & \quad + t^* (1-\prob)^{t^*-1}\\
        &= \frac{(1-\prob)^{t^*} - (1-\prob)^T}{\prob} +  \frac{T \cdot((1-\prob)^T - (1-\prob)^{t^*})}{\prob \lceil \frac{\market}{\prob} \rceil} + \frac{T \cdot((1-\prob)^{t^*} - (1-\prob)^T)}{\prob \lceil \frac{\market}{\prob} \rceil} - T (1-\prob)^{t^* - 1} \\
        &\quad - \frac{(1-\prob)^{t^*} - (1-\prob)^T}{\prob} + \lceil \frac{\market}{\prob} \rceil (1-\prob)^{t^* - 1} + t^* (1-\prob)^{t^* - 1}\\
        &= t^* (1-\prob)^{t^* - 1} - (T - \lceil \frac{\market}{\prob} \rceil)(1-\prob)^{t^* - 1}\\
        &= 0.
    \end{align*}
    With this equality, \ref{eqn: AdvTheoremD4Feasibility} is proven and the feasibility of the solution follows.

\paragraph{Computing the objective value.}
    We now show that the objective value of our feasible solution  matches the bound in the lemma statement. Using equations \eqref{4a} and \eqref{1b}, we compute the objective as  $\delta = -\sum_{s=0}^{T} \alpha_s = -\alpha_T -\sum_{s=0}^{t^*-1} \alpha_s$. Our proof here relies on \Cref{lemma: AdvTheoremProofSum} (b) and (c), which respectively state
    \begin{align*}
            \sum_{s=0}^{t^*-2} \alpha_s = -\frac{1}{1 + \market} - \frac{\market}{1 + \market} \left( \frac{\market(1-\prob)}{\market +\prob} \right)^{t^* - 1} +  \alpha_{t^*-1} \left( \frac{\market - (\market + \prob)(\frac{\market}{\market + \prob})^{t^*}}{\prob} \right) + (1-\prob)^{t^*-1},
    \end{align*}
    \begin{align*}
        \text{and} &\quad -  \alpha_T - \alpha_{t^*-1} \left( 1 + \frac{\market - (\market + \prob)(\frac{\market}{\market + \prob})^{t^*}}{\prob} \right) - (1-\prob)^{t^*-1}\\ &= - \frac{(1-\prob)^T + (1-\prob)^{T - \lceil \frac{\market}{\prob}\rceil - 1} (\prob ( \lceil \frac{\market}{\prob} \rceil+1) -1)}{ \prob \lceil \frac{\market}{\prob}\rceil} \left(\frac{\market}{\market + \prob} \right)^{T - \lceil \frac{\market}{\prob}\rceil}.
    \end{align*}
    With these, we derive 
    \begin{align*}
        \delta &= - \alpha_T - \alpha_{t^*-1} - \sum_{s=0}^{t^*-2} \alpha_s \\
        &= - \alpha_T - \alpha_{t^*-1} \left( 1 + \frac{\market - (\market + \prob)(\frac{\market}{\market + \prob})^{t^*}}{\prob} \right) +\frac{1}{1 + \market} + \frac{\market}{1 + \market} \left( \frac{\market(1-\prob)}{\market +\prob} \right)^{t^* - 1} - (1-\prob)^{t^*-1}\\
        &= \frac{1}{1 + \market} + \frac{\market}{1 + \market} \left( \frac{\market(1-\prob)}{\market +\prob} \right)^{T - \lceil \frac{\market}{\prob}\rceil - 1} -  \alpha_T - \alpha_{t^*-1} \left( 1 + \frac{\market - (\market + \prob)(\frac{\market}{\market + \prob})^{t^*}}{\prob}  \right) - (1-\prob)^{t^*-1}\\
        &= \frac{1}{1 + \market} + \frac{\market}{1 + \market} \left( \frac{\market(1-\prob)}{\market +\prob} \right)^{T - \lceil \frac{\market}{\prob}\rceil - 1} - \frac{(1-\prob)^T + (1-\prob)^{T - \lceil \frac{\market}{\prob}\rceil - 1} (\prob ( \lceil \frac{\market}{\prob} \rceil+1) -1)}{ \prob \lceil \frac{\market}{\prob}\rceil} \left(\frac{\market}{\market + \prob} \right)^{T - \lceil \frac{\market}{\prob}\rceil}.
    \end{align*}


For each case we provided a feasible solution and computed the corresponding objective value. Given that $\DLP$ is a maximization problem, the value of $\DLP$ is bounded from below by the objective value of these feasible solutions, as stated in \Cref{prop: AdvCR}. This establishes our lower bounds.
\hfill\Halmos 

The remainder of this section is dedicated to proving the lemmas that were omitted in the proof of Case~2.
\subsection{Proof of \Cref{lemma: AdvTheoremDConstraints}}
\label{proof: AdvTheoremDConstraints}
Our proof relies on the following claim which we prove at the end of this section.

\begin{restatable}{claim}{AdvTheoremBetaRecurrence}
Let $(\alphabf, \betabf)$ as in (\ref{1a})-(\ref{1d}) and (\ref{2a})-(\ref{2d}). Then, we have
\begin{equation}
    \label{eq-proof: AdvRecurrenceRelation}
    \beta_{j} = \beta_{j+1} + \frac{\prob}{\market} \left( (1-\prob)^j + \sum_{s=j}^T \alpha_s \right), \quad \forall j = 1, \ldots, T-1.
\end{equation}
\end{restatable}


\AdvTheoremDConstraints*
    We proceed by induction. It is immediate that the lemma statement holds for $j=T$. We now suppose that it holds for  $j+1 \leq T$ and  prove that it holds for $j$.  As we assume that $(\alphabf,\betabf)$ satisfy (\ref{1a})-(\ref{1d}) and (\ref{2a})-(\ref{2d}),  Claim \eqref{eq-proof: AdvRecurrenceRelation} guarantees that $\beta_{j} = \beta_{j+1} + \frac{\prob}{\market}\left( (1-\prob)^j + \sum_{s=j}^T \alpha_s \right)$ for $j = 1, \ldots, T-1$. With that substitution we obtain the first equality in the following derivation
    \begin{align*}
         &\quad \frac{1-(1-\prob)^j}{\prob} + \frac{\market}{\prob} \beta_j + j \sum_{s=j}^{T} \alpha_s + \sum_{s=0}^{j-1} (s+1) \alpha_s\\ &= \frac{1-(1-\prob)^j}{\prob} + \frac{\market}{\prob} \left( \beta_{j+1} + \frac{\prob}{\market} \left( (1-\prob)^j + \sum_{s=j}^T \alpha_s \right)\right) + j \sum_{s=j}^{T} \alpha_s + \sum_{s=0}^{j-1} (s+1) \alpha_s\\
         &= \frac{1-(1-\prob)^j}{\prob} + (1-\prob)^j + \frac{\market}{\prob} \beta_{j+1} + (j+1) \sum_{s=j}^T \alpha_s + \sum_{s=0}^{j-1}(s+1) \alpha_s\\
         &= \frac{1-(1-\prob)^j + \prob (1-\prob)^j}{\prob}  + \frac{\market}{\prob} \beta_{j+1} + (j+1) \sum_{s=j+1}^T \alpha_s + (j+1)\alpha_j +  \sum_{s=0}^{j-1}(s+1) \alpha_s\\
         &= \frac{1 - (1-\prob)^j(1-\prob)}{\prob} + \frac{\market}{\prob} \beta_{j+1} + (j+1) \sum_{s=j+1}^T \alpha_s + \sum_{s=0}^{j} (s+1) \alpha_s\\
         &= \frac{1 - (1-\prob)^{j+1}}{\prob} + \frac{\market}{\prob} \beta_{j+1} + (j+1) \sum_{s=j+1}^T \alpha_s + \sum_{s=0}^{j} (s+1) \alpha_s\\
         &= \frac{1- (1-\prob)^T}{\prob} + \frac{\market}{\prob} \beta_T + T \alpha_T + \sum_{s=0}^{T-1} (s+1) \alpha_s.
    \end{align*}
    The remaining equalities are algebraic manipulations except for the last one, in which we use the inductive hypothesis. Thus, to complete the proof of the lemma we only need to prove Claim \eqref{eq-proof: AdvRecurrenceRelation}.

\textbf{Proof of Claim \eqref{eq-proof: AdvRecurrenceRelation}.}
We distinguish between the cases $j \geq t^*$ and $j < t^*$.


\emph{Case 1: $j\in\{t^*, \ldots, T-1\}$.} We first recall that, for such $j$, \eqref{1b} and \eqref{2c} guarantee that $$\sum_{s=j}^T \alpha_s = \alpha_T\quad\text{ and }\quad\beta_j = \frac{(1-\prob)^j - (1-\prob)^T + (T-j) \cdot \mu \alpha_T}{\market} + \sum_{s=0}^{T-1} \alpha_s.$$ 
We remark that, due to \eqref{2d}, the latter equality also holds for $j=T$.
From these we obtain the first equality in the following derivation:
        \begin{align*}
            \beta_{j+1} + \frac{\prob}{\market} \left( (1-\prob)^j + \sum_{s=j}^T \alpha_s \right) &=  \frac{(1-\prob)^{j+1} - (1-\prob)^T + (T-j-1) \cdot \mu \alpha_T}{\market} + \sum_{s=0}^{T-1} \alpha_s  + \frac{\prob}{\market} \left( (1-\prob)^j + \alpha_T \right)\\
            &= \frac{(1-\prob)^{j+1} + \prob (1-\prob)^j - (1-\prob)^T + (T-j) \cdot \mu \alpha_T}{\market}+  \sum_{s=0}^{T-1} \alpha_s\\
            &=  \frac{(1-\prob)^{j} ((1-\prob) + \prob) - (1-\prob)^T + (T-j) \cdot \mu \alpha_T}{\market}+  \sum_{s=0}^{T-1} \alpha_s\\
            &= \frac{(1-\prob)^{j} - (1-\prob)^T + (T-j) \cdot \mu \alpha_T}{\market}+  \sum_{s=0}^{T-1} \alpha_s\\
            &= \beta_j.
        \end{align*}
        The remaining equalities are algebraic manipulations, except for the last which substitutes the definition of $\beta_j$. This concludes the first case.

\emph{Case 2: $j \in \{1, \ldots, {t^*-1}\}$.} For such $j$ \eqref{2b} sets $\beta_j = \sum_{s=0}^{j-1} \alpha_s$. Note that when $j=t^*-1$ the term $\beta_{t^*}$ appears in the right hand side of the claim. We claim that we also have that $\beta_{t^*} = \sum_{s=0}^{t^*-1}\alpha_s$. This is easily verified by applying \eqref{2c}, substituting the value of $\alpha_T$ from \eqref{1a} and $(T-t)^*=\frac{\market}{\prob}$, and then cancelling out terms in the below derivation:
        \begin{align*}
            \beta_{t^*} &= \frac{(1-\prob)^{t^*} - (1-\prob)^T + (T-t^*) \cdot \mu \alpha_T}{\market} + \sum_{s=0}^{T-1} \alpha_s\\
            &= \frac{(1-\prob)^{t^*} - (1-\prob)^T + \lceil \frac{\market}{\prob} \rceil \cdot \mu \cdot \frac{ (1-\prob)^T - (1-\prob)^{t^*} }{\prob \lceil \frac{\market}{\prob}\rceil}}{\market}  + \sum_{s=0}^{T-1} \alpha_s\\
            &=   \sum_{s=0}^{T-1} \alpha_s= \sum_{s=0}^{t^*-1} \alpha_s.
        \end{align*}
        
        With this in mind, we want to prove
        \begin{align*}
            & & \quad \beta_{j} &= \beta_{j+1} + \frac{\prob}{\market} \left( (1-\prob)^j + \sum_{s=j}^T \alpha_s \right), & \forall j = 1, \ldots, t^*-1\\
            \Longleftrightarrow & & \sum_{s=0}^{j-1} \alpha_s &= \sum_{s=0}^{j} \alpha_s + \frac{\prob}{\market} \left( (1-\prob)^j + \sum_{s=j}^T \alpha_s \right), & \forall j = 1, \ldots, t^*-1\\
            \Longleftrightarrow & & -\alpha_j &= \frac{\prob}{\market} \left( (1-\prob)^j + \sum_{s=j+1}^T \alpha_s + \alpha_j \right), & \forall j = 1, \ldots, t^*-1\\
            \Longleftrightarrow & & \alpha_j &= -\frac{\prob}{\market +\prob} \left( (1-\prob)^j + \sum_{s=j+1}^T \alpha_s \right), & \forall j = 1, \ldots, t^*-1.
        \end{align*}
        The first equivalence comes from $\beta_j = \sum_{s=0}^{j-1} \alpha_s$, the second from subtracting $\sum_{s=0}^{j} \alpha_s$ on both sides, and the third from further algebraic manipulations. Thus, to complete the proof we can equivalently show that for $j = 1, \ldots, t^*-1$

    \begin{equation}
        \label{eqn: AdvTheoremAlphaRecurrence}
        \alpha_j = -\frac{\prob}{\market+\prob} \left( (1-\prob)^j + \sum_{s=j+1}^T \alpha_s \right).
    \end{equation}

We proceed inductively. For the base case, consider $j = t^*-1$. In this case, using that $\alpha_{s} = 0$ for $s=t^*, \ldots, T-1$, equation (\ref{eqn: AdvTheoremAlphaRecurrence}) becomes
    \begin{equation*}
        \alpha_{t^*-1} = -\frac{\prob}{\market +\prob} \left( (1-\prob)^{t^*-1} + \alpha_T \right).
    \end{equation*}
    We prove this via the following derivation
    \begin{align*}
        -\frac{\prob}{\market +\prob} \left( (1-\prob)^{t^*-1} + \alpha_T \right) &= -\frac{\prob}{\market +\prob} \left( (1-\prob)^{t^*-1} + \frac{ (1-\prob)^T - (1-\prob)^{t^*} }{\prob \lceil \frac{\market}{\prob}\rceil}\right)\\
        &= - \frac{ (1-\prob)^T - (1-\prob)^{t^*} + \prob \lceil \frac{\market}{\prob} \rceil  (1-\prob)^{t^*-1} }{\lceil \frac{\market}{\prob} \rceil \cdot (\market+\prob)}\\
        &= \frac{ (1-\prob)^{t^*-1}( (1-\prob) - \prob \lceil\frac{\market}{\prob} \rceil) - (1-\prob)^T}{\lceil \frac{\market}{\prob} \rceil \cdot (\market+\prob)}\\
        &= \frac{(1-\prob)^{t^*-1} \left(1 - (\lceil \frac{\market}{\prob}\rceil+1)\prob \right) - (1-\prob)^T}{\lceil \frac{\market}{\prob} \rceil \cdot (\market+\prob)}\\
        &= \alpha_{t^*-1},
    \end{align*}
    where the first equality substitutes the definition of $\alpha_T$ from \eqref{1a}, the last equality substitutes the definition of $\alpha_{t^*-1}$ from \eqref{1c}, and the other equalities follow by rearranging terms.  This yields the base case $j = t^*-1$.

    To prove the inductive step, we use the following equivalence
    \begin{equation}
    \label{eqn-proof: AdvTheoremAlphaDifference}
        \alpha_j - \frac{\market}{\market+\prob}\alpha_{j+1} = - \frac{\prob^2}{\market+\prob} (1-\prob)^j, \quad \forall j = 0, \ldots, t^*-2,
    \end{equation}
    which we verify using the definition of $\alpha_j$ from \eqref{1d}. Fixing $j \in \{0, \ldots, t^*-2\}$ we have 
    \begin{align*}
        \alpha_j - \frac{\market}{\market+\prob}\alpha_{j+1} &= \left(\frac{\market}{\market+\prob}\right)^{t^* - 1 - j} \alpha_{t^* -1} - \frac{\prob (1-\prob)^{j}}{ 1+ \market} \left( 1 - \left( \frac{\market(1-\prob)}{\market +\prob} \right)^{t^* - 1 - j} \right)\\
        &\quad - \frac{\market}{\market+\prob} \left( \left(\frac{\market}{\market+\prob}\right)^{t^* - 2 - j} \alpha_{t^* -1} - \frac{\prob (1-\prob)^{j+1}}{ 1+ \market} \left( 1 - \left( \frac{\market(1-\prob)}{\market +\prob} \right)^{t^* - 2 - j} \right)\right)\\
        &=   \frac{\prob \market (1-\prob)^{j+
        1}}{(1+\market) (\market+\prob)} \left( 1 - \left( \frac{\market(1-\prob)}{\market+\prob} \right)^{t^* - 2 - j} \right)  - \frac{\prob (1-\prob)^{j}}{1+ \market} \left( 1 - \left( \frac{\market(1-\prob)}{\market+\prob} \right)^{t^* - 1 - j} \right) \\
        &= \frac{\prob(1-\prob)^j}{1+\market} \left( \frac{\market(1-\prob)}{\market+\prob} - \left( \frac{\market(1-\prob)}{\market+\prob} \right)^{t^* - 1 - j} - 1 + \left( \frac{\market(1-\prob)}{\market+\prob} \right)^{t^* - 1 - j}\right)\\
        &= \frac{\prob(1-\prob)^j}{1+\market} \cdot \frac{-(1+\market)\prob}{\market+\prob}\\
        &= -\frac{\prob^2}{\market+\prob} (1-\prob)^j.
    \end{align*}
    
    We now finish the proof by completing the inductive step.  Suppose \eqref{eqn: AdvTheoremAlphaRecurrence} holds for some $j \in \{2, \ldots, t^*-1\}$. Equivalently, we have
    \begin{equation}
        \label{eqn: proof-AdvAlphaRecurrence2}
        - \frac{\market+\prob}{\prob}\alpha_j - (1-\prob)^j = \sum_{s=j+1}^T \alpha_s.
    \end{equation}
We can then complete the induction step by showing \eqref{eqn: AdvTheoremAlphaRecurrence} for $j-1$: 
    \begin{align*}
        -\frac{\prob}{\market+\prob} \left( (1-\prob)^{j-1} + \sum_{s=j}^T \alpha_s \right) &= -\frac{\prob}{\market+\prob} \left( (1-\prob)^{j-1} + \alpha_j + \sum_{s=j+1}^T \alpha_s \right)\\
        &= -\frac{\prob}{\market+\prob} \left( (1-\prob)^{j-1} + \alpha_j - \frac{\market+\prob}{\prob}\alpha_j - (1-\prob)^j \right)\\
        &= -\frac{\prob}{\market+\prob} \left( (1-\prob)^{j-1} (1- (1-\prob))  + \alpha_j \left( 1 - \frac{\market+\prob}{\prob} \right)\right)\\
        &= -\frac{\prob}{\market+\prob} \left( (1-\prob)^{j-1} (1- (1-\prob))  - \alpha_j  \frac{\market}{\prob} \right)\\
        &= -\frac{\prob^2}{\market+\prob} (1-\prob)^{j-1} + \frac{\market}{\market+\prob} \alpha_j\\
        &= \alpha_{j-1}.
    \end{align*}
    Here, the second equalitaty used \eqref{eqn: proof-AdvAlphaRecurrence2}, and the last equality used \eqref{eqn-proof: AdvTheoremAlphaDifference}. Thus, we have shown \eqref{eqn: AdvTheoremAlphaRecurrence} for $j=1,\ldots,t^*-1$ which completes the proof of the claim.   \hfill\Halmos

\subsection{Proof of \Cref{lemma: AdvTheoremProofSum}}
\label{proof: AdvTheoremProofSum}
We begin with a claim that proves useful in our computations; we include the proof at the end of this section.
\begin{restatable}{claim}{AdvArithmeticoGeometric}
 \label{claim: AdvArithmeticoGeometric}
    We have
        \begin{align*}
            \sum_{u=1}^{t^*-1} u \left(\frac{\market}{\market+\prob}\right)^{u} &= -\frac{(\market + \mu) \left( \market \left(  \left( \frac{\market}{\market + \mu} \right)^{t^*} -1 \right) + \mu t^* \left( \frac{\market}{\market + \mu} \right)^{t^*} \right)}{\mu^2},\\
            \sum_{s=0}^{t^*-2} s (1-\prob)^s &= \frac{-\prob^2 + (1-\prob)^{t^*} + \prob(t^*(1-\prob)^{t^*} - 2(1-\prob)^{t^*} + 2) - 1}{(\prob - 1)\prob^2}.
        \end{align*}
        Furthermore, these results imply
        \begin{align*}
            \sum_{s=0}^{t^*-2} \left(s+1+\frac{\market}{\prob} \right) \left(\frac{\market}{\market+\prob}\right)^{t^* - 1 - s} &= \frac{\market(t^*-1)}{\prob}, \tag{AM-GM1} \label{eq-proof: AMGM1}\\
            \sum_{s=0}^{t^*-2} \left(s+1+\frac{\market}{\prob} \right) (1-\prob)^s &= \frac{(1 - \prob)^{t^* - 1} \left( -(\market + 1 + \prob (t^* - 1)) \right) + \market + 1}{\prob^2}. \tag{AM-GM2} \label{eq-proof: AMGM2}
        \end{align*}
\end{restatable}

We now restate and prove the lemma. 

\AdvTheoremProofSum*

\emph{Proof of \Cref{lemma: AdvTheoremProofSum}.}
    The proof is organized sequentially, proving (1.) first, then (2.) and lastly (3.).
    \begin{enumerate}
        \item We prove the first result, using  equations \ref{eq-proof: AMGM1} and \ref{eq-proof: AMGM2} from the claim:
        %
    \begin{align*}
        \sum_{s=0}^{t^*-2} \left(s+1+\frac{\market}{\prob} \right) \alpha_s &= \sum_{s=0}^{t^*-2} \left(s+1+\frac{\market}{\prob} \right) \left(\left(\frac{\market}{\market+\prob}\right)^{t^* - 1 - s} \alpha_{t^* -1} - \frac{\prob (1-\prob)^{s}}{ 1+ \market} \left( 1 - \left( \frac{\market(1-\prob)}{\market +\prob} \right)^{t^* - 1 - s} \right)\right)\\ 
        &= \alpha_{t^* -1}\sum_{s=0}^{t^*-2} \left(s+1+\frac{\market}{\prob} \right) \left(\frac{\market}{\market+\prob}\right)^{t^* - 1 - s}\\
        &\quad - \sum_{s=0}^{t^*-2} \left(s+1+\frac{\market}{\prob} \right)\frac{\prob (1-\prob)^{s}}{1+\market} \left( 1 - \left( \frac{\market(1-\prob)}{\market+\prob} \right)^{t^* - 1 - s} \right)\\
        &=\alpha_{t^* -1}\sum_{s=0}^{t^*-2} \left(s+1+\frac{\market}{\prob} \right)\left(\frac{\market}{\market+\prob}\right)^{t^* - 1 - s} - \frac{\prob}{1+\market} \sum_{s=0}^{t^*-2} \left(s+1+\frac{\market}{\prob} \right) (1-\prob)^s\\
        & \quad + \frac{\prob (1-\prob)^{t^*-1}}{1+\market} \sum_{s=0}^{t^*-2} \left(s+1+\frac{\market}{\prob} \right) \left(\frac{\market}{\market+\prob}\right)^{t^* - 1 - s}\\
        &= \alpha_{t^*-1} \cdot \frac{\market(t^*-1)}{\prob} - \frac{\prob}{1+\market} \cdot \frac{(1 - \prob)^{t^* - 1} \left( -(\market + 1 + \prob (t^* - 1)) \right) + \market + 1}{\prob^2}\\
        &\quad + \frac{\prob (1-\prob)^{t^*-1}}{1+\market} \cdot \frac{\market(t^*-1)}{\prob}\\
        &= \alpha_{t^*-1} \cdot \frac{\market(t^*-1)}{\prob} + \frac{(1-\prob)^{t^*-1} (\prob \market (t^*-1) + \prob (t^*-1) + \market + 1) - (\market + 1) }{\prob(1+\market)} \\
        &= \alpha_{t^*-1} \cdot \frac{\market(t^*-1)}{\prob} + \frac{(1-\prob)^{t^*-1} (\prob(t^*-1) (\market + 1) + \market + 1) - (\market + 1) }{\prob(1+\market)}\\
            &= \alpha_{t^*-1} \cdot \frac{\market(t^*-1)}{\prob} - \frac{1}{\prob} (1 - (1-\prob)^{t^*} - t^* \prob (1-\prob)^{t^*-1}).
    \end{align*}
    Where every equality is algebraic manipulation except the fourth one, in which we replaced the aforementioned sums (\ref{eq-proof: AMGM1} and \ref{eq-proof: AMGM2}).
    \item We prove the second equation. This simply follows from the definition of $\alpha_s$ when $s \leq t^*-2$ (Equation \ref{1d}) and identifying geometric sums:
    \begin{align*}
         \sum_{s=0}^{t^*-2} \alpha_s  &=  \sum_{s=0}^{t^*-2} \left(\left(\frac{\market}{\market+\prob}\right)^{t^* - 1 - s} \alpha_{t^* -1} - \frac{\prob (1-\prob)^{s}}{ 1+ \market} \left( 1 - \left( \frac{\market(1-\prob)}{\market +\prob} \right)^{t^* - 1 - s} \right) \right)\\
        &=  \alpha_{t^*-1} \sum_{s=0}^{t^*-2} \left(\frac{\market}{\market+\prob}\right)^{t^* - 1 - s} - \frac{\prob}{1 + \market} \left( \sum_{s=0}^{t^*-2} (1-\prob)^s - \sum_{s=0}^{t^*-2} (1-\prob)^s \left( \frac{\market(1-\prob)}{\market +\prob} \right)^{t^* - 1 - s}  \right)\\
        &=  \alpha_{t^*-1} \left( \frac{\market - (\market + \prob)(\frac{\market}{\market + \prob})^{t^*}}{\prob} \right) - \frac{\prob}{1 + \market} \left( \frac{1 - (1-\prob)^{t^*-1}}{\prob} - (1-\prob)^{t^*-1} \sum_{s=0}^{t^*-2} \left(\frac{\market}{\market+\prob} \right)^{t^*-1-s} \right) \\
        &= \alpha_{t^*-1} \left( \frac{\market - (\market + \prob)(\frac{\market}{\market + \prob})^{t^*}}{\prob} \right) - \frac{\prob}{1+ \market} \left( \frac{1 - (1-\prob)^{t^*-1}}{\prob} - (1-\prob)^{t^*-1}\cdot \left( \frac{\market - (\market + \prob)(\frac{\market}{\market + \prob})^{t^*}}{\prob} \right) \right) \\
        &= \alpha_{t^*-1} \left( \frac{\market - (\market + \prob)(\frac{\market}{\market + \prob})^{t^*}}{\prob} \right) -  \frac{1-(1+\market)(1-\prob)^{t^*-1} + (1-\prob)^{t^*-1} (\market + \prob) (\frac{\market}{\market + \prob})^{t^*}}{1 + \market}\\
        &= -\frac{1}{1 + \market} - \frac{\market}{1 + \market} \left( \frac{\market(1-\prob)}{\market +\prob} \right)^{t^* - 1} +  \alpha_{t^*-1} \left( \frac{\market - (\market + \prob)(\frac{\market}{\market + \prob})^{t^*}}{\prob} \right) + (1-\prob)^{t^*-1}.
    \end{align*}
    \item We prove the last equation which is a simple algebraic result once replacing the values of $\alpha_T$ and $\alpha_{t^*-1}$ (given in equations \ref{1a} and \ref{1c} respectively). Hence,
    \begin{align*}
        &\quad  -  \alpha_T - \alpha_{t^*-1} \left( 1 + \frac{\market - (\market + \prob)(\frac{\market}{\market + \prob})^{t^*}}{\prob} \right) - (1-\prob)^{t^*-1}\\
        &=- \frac{ (1-\prob)^T - (1-\prob)^{t^*} }{\prob \lceil \frac{\market}{\prob}\rceil} - \frac{(1-\prob)^{t^*-1} \left(1 - (\lceil \frac{\market}{\prob}\rceil+1)\prob \right) - (1-\prob)^T}{\lceil \frac{\market}{\prob} \rceil \cdot (\market+\prob)} \cdot \frac{(\market+\prob)- (\market + \prob)(\frac{\market}{\market + \prob})^{t^*}}{\prob} \\
        &\quad - (1-\prob)^{t^*-1}\\
        &=  \frac{(1-\prob)^{t^*} - (1-\prob)^T - \left(1- (\frac{\market}{\market + \prob})^{t^*}\right) \left((1-\prob)^{t^*-1} \left(1 - (\lceil \frac{\market}{\prob}\rceil+1)\prob \right) - (1-\prob)^T \right)}{\prob \lceil \frac{\market}{\prob}\rceil} - (1-\prob)^{t^*-1}\\
        &= \frac{(1-\prob)^{t^*} - (1-\prob)^T - \left( (1-\prob)^{t^*} - \lceil \frac{\market}{\prob} \rceil (1-\prob)^{t^*-1} - (1-\prob)^T \right)}{\prob \lceil \frac{\market}{\prob}\rceil}\\
        &\quad  + \frac{ (\frac{\market}{\market + \prob})^{t^*} \left((1-\prob)^{t^*-1} \left(1 - (\lceil \frac{\market}{\prob}\rceil+1)\prob \right) - (1-\prob)^T \right)}{\prob \lceil \frac{\market}{\prob} \rceil} - (1-\prob)^{t^*-1}\\
        &= \frac{ (\frac{\market}{\market + \prob})^{t^*} \left((1-\prob)^{t^*-1} \left(1 - (\lceil \frac{\market}{\prob}\rceil+1)\prob \right) - (1-\prob)^T \right)}{\prob \lceil \frac{\market}{\prob} \rceil}\\
        &= - \frac{(1-\prob)^T + (1-\prob)^{T - \lceil \frac{\market}{\prob}\rceil - 1} (\prob ( \lceil \frac{\market}{\prob} \rceil+1) -1)}{ \prob \lceil \frac{\market}{\prob}\rceil} \left(\frac{\market}{\market + \prob} \right)^{T - \lceil \frac{\market}{\prob}\rceil}. 
    \end{align*}
    \hfill \Halmos
    \end{enumerate}

\subsubsection*{Proof of Claim.}

        We use the closed form expression of the arithmetico-geometric sum for $r \not= 1$:
        \begin{equation}
        \label{eq: arithmetico-geometric}
        \sum_{k=1}^{n} k r^k = \frac{r(1-r^n)}{(1-r)^2} - \frac{n r^{n+1}}{1-r}
        \end{equation}
        Hence,
        \begin{align*}
            \sum_{u=1}^{t^*-1} u \left(\frac{\market}{\market+\prob}\right)^{u} &= \frac{ \frac{\market}{\market + \prob} (1 - ( \frac{\market}{\market + \prob})^{t^*-1})}{(1 - \frac{\market}{\market + \prob})^2} - \frac{(t^*-1) (\frac{\market}{\market + \prob})^{t^*}}{(1 - \frac{\market}{\market + \prob})}\\
            &= \left(\frac{\market + \prob}{\prob}\right)^2 \left( \frac{\market}{\market + \prob} - \left( \frac{\market}{\market + \prob}\right)^{t^*} - \frac{\prob}{\market + \prob} (t^*-1) \left(\frac{\market}{\market + \prob}\right)^{t^*} \right)\\
            &= \left(\frac{\market + \prob}{\prob}\right)^2 \left( \frac{\market}{\market + \prob} - \left(1 +  \frac{\prob}{\market + \prob} (t^*-1) \right) \left(\frac{\market}{\market + \prob}\right)^{t^*} \right)\\
            &= \frac{\market + \prob}{\prob^2} \left( \market - (\market + \prob t^*) \left(\frac{\market}{\market + \prob}\right)^{t^*} \right)\\
            &= -\frac{(\market + \mu) \left( \market \left(  \left( \frac{\market}{\market + \mu} \right)^{t^*} -1 \right) + \mu t^* \left( \frac{\market}{\market + \mu} \right)^{t^*} \right)}{\mu^2}.
        \end{align*}
        Where the first equality used the closed form of the arithmetico-geometric sum (\ref{eq: arithmetico-geometric}), and the rest is algebraic manipulation.  With this result in mind, we prove \ref{eq-proof: AMGM1}:
        \begin{align*}
         & \sum_{s=0}^{t^*-2} \left(s+1+\frac{\market}{\prob} \right) \left(\frac{\market}{\market+\prob}\right)^{t^* - 1 - s}\\
         &= \sum_{s=0}^{t^*-2} s \left(\frac{\market}{\market+\prob}\right)^{t^* - 1 - s} + \left(1 + \frac{\market}{\prob}\right) \sum_{s=0}^{t^*-2} \left(\frac{\market}{\market+\prob}\right)^{t^* - 1 - s}\\
         &= \sum_{u=1}^{t^*-1} (t^*-1-u) \left(\frac{\market}{\market+\prob}\right)^{u}  + \left(1 + \frac{\market}{\prob}\right) \sum_{u=1}^{t^*-1} \left(\frac{\market}{\market+\prob}\right)^{u}\\
         &= \left(t^* + \frac{\market}{\prob}\right) \sum_{u=1}^{t^*-1} \left(\frac{\market}{\market+\prob}\right)^{u} - \sum_{u=1}^{t^*-1} u \left(\frac{\market}{\market+\prob}\right)^{u}\\
         &= \left(t^* + \frac{\market}{\prob}\right) \cdot \frac{\market -(\market+\prob) (\frac{\market}{\market + \prob})^{t^*}}{\prob} + \frac{(\market + \mu) \left( \market \left(  \left( \frac{\market}{\market + \mu} \right)^{t^*} -1 \right) + \mu t^* \left( \frac{\market}{\market + \mu} \right)^{t^*} \right)}{\mu^2}\\
         &= \frac{(\prob t^* + \market) \market - (\prob t^* + \market) (\market + \prob)(\frac{\market}{\market + \prob})^{t^*} + (\market + \prob) \left( (\prob t^* + \market)\left( \frac{\market}{\market + \mu} \right)^{t^*} -\market \right)}{\prob^2}\\
         &= \frac{(\prob t^* + \market) \market -\market (\market + \prob)}{\prob^2}\\
         &= \frac{\market(t^*-1)}{\prob}.
    \end{align*}
    
        On the other hand, we compute
        \begin{align*}
            \sum_{s=0}^{t^*-2} s (1-\prob)^s &= \frac{(1-\prob) (1-(1-\prob)^{t^*-2})}{\prob^2} - \frac{(t^*-2) (1-\prob)^{t^*-1}}{\prob}\\
            &= \frac{(1-\prob)}{\prob^2} (1-(1-\prob)^{t^*-2} - \prob(t^*-2)(1-\prob)^{t^*-2})\\
            &= \frac{(1-\prob)}{\prob^2} (1 - (1-\prob)^{t^*-2}(1 + \prob (t^*-2) ) )\\
            &= \frac{(1-\prob)^2 - (1-\prob)^{t^*}(1 + \prob (t^*-2) )}{\prob^2 (1-\prob)}\\
            &= \frac{1 - 2\prob + \prob^2 - (1-\prob)^{t^*} - \prob(t^* -2 )(1-\prob)^{t^*} }{\prob^2 (1-\prob)}\\
            &= \frac{\prob^2 - (1-\prob)^{t^*} - \prob((t^* -2 )(1-\prob)^{t^*} + 2) +1 }{\prob^2 (1-\prob)}\\
             &= \frac{-\prob^2 + (1-\prob)^{t^*} + \prob(t^*(1-\prob)^{t^*} - 2(1-\prob)^{t^*} + 2) - 1}{(\prob - 1)\prob^2}.
        \end{align*}
        Where, again, the first equality used the closed form of the arithmetico-geometric sum (\ref{eq: arithmetico-geometric}), and the rest is algebraic manipulation. Finally, we compute \ref{eq-proof: AMGM2} using the previous result:
        \begin{align*}
        & \sum_{s=0}^{t^*-2} \left(s+1+\frac{\market}{\prob} \right) (1-\prob)^s\\
        &=  \sum_{s=0}^{t^*-2} s (1-\prob)^s + \left(1 + \frac{\market}{\prob}\right) \sum_{s=0}^{t^*-2} (1-\prob)^s\\
        &= \frac{-\prob^2 + (1-\prob)^{t^*} + \prob(t^*(1-\prob)^{t^*} - 2(1-\prob)^{t^*} + 2) - 1}{(\prob - 1)\prob^2}\\
        &\quad + \left(1 + \frac{\market}{\prob}\right) \frac{ (1-\prob)^{t^*} + \prob - 1}{(\prob-1)\prob}\\
        &=  \frac{(\market +\prob)((1-\prob)^{t^*} + \prob - 1) -\prob^2 + (1-\prob)^{t^*} + \prob(t^*(1-\prob)^{t^*} - 2(1-\prob)^{t^*} + 2) - 1}{(\prob - 1)\prob^2}\\
        &=  \frac{(\market +\prob)(1-\prob)^{t^*} + (\market + \prob)(\prob - 1) -\prob^2 + (1-\prob)^{t^*} + \prob(t^*(1-\prob)^{t^*} - 2(1-\prob)^{t^*} + 2) - 1}{(\prob - 1)\prob^2}\\
        &= \frac{(\market + \prob)(1-\prob)^{t^*} + \market (\prob-1) + \prob + (1-\prob)^{t^*} + \prob(t^*(1-\prob)^{t^*} - 2(1-\prob)^{t^*}) - 1 }{(\prob - 1)\prob^2}\\
        &= \frac{(\market - \prob)(1-\prob)^{t^*} + \market (\prob-1)+ (1-\prob)^{t^*} + \prob t^*(1-\prob)^{t^*} + \prob - 1 }{(\prob - 1)\prob^2}\\
        &= \frac{-(\market - \prob)(1-\prob)^{t^*-1} + \market - (1-\prob)^{t^*-1} - \prob t^*(1-\prob)^{t^*-1} + 1 }{\prob^2}\\
        &= \frac{(1 - \prob)^{t^* - 1} \left( -(\market + 1 + \prob (t^* - 1)) \right) + \market + 1}{\prob^2}. 
    \end{align*}
    \hfill\Halmos

\subsection{Proof of \Cref{lemma: AdvTheoremNegativeBeta}}
\label{proof: AdvTheoremNegativeBeta}
Let $\prob \in (0,1]$. We prove the lemma by showing that the left-hand side of the inequality (i) evaluates to zero for $j = t^*$ and $j = T$, and (ii) is a convex function of $j$. As a result, for all values $j\in[t^*,T]$, the left-hand side evaluates to a non-positive number and the inequality holds for such values.  
  
For (i), with $j=T$, the terms $(T-j)$ and $(1-\prob)^j-(1-\prob)^T$ both evaluate to 0; with  $j=t^*$, the left-hand side is
  \begin{equation*}
      (1-\prob)^{t^*} - (1-\prob)^T + (T-t^*) \cdot \prob \alpha_T = (1-\prob)^{t^*} - (1-\prob)^T + (T-t^*) \cdot \frac{(1-\prob)^{t^*} - (1-\prob)^T}{\lceil \frac{\market}{\prob} \rceil} = 0,
  \end{equation*}
  where we used the definition of $\alpha_T$ in \eqref{1a} and that $T-t^* = \lceil \frac{\market}{\prob} \rceil$. This shows (i). 
  
  For (ii), we compute the second derivative of the left-hand side with respect to $j$ and show that it is non-negative: 
  \begin{equation*}
      \frac{\partial^2}{\partial j^2} \left( (1-\prob)^j - (1-\prob)^T + (T-j) \cdot \mu \alpha_T \right) = \frac{\partial^2}{\partial j^2} ((1-\prob)^j) = (1-\prob)^j \log^2 (1-\prob) \geq 0.\hfill\Halmos
  \end{equation*}

\section{Proofs of Upper Bounds in Section \ref{sec: overview_results} (Propositions \ref{prop: adversarial CR UB} and \ref{prop: stochastic CR UB})}
\label{section: impossibility}
 In this section, we present upper bounds on the performance of any delayed algorithm, even randomized, within both adversarial and stochastic arrival models. 
 
\subsection*{Proof of \Cref{prop: adversarial CR UB} (upper bound for adversarial arrivals)}
\label{ssec: adversarial_impossibility}
The proof first consists of describing a family of instances along with a distribution over these instances. We then identify that (i) these instances are $\market$-oversupplied if $\market \leq 1$ or $\market$-undersupplied if $\market > 1$, and (ii) we characterize an upper bound on the performance of delayed deterministic algorithms on these instances. With these components, we conclude using Yao's Lemma \citep{yao1977probabilistic}, which allow us to upper bound the CR of any delayed algorithm by the presented bound. 

We start by introducing the family of instances we consider, which are parameterized by $s \in \mathbb{Z}_{>0}$. Let $\market > 0$. By our assumption that $\market$ is rational, there exist integers $ p $ and $ q $ such that $ \market = \frac{p}{q} $. Throughout this subsection, we define $L = s! \cdot q $ and $ \prob = \frac{1}{L!}=\frac{1}{(s!q)!} $, where $ n!$ denotes the factorial of $ n $. We set the number of arrival periods to $ T = \frac{\market}{\prob} L^2 $, which is an integer by construction. The set of supply nodes $\supply$ contains $L^2$ nodes, which we categorize into \textit{types}. Given an enumeration $v_1, \ldots, v_{L^2}$ of the supply nodes, we define the $j$-th supply type $ \supply_j $ as $ \{ v_{(j-1)L+1}, \ldots, v_{jL} \} $ for $ j = 1, \ldots, L $. That is, each set of supply types $\supply_j$ consists of $L$ supply nodes.  The set of demand nodes $V$ is defined as the union of $L$ groups of nodes $V_1, \ldots, V_L$, where $ V_i $ contains $\market/\prob \cdot L $ demand nodes and every node in $ V_i $ has edges only to nodes in $ \supply_i, \ldots, \supply_L$. The $L$ groups of vertices in $V$ arrive online in numerical order; internal to a group, the vertices arrive in arbitrary order which is described by $\sigmabf$. Before any arrival, the supply types are permuted uniformly at random. In particular, we let $\mathcal{S}_s[\market]$ the set of all the instances that can be made with permutations on the supply types and $\mathcal{D}_s[\market]$ a random variable that selects an instance in $\mathcal{S}_s[\market]$ 
uniformly at random.


The next lemma characterizes an upper bound on the performance of any delayed deterministic algorithm 
We prove this result in \Cref{proof: AdvWaterFillingOPT}.
\begin{restatable}{lemma}{AdvWaterFillingOPT} 
\label{claim: AdvWaterFillingOPT} 
For any deterministic delayed algorithm $\alg$ that knows the instance-generating process described above we have that 
    \begin{equation*}
     \mathbb E[\alg(\inst, \sigmabf)] \leq 
    L \sum_{\ell=1}^L \left( 1 - (1-\prob)^{\frac{\market}{\prob} (H_L - H_{L-\ell})} \right).
    \end{equation*}
\end{restatable}
Since the deterministic linear program is unaffected by the permutation of supply types, the following result is straightforward; the optimal matching pairs nodes in $V_i$ with nodes in $\supply_i$. Furthermore, the instances in $\mathcal S_s[\market]$ are $\market$-imbalanced. The proof is provided in \Cref{proof: AdvOfflineInstCL}.

\begin{restatable}{lemma}{AdvOfflineInstCL}
    \label{claim: AdvOfflineInstCL} 
     Fix $\market > 0$. We have, for any $\inst \in \mathcal{S}_s[\market]$
     \begin{equation*}
        \offI(\inst) = \min \{1, \market \} \cdot L^2. 
    \end{equation*}
    In particular, $\inst$ is $\market$-undersupplied for $\market \geq 1$ and $\market$-oversupplied for $\market \leq 1$.
\end{restatable}
To proceed we require one last technical result, the proof of which can be found in \Cref{proof: AdvCRLimit}.

\begin{restatable}{lemma}{AdvCRLimit}
    \label{claim: AdvCRLimit}
    \begin{equation*}
        \lim_{L \rightarrow \infty} \left( 1-\frac{\sum_{\ell=0}^{L-1} \exp(\market H_{\ell})}{L \exp(\market H_L)} \right)= \frac{\market}{\market + 1}.
    \end{equation*}
\end{restatable}


We now upper bound the performance of any algorithm on $S_s[\market]$ and conclude the proof. Suppose that
$\market \leq 1$ which implies that $\mathcal{S}_s[\market]$ is a set of $\market$-undersupplied instances by \Cref{claim: AdvOfflineInstCL}. Denote $\optD$ the best (possibly randomized) delayed algorithm that achieves the highest CR for instances in $\mathcal{S}_s[\market]$. By applying Yao's Lemma we have that its expected reward on the worst-instance in $\mathcal{S}_s[\market]$ for this algorithm is bounded by the right-hand term in \Cref{claim: AdvWaterFillingOPT}. Combining this with \Cref{claim: AdvOfflineInstCL} we have
\begin{equation*}
    \min_{\inst \in \mathcal{S}_s[\market]} \frac{\mathbb E[\optD(\inst, \sigmabf)]}{\offI(\inst,\kappa)} \leq \frac{\sum_{\ell=1}^L \left( 1 - (1-\prob)^{\frac{\market}{\prob} (H_L - H_{L-\ell})} \right)}{\market L} = \frac{1}{\market} \left( 1 - \frac{\sum_{\ell=1}^L  (1-\prob)^{\frac{\market}{\prob} (H_L - H_{L-\ell})}}{L} \right).
\end{equation*}
Note that for $\market \geq 1$, the above set of inequalities hold without the factor of $1/\market$. Therefore, we will bound the left hand side term by $\frac{1}{\market + 1}$ as the proof for $\market$-oversupplied instances will follow without considering this factor.

Let $\varepsilon >0$. Fix two positive numbers $\varepsilon_1$ and $\varepsilon_2$ such that $(\varepsilon_1 + \varepsilon_2)/\market \leq \varepsilon$. Recall the fact that $(1-\prob)^{1/\prob}$ converges from above to $1/e$ as $\prob \rightarrow 0^+$. Hence by continuity of the sum and the definition of limit as $\prob \xrightarrow[]{} 0^+$, for every $\varepsilon_1 > 0$ there exists $\prob(\varepsilon_1)>0$ sufficiently small (and hence $s(\varepsilon_1)$ sufficiently large) such that
    \begin{align*}
         \min_{\inst \in \mathcal{S}_{s(\varepsilon_1)}[\market]} \frac{\mathbb E[\optD(\inst, \sigmabf)]}{\offI(\inst,\kappa)} 
        &\leq \frac{1}{\market} \left( 1-\frac{\sum_{\ell=1}^L  \exp \left( - \market(H_L - H_{L-\ell}) \right)}{L} + \varepsilon_1 \right)\\
        &= \frac{1}{\market} \left( 1-\frac{\sum_{\ell=0}^{L-1} \exp(\market H_{\ell})}{L \exp( \market H_L)} + \varepsilon_1 \right).
    \end{align*}
    Finally, by \Cref{claim: AdvCRLimit} there exists $L(\varepsilon_2)$ sufficiently large (and hence $s(\varepsilon_2)$ sufficiently large) such that
    \begin{equation*}
        \label{ineq: k/k+1 + eps}
        \min_{\inst \in \mathcal{S}_{\max\{s(\varepsilon_1), s(\varepsilon_2) \}}[\market]} \frac{\mathbb E[\optD(\inst, \sigmabf)]}{\offI(\inst,\kappa)}  \leq \frac{1}{\market} \left( \frac{\market}{\market+1} + \varepsilon_1 + \varepsilon_2 \right)\leq \frac{1}{\market+1} + \varepsilon.\hfill\Halmos
    \end{equation*}

\subsection*{Proof of \Cref{prop: stochastic CR UB} (upper bound for stochastic arrivals)}
\label{ssec: stochastic_upper_bound}
Let $\market > 0$; we construct an instance $\inst[\market]$ consisting of one unique demand type $v$ connected to a single supply node $u$. We set the probability of consumption for $(u,v)$ as $\prob= \frac{\market}{n}$, where $n$ denotes an integer satisfying $n > \market$, and let   $T = \frac{\market}{\prob}=n$ be the number of arrivals.

Let $\alg$ be the algorithm that matches all the demand arrivals to $u$ and $Z_u$ be the random indicator variable such that $Z_u = 1$ if and only if $u$ is consumed at the end of the time horizon. We have
\begin{align*}
    \mathbb E[\alg(\inst[\market])] =  \mathbb E[Z_u] = \mathbb P[\text{$Z_u = 1$}]
    = 1-(1-\prob)^{\frac{\market}{\prob}},
\end{align*}
Hence, the performance of any algorithm in this instance is bounded by $1-(1-\prob)^{\frac{\market}{\prob}}$. 

For the benchmark problem, we analyze the instance for $\market \leq 1$ and $\market> 1$, and show that it is $\market$-oversupplied and $\market$-undersupplied respectively. First, observe that $\offI(\inst[\market]) = \min \{1,\market\}$ holds as $x_{u,v} = \min\{1,\market\}/\prob$ is optimal for the benchmark problem (see constraints \eqref{stochastic_lp: kappa-constraint} for $\market = 1$ and \eqref{stochastic_lp: demand_constraint}). Next, observe that  $\offI(\inst[\market], \bar{\market}) = \min\{\bar{\market},\market\}$ which follows by noting that it is optimal to set $x_{u,v} = \min\{\bar{\market},\market\}/\prob$ (see constraints \eqref{stochastic_lp: kappa-constraint} and \eqref{stochastic_lp: demand_constraint}). Armed with these results, we can see that: (1) for $\market \leq 1$ we have that $\offI(\inst[\market]) = \market$ and hence the equality $\offI(\inst[\market], \bar{\market}) = \offI(\inst[\market])$ holds if and only if $\bar{\market} \geq \market$, making $\inst[\market]$ an $\market$-oversupplied instance and (2) for $\market > 1$ we have that $\offI(\inst[\market]) = 1$ and hence the equality $\offI(\inst[\market], \bar{\market}) = \bar{\market} \cdot \offI(\inst[\market])$ holds if and only if $\bar{\market} \leq \market$, making $\inst[\market]$ an $\market$-undersupplied instance.

Combining the above we find that, for any delayed algorithm $\alg_D$ in an instance of this type,
\begin{equation*}
    \frac{\mathbb E[\alg_D(\inst[\market])]}{\offI(\inst[\market])} \leq \frac{1-(1-\prob)^{\market/\prob}}{\min\{1,\market\}} \xrightarrow[\prob \rightarrow 0^+]{} \begin{cases}
        1- e^{-\market} & \text{if } \market > 1,\\
        \frac{1- e^{-\market}}{\market} & \text{if } \market \leq 1.\\
    \end{cases}
\end{equation*}
\hfill\Halmos 

\subsection{Remaining Proofs of \Cref{section: impossibility}}

\subsubsection[]{Proof of \Cref{claim: AdvWaterFillingOPT}.}
\label{proof: AdvWaterFillingOPT}
First, observe that for any given optimal delayed deterministic algorithm (i.e., one delayed algorithm that maximizes the left-hand side of the inequality in the lemma), it is necessary to balance the allocation of matches within a specific group $V_i$ (meaning that for any $u_1$, $u_2 \in V_i$ the number of matches to $u_1$ differs at most of one to the number of matches to $u_2$). This is due to the concavity of $1-(1-\prob)^x$, which implies that an algorithm that balances the demand across the supply nodes in an specific supply type leads to a strictly better objective.

Second, we claim that after the last arrival of some group $V_i$, any optimal delayed deterministic algorithm must also balance\footnote{Balancing here means the following: pick two nodes in the different supply types that have distinct number of matches, and modify the algorithm such that it balances the matches across them.} the matches done across supply types. For the sake of contradiction, suppose an optimal deterministic delayed algorithm does not proceed in this way. Since each supply type is indistinguishable from the others, in the next step, the least-matched supply type could be dropped (i.e., not have edges with the following demand groups) with equal probability as the supply type with the most-matched number of demand nodes. A modified algorithm that balances the number of demand nodes in these two supply types achieves a strictly higher expected number of consumed nodes, again, by concavity of $1-(1-\prob)^{x}$. This proves the stated claim.

The rest of the proof consists on counting how many demand nodes get matched to a particular supply node of any given supply type, following the delayed algorithm that balances demand. We claim that every node in the $\ell$-th supply type is matched to $\frac{\market}{\prob} (H_L - H_{L-\ell})$ demand nodes, for $\ell = 1, \ldots, L$. We prove this by induction. For $\ell = 1$ this is true since there are $\frac{\market}{\prob} L$ arrivals that get divided into $L$ supply types, and every type consists of $L$ nodes, yielding $\frac{\market}{\prob} L \cdot \frac{1}{L^2} = \frac{\market}{\prob} (H_L - H_{L-1})$ matched demand nodes per supply node. Suppose the claim is true for $\ell \geq 1$, and after the $(\ell+1)$-th arrival group the algorithm assigns $\frac{\market}{\prob} L$ arrivals to $L-(\ell+1)$ resource types each of those having $L$ nodes, and hence any node of the dropped type has exactly
\begin{equation*}
    \frac{\market}{\prob} (H_L - H_{L-\ell}) + \frac{\market}{\prob} \cdot \frac{1}{L-(\ell+1)} = \frac{\market}{\prob} (H_L - H_{L-(\ell+1)})
\end{equation*}
demand nodes matched to it and the induction steps follows. It follows that in expectation any supply node from a resource type that was dropped on the $\ell$th step is sucessful with probability $ 1 - (1-\prob)^{\frac{\market}{\prob} (H_L - H_{L-\ell})}$. Finally, we conclude by the claim and using that there are $L$ nodes per resource type.
\hfill\Halmos 

\subsubsection[]{Proof of \Cref{claim: AdvOfflineInstCL}.}
\label{proof: AdvOfflineInstCL}
Let $\inst \in \mathcal{S}_s[\market]$. We first prove that $\offI(\inst) = \min\{1,\market\} \cdot L$ and with this result we show that the instance is imbalanced when $\market \neq 1$. Note the following feasible solution to the offline benchmark; match the $\frac{\market}{\prob} L$ demand arrivals of a given type $V_i$ to the supply nodes of the resource type $\supply_i$. This assignment will consume $\min\{1, \market \} \cdot L$ nodes. Since we have $L$ groups, the total number of consumed nodes under this assignment is $\min\{1, \market \} \cdot L^2$. Observe that this assignment is optimal; denote $\xbf^*$ the previously defined assignment and by adding the constraints of type \eqref{adversarial_lp: demand_constraint} over $t \in V_i$ we have that any feasible solution $\xbf$ must satisfy (recall that $|V_i| = \market/\prob \cdot L$)
\begin{equation*}
    \sum_{u: (u,t) \in E,\, t \in V_i} x_{u,t} \leq \frac{\market}{\prob} L.
\end{equation*}
Summing the above equation for all $i \in [L]$ we arrive at
\begin{equation*}
    \sum_{i=1}^L \sum_{u: (u,t) \in E,\, t \in V_i} x_{u,t} = \sum_{u: (u,t) \in E} x_{u,t} \leq \frac{\market}{\prob} L^2.
\end{equation*}
In other words, we have $\sum_{(u,t) \in E} \prob \cdot x_{u,t} \leq \market \cdot L^2$. Furthermore, summing over all nodes $u \in \supply$ constraints of type \eqref{adversarial_lp: kappa-constraint} (for $\market=1$) we have $\sum_{(u,t) \in E} \prob \cdot x_{u,t} \leq L^2$ (recall that $|\supply| = L^2$). Hence, by these two inequalities the objective of the LP benchmark is at most $\min\{1, \market \} \cdot L^2$. Since $\xbf^*$ achieves this value, it is optimal and then this proves $\offI(\inst) = \min\{1,\market\} \cdot L$.

We now prove that the instance is imbalanced when $\market \not= 1$. Consider $\offI(\inst, \bar{\market})$ and by adding constraints of type \eqref{adversarial_lp: kappa-constraint} over all nodes $u \in \supply$ we have $\sum_{(u,t) \in E} \prob \cdot x_{u,t} \leq \bar{\market} L^2$. Since constraints of type \eqref{adversarial_lp: demand_constraint} also hold by the previous argument we have $\offI(\inst, \bar{\market}) \leq \min \{\market, \bar{\market} \} \cdot L^2$. Now, consider a solution $\xbf^*$ to $\offI(\inst, \bar{\market})$ that matches demand nodes from $V_i$ to supply nodes of type $\supply_i$, with the total mass of this solution on these edges being exactly $\min \{\market, \bar{\market} \}/\prob \cdot L$, i.e., let $\xbf^*$ such that
\begin{equation*}
    \sum_{u: (u,t) \in E,\, u \in \supply_i, \, t \in V_i} x^*_{u,t} = \frac{\min \{\market, \bar{\market} \}}{\prob} L.
\end{equation*}
This can be done since each $V_i$ consist of $\market/\prob \cdot L$ demand nodes and each $\supply_i$ of $L$ supply nodes with capacity $\bar{\market}/\prob$. By summing over $i \in [L]$, it follows that $\offI(\inst, \bar{\market}) \geq \min \{\market, \bar{\market} \} \cdot L^2$ and hence $\offI(\inst, \bar{\market}) = \min \{\market, \bar{\market} \} \cdot L^2$. Armed with these results, we can see that: (1) for $\market \leq 1$ we have that $\offI(\inst) = \market \cdot L^2 $ and hence the equality $\offI(\inst, \bar{\market}) = \offI(\inst)$ holds if and only if $\bar{\market} \geq \market$, making $\inst$ an $\market$-oversupplied instance and (2) for $\market > 1$ we have that $\offI(\inst) = L^2$ and hence the equality $\offI(\inst, \bar{\market}) = \bar{\market} \cdot \offI(\inst)$ holds if and only if $\bar{\market} \leq \market$, making $\inst$ an $\market$-undersupplied instance.
\hfill\Halmos 

\subsubsection[]{Proof of \Cref{claim: AdvCRLimit}.}
\label{proof: AdvCRLimit}


By algebra of limits, it suffices to show that
    \begin{equation*}
        \lim_{L \rightarrow \infty} \frac{\sum_{\ell = 0}^{L-1} \exp(\market H_{\ell})}{L \exp(\market H_L)} = \frac{1}{\market + 1}.
    \end{equation*}
    Let $\varepsilon_1, \varepsilon_2 > 0$. We will use the known fact that
\begin{equation*}
    \lim_{n \rightarrow \infty} \left(H_n - \log(n)\right) = \gamma,
\end{equation*}
where $\gamma \approx 0.577$ is the Euler-Mascheroni constant. Denote $H_\ell - \log(\ell) = c(\ell)$, for $\ell \geq 0$. Then, we have $H_\ell - H_L = \log(\ell/L) + c(\ell) - c(L)$ and we can derive
    \begin{equation}
        \label{eq: lemma7first}
         \frac{\sum_{\ell=0}^{L-1} \exp( \market H_{\ell})}{L \exp(\market H_L)} = \frac{1}{L} \sum_{\ell=0}^{L-1} \exp(\market (H_\ell - H_L)) = \frac{1}{L} \sum_{\ell=0}^{L-1} \left(\frac{\ell}{L}\right)^{\market} \exp(\market (c(\ell)- c(L))).
    \end{equation}
    By convergence of $c(\ell)$ as $\ell \rightarrow \infty$ and continuity of the exponential function, there exists some $\ell_0 > 0$ such that for all $\ell, L \geq \ell_0$ we have
    \begin{equation}
        \label{eq: lemma7second}
        1- \varepsilon_1 \leq \exp(\market(c(\ell)- c(L))) \leq 1 + \varepsilon_1.
    \end{equation}
    Using the right hand side of \eqref{eq: lemma7second} in \eqref{eq: lemma7first} we get
    \begin{equation*}
        \frac{1}{L} \sum_{\ell=\ell_0}^{L-1} \left(\frac{\ell}{L}\right)^{\market} \exp(\market (c(\ell)- c(L))) \leq  \frac{1+\varepsilon_1}{L^{\market + 1}} \sum_{\ell=\ell_0}^L \ell^{\market} \leq \frac{1+\varepsilon_1}{L^{\market + 1}} \left( \int_{\ell_0}^L x^{\market} dx + \ell_0^{\market} \right) = \frac{1+\varepsilon_1}{L^{\market + 1}} \cdot \left( \frac{L^{\market + 1}}{\market + 1} + \ell_0^{\market} \right)
    \end{equation*}
    In a similar way, using the left hand side of \eqref{eq: lemma7second} in \eqref{eq: lemma7first} arrive at
    \begin{equation*}
        \frac{1}{L} \sum_{\ell=\ell_0}^{L-1} \left(\frac{\ell}{L}\right)^{\market} \exp(\market (c(\ell)- c(L))) \geq  \frac{1-\varepsilon_1}{L^{\market + 1}} \sum_{\ell=\ell_0}^L \ell^{\market} \geq \frac{1-\varepsilon_1}{L^{\market + 1}} \int_{\ell_0}^L x^{\market} dx = \frac{1-\varepsilon_1}{L^{\market + 1}} \cdot \left( \frac{L^{\market + 1}}{\market + 1} - \frac{\ell_0^{\market+1}}{\market+1} \right).
    \end{equation*}
    In summary, 
    \begin{equation}
        \label{eq: lemma7summary}
        \frac{1-\varepsilon_1}{L^{\market + 1}} \cdot \left( \frac{L^{\market + 1}}{\market + 1} - \frac{\ell_0^{\market+1}}{\market+1} \right) \leq \frac{1}{L} \sum_{\ell= \ell_0}^{L-1} \left(\frac{\ell}{L}\right)^{\market} \exp(\market (c(\ell)- c(L))) \leq  \frac{1+\varepsilon_1}{L^{\market + 1}} \cdot \left( \frac{L^{\market + 1}}{\market + 1} + \ell_0^{\market} \right).
    \end{equation}
    On the other hand, for sufficiently big $L$ we must have that
    \begin{equation*}
        - \varepsilon_2 \leq \frac{1}{L} \sum_{\ell=0}^{\ell_0} \left(\frac{\ell}{L}\right)^{\market} \exp(\market (c(\ell)- c(L))) \leq \varepsilon_2 \quad \text{and} \quad -\varepsilon_1 \leq -\frac{(1+\varepsilon_1) \ell_0^{\market+1}}{L^{\market+1}(\market+1)} \quad \text{and} \quad \frac{1+\varepsilon_1}{L^{\market+1}} \ell_0^{\market}\leq \varepsilon_1.
    \end{equation*}
    Using the above conditions along with \eqref{eq: lemma7first} and \eqref{eq: lemma7summary} we can derive
    \begin{align*}
        \frac{\sum_{\ell=0}^{L-1} \exp( \market H_{\ell})}{L \exp(\market H_L)} &= \frac{1}{L} \sum_{\ell=0}^{L-1} \left(\frac{\ell}{L}\right)^{\market} \exp(\market (c(\ell)- c(L)))\\
        &=  \frac{1}{L} \sum_{\ell=0}^{\ell_0} \left(\frac{\ell}{L}\right)^{\market} \exp(\market (c(\ell)- c(L))) +  \frac{1}{L} \sum_{\ell=\ell_0}^{L-1} \left(\frac{\ell}{L}\right)^{\market} \exp(\market (c(\ell)- c(L)))\\
        &\leq \varepsilon_2 + \frac{1+\varepsilon_1}{1 + \market} + \frac{1+\varepsilon_1}{L^{\market+1}} \ell_0^{\market}\\
        &\leq \varepsilon_1 + \varepsilon_2 + \frac{1+\varepsilon_1}{1+\market}.
    \end{align*}
    In a similar way, using the above conditions as before with \eqref{eq: lemma7first} and \eqref{eq: lemma7summary}  we can derive the lower bounds for the above expression and get
    \begin{equation*}
        \frac{1-\varepsilon_1}{\market + 1}- (\varepsilon_1 + \varepsilon_2)
        \leq \frac{\sum_{\ell=0}^{L-1} \exp( \market H_{\ell})}{L \exp(\market H_L)} \leq \varepsilon_1 + \varepsilon_2 + \frac{1+\varepsilon_1}{1+\market}.
    \end{equation*}
    Since $\varepsilon_1$ and $\varepsilon_2$ are arbitrary, we conclude the desired limit.
\hfill\Halmos 

\section{Results and proofs for our generalized imbalance definition}
\setcounter{proposition}{0}
\label{sec: alternative_def}
In this section we begin by presenting our result on how to compute the $\market$-imbalance pair $(\undersup, \oversup)$ (\Cref{def: new_over_undersupplied}) for a known instance $\inst$. \bbcomment{either here or in main body: how does a platform, in practice, know which ones are under-/oversupplied.} Thereafter, we introduce the necessary lemmas to proceed with the proof of \Cref{theo: CRguarantee_extended_def}.

 The computation of the imbalance pair relies on a classifier that outputs a partition of $\supply$ into $(\undersup, \mathcal{O})$ in an offline scenario. The classifier is based on the following optimization formulation, which (following \citet{feng2024two}) aims to achieve an optimal allocation that spreads demand out across supply nodes as evenly as possible:
\begin{align*}
             \min_{\mathbf{x}\geq \mathbf{0}} \quad & \sum_{u \in \supply}{ \exp\left( 1- \sum_{t: (u, t) \in E} \prob \cdot x_{u, t} \right)} \tag{\(\mathcal{P}^{\mathrm{convex}}(\inst)\)} \label{prob: convex}\\
    \textrm{s.t.} \quad & \sum_{t: (u,t) \in E } \prob \cdot  x_{u, t} \leq 1, \quad && \forall u \in \supply \\
    & \sum_{u: (u,t) \in E} x_{u, t} \leq 1, \quad && \forall t \in [T].
\end{align*}

Let $\xbf^*$ be any optimal solution of \ref{prob: convex}. We define the following sets\footnote{Though $\xbf^*$ is not unique, it is easily visible that the resulting sets $\classU$ and $\classO$ are independent of the choice of~$\xbf^*$; as our results do not require this property, we do not explicitly prove it.} that partition $\supply$:
\begin{equation*}
    \classU = \left\{ u \in \supply : \sum_{t: (u,t) \in E} \prob 
 \cdot x^*_{u,t} = 1  \right\} \quad \text{and} \quad \classO =  \left\{ u \in \supply : \sum_{t: (u,t) \in E} \prob 
 \cdot x^*_{u,t} < 1  \right\}.
\end{equation*}
Similarly, we define a partition of 
$[T]$ given by $T^{\oversup} = \{ t \in [T]: \exists u \in \classO, (u,t) \in E \}$ and $T^{\undersup} = [T] \setminus T^{\oversup}$.
Our next result states that, given $\xbf^*$, $\classU$ and $\classO$ recover a $\market$-imbalanced pair. We defer its proof to \Cref{proof: k-imbalanced and sol_does_not_use_edges}.
\begin{proposition}
    \label{prop: k-imbalanced result}
    {If} 
    $\inst$ admits a $\market$-imbalanced pair $(\undersup, \oversup)$ with $\market > 1$ then $(\classU, \classO) = (\undersup, \oversup)$. 
\end{proposition}


In the rest of this section, we prove \Cref{theo: CRguarantee_extended_def}. To this end, we first introduce an extension of Algorithm \ref{alg: adv_greedy_alg}, Algorithm \ref{alg: new_greedyD}, which takes as input two sets partitioning  $\supply$. This algorithm leverages the $\market$-imbalance pair structure by prioritizing matches to supply vertices in $\undersup$ whenever possible; if no such match is feasible, it resorts to matching the arrival to vertices in $\oversup$. Though \Cref{theo: CRguarantee_extended_def} applies to a much wider class of instances than \Cref{theorem: adversarial CR LB},
Algorithm \ref{alg: new_greedyD} thus requires knowledge of whether each supply node is, respectively, in the under- and oversupplied set -- without this knowledge, the guarantee need not hold (see {\Cref{ssec: example_newgreedy}}).

\RestyleAlgo{ruled}
\begin{algorithm}[h]
\caption{$\altgreedy$}
\label{alg: new_greedyD}
\textbf{Input:} An instance $\inst$, a sequence of arrivals $\sigmabf$ and two sets that partition $\supply$, $(\undersup, \oversup)$. 

\textbf{Output:} A sequence of decisions for the online problem.

\For{$t= 1$ \KwTo $T$}{
Observe the arrival $t$.

\If{$N(t) \cap \undersup \not= \emptyset$}{
Choose $u^* \in  \argmin_{u \in N(t) \cap \undersup} n^g_u$, breaking ties arbitrarily.
}

\Else{Choose $u^* \in  \argmin_{u \in N(t)} n^g_u$, breaking ties arbitrarily.}{}

Assign arrival $t$ to node $u^*$.

Update the number of matches to $u^*$: $n^g_{u^*} \leftarrow n^g_{u^*} + 1$.
}

\end{algorithm}

We now state the final two lemmas required for the proof; the first shows that there is a solution to the offline problem that does not use edges between $\undersup$ and $\Toversup$. Its proof is provided in \Cref{proof: k-imbalanced and sol_does_not_use_edges}.


\begin{lemma}
    \label{lemma: sol_does_not_use_edges}
    If an instance $\inst$ admits a $\market$-imbalanced pair $(\undersup, \oversup)$ with $\market > 1$, then there exists an optimal solution $\xbf$ to $\offI(\inst)$ with (i) $x_{t, u} = 0$ for any $u \in \undersup$ and $t \in \Toversup$, i.e., $\xbf$ uses no edge between $\undersup$ and $\Toversup$, and (ii) $x_{t, u} = 0$ for any $u \in \oversup$ and $t \in \Tundersup$, i.e., $\xbf$ uses no edge between $\oversup$ and $\Tundersup$.
\end{lemma}
The second lemma states that the instances induced\footnote{Given $\supply' \subseteq \supply$ and $D \subseteq [T]$, induced here means that we consider a new instance with arrivals $D$, supply set $\supply'$ and edges $E' = \{(u,t) \in E: u \in \supply', t \in D\}$. For the corresponding arrival vector $\sigmabf$, the induced instance considers the subvector that only includes components of $D$.} by $(\undersup, \Tundersup)$ and $(\oversup, \Toversup)$ are $\market$-undersupplied and $1/\market$-oversupplied respectively. We defer its proof to \Cref{proof: U and O are under/oversupplied}.

\begin{lemma}
    \label{lemma: U and O are under/oversupplied}
    {If an instance $\inst$ admits a $\market$-imbalanced pair $(\undersup, \oversup)$ with $\market > 1$, then the instance induced by 
    $(\undersup, \Tundersup)$ is $\market$-undersupplied and the instance induced by 
    $(\mathcal O, T^{\oversup})$ is $1/\market$-oversupplied.}
    
\end{lemma}

With these lemmas and definitions established, we now proceed with the proof of \Cref{theo: CRguarantee_extended_def}.

\noindent{\textbf{Proof of \Cref{theo: CRguarantee_extended_def}.} {Suppose $\inst$ is} an instance that admits a $\market$-imbalanced pair $(\undersup, \oversup)$. We will prove that the CR of Algorithm \ref{alg: new_greedyD} with inputs $\inst$, $\sigmabf$ and $(\undersup, \oversup)$ is at least $\frac{\market}{1+\market}$, for any arrival sequence $\sigmabf$, which yields the result. Consider the induced instances $(\undersup, \Tundersup)$ and $(\mathcal O, T^{\oversup})$. Let $\xbf$ be an optimal solution as in \Cref{lemma: sol_does_not_use_edges};
denoting by $\offI(\inst |\, \undersup, T^{\undersup} )$ and $\offI(\inst |\, \oversup, \Toversup )$ the optimal value in the induced instances by $(\undersup, \Tundersup)$ and $(\mathcal O, T^{\oversup})$, we can then decompose the objective into two terms:
    \begin{align*}
        \offI(\inst) = \sum_{(u,t) \in E} \prob \cdot x_{u,t} &= \sum_{(u,t) \in E: u \in \undersup, t \in T^{\undersup}} \prob \cdot x_{u,t} + \sum_{(u,t) \in E: u \in \mathcal O, t \in \Toversup} \prob \cdot x_{u,t}\\
        &= \offI(\inst |\, \undersup, T^{\undersup} ) + \offI(\inst |\, \oversup, \Toversup ).
    \end{align*}
    Here, the first equality holds because $\xbf$ is optimal for $\offI(\inst)$, the second holds because $\xbf$ does not use edges between $\undersup$ and $\Toversup$ or between $\oversup$ and $\Tundersup$ (\Cref{lemma: sol_does_not_use_edges}).
    For the third equality, notice that $\offI(\inst)\geq \offI(\inst |\, \undersup, T^{\undersup} ) + \offI(\inst |\, \oversup, \Toversup )$, because the optimization problems on the right-hand side are more constrained (they do not allow edges between $\undersup$ and $\Toversup$ or between $\oversup$ and $\Tundersup$); moreover, as projecting $\xbf$ onto the induced instances yields feasible solutions to those problems, we also find that $\offI(\inst |\, \undersup, T^{\undersup} ) + \offI(\inst |\, \oversup, \Toversup )\geq \sum_{(u,t) \in E: u \in \undersup, t \in T^{\undersup}} \prob \cdot x_{u,t} + \sum_{(u,t) \in E: u \in \mathcal O, t \in \Toversup} \prob \cdot x_{u,t}$; the equality follows.
    
    Consider the execution of $\altgreedy(\inst,\sigmabf, (\undersup,\oversup))$. Observe that, constrained to each subinstance, Algorithm \ref{alg: new_greedyD} acts exactly as Algorithm \ref{alg: adv_greedy_alg} ($\greedy$) on $(\undersup, \Tundersup)$ and $(\mathcal O, T^{\oversup})$. Hence, we can also separate the performance of $\altgreedy(\inst,\sigmabf, (\undersup,\oversup))$ in two terms corresponding each induced instance, which we denote by $\greedy(\inst, \sigmabf |\, \undersup, T^{\undersup})$ and $\greedy(\inst, \sigmabf |\, \oversup, \Toversup)$. By \Cref{lemma: U and O are under/oversupplied} the instance induced by the supply-arrival tuple $(\undersup, \Tundersup)$ is $\market$-undersupplied. Similarly, the instance induced by the supply-arrival tuple $(\mathcal O, T^{\oversup})$ is $1/\market$-undersupplied and by invoking \Cref{theorem: adversarial CR LB} we get that
        \begin{equation*}
    \frac{\mathbb E[ \greedy(\inst, \sigmabf |\, \undersup, T^{\undersup})]}{\offI(\inst |\, \undersup, T^{\undersup})} \geq \frac{\market}{\market + 1} \quad \text{and} \quad 
    \frac{\mathbb E[ \greedy(\inst | \, \oversup, \Toversup)]}{\offI(\inst |\, \oversup, \Toversup)} \geq \frac{1}{1/\market + 1} = \frac{\market}{\market + 1}.
\end{equation*}

We conclude the theorem by noting that
\begin{align*}
    \mathbb E[ \altgreedy(\inst, \sigmabf, (\undersup, \oversup))] &= \mathbb E[ \greedy(\inst, \sigmabf |\, \undersup, T^{\undersup})] + \mathbb E[ \greedy(\inst, \sigmabf |\, \oversup, \Toversup)]\\
    &\geq \frac{\market}{\market+ 1} \cdot \offI(\inst |\, \undersup, T^{\undersup}) + \frac{\market}{\market+ 1} \cdot \offI(\inst |\, \oversup, \Toversup)\\
    &= \frac{\market}{\market+ 1}\offI(\inst).
\end{align*}
\hfill
\Halmos

\subsection{Proof of \Cref{prop: k-imbalanced result} and \Cref{lemma: sol_does_not_use_edges}}
\label{proof: k-imbalanced and sol_does_not_use_edges}

{Our proofs of  \Cref{prop: k-imbalanced result} and \Cref{lemma: sol_does_not_use_edges} rely on one proposition and three supporting lemmas; we defer the proofs of the latter to \Cref{ssec: auxiliary results}.  We begin by presenting and proving the following proposition, which can be viewed as a special case of the matching skeleton characterization from \citet{goel2012communication} and \citet{feng2024two}. 
The proposition provides a characterization of optimal solutions to \ref{prob: convex}.
} 

\begin{proposition}
    \label{cor: two-stage corollary}
    Denote by $\xbf^*$ an optimal solution to \ref{prob: convex}. We have
    \begin{enumerate}[(i)]
        \item There are no edges in $E$ between $\Tundersup$ and $\classO$.
        \item All demand nodes in $\Toversup$ are fully matched by $\xbf^*$, i.e., $\sum_{u: (u,t) \in E} x_{u,t}^* = 1$, $\forall t \in \Toversup$.
        \item The solution $\xbf^*$ does not use edges between $T^{\oversup}$ and $\classU$, i.e. $x^*_{u,t} = 0,$ $\forall t \in T^{\oversup}$, $\classU$.
    \end{enumerate}
\end{proposition}

\noindent \textbf{Proof of \Cref{cor: two-stage corollary}.} {(i) follows directly from the definition of $T^\undersup$ since we have $T^{\undersup}=\{ t \in [T]: \forall u \in \classO, (u,t) \not\in E \}$. For (ii), suppose a node $t \in T^\oversup$ is not fully matched, i.e., $\sum_u x_{u,t}<1$. By definition, all  nodes $u \in \oversup \cap N(t)$ have non-binding supply constraints, i.e. $\sum_{t':(u,t') \in E} \prob \cdot x_{u,t'}^* < 1$. Hence, we could increase $x^*_{u,t}$ for $u \in \classO\cap N(t)$ to obtain a feasible solution with a better objective; this contradicts optimality of $\xbf^*$. For (iii), suppose $x^*_{u,t} > 0$ for some $u \in \classU$ and $t \in T^\oversup$. As $t$ has at least one edge to a node $o \in \classO$ (definition of $T^\oversup$), the strict convexity of $\exp(1-x)$ implies that we can  improve the solution $\xbf^*$ by transferring some mass from the edge $(u, t)$ to the edge $(o, t)$, which contradicts optimality of $\xbf^*$. \hfill\Halmos
        

}

{Our first auxiliary lemma expresses that any optimal solution to \ref{prob: convex} is also optimal in $\offI(\inst)$. 
Intuitively, the balanced allocation provided by \ref{prob: convex} utilizes the maximum arrival mass and allocates it as evenly as possible; however, since both problems share the same feasible set, the solution allocates the same arrival mass as $\offI(\inst)$ does. We note the similarity of this result to Theorem EC.1 in \citet{feng2024two}, where it is shown that any optimal solution to \ref{prob: convex} remains unchanged when the exponential function in the objective is replaced by any differentiable, increasing, and strictly convex function. In our case, only a minor adaptation of their proof is required, as the objective in $\offI(\inst)$ is the identity function, which is not strictly convex. For completeness, we provide our own proof based on their techniques. While less general, it is simpler and adjusted to our setting.
}

\begin{lemma}
    \label{claim: PconvexOptSol}
    Any optimal solution $\xbf^*$ of \ref{prob: convex} is also optimal in $\offI(\inst)$.
\end{lemma}

{Our second lemma establishes that every node in $\classU$ must have its supply constraint tight in every optimal solution to $\offI(\inst)$ and that a supply node must belong to $\classU$ if its corresponding supply constraint is tight in every optimal solution to the offline problem.}

\begin{lemma}
    \label{claim: convexProb1}
    Let $\xbf^*$ be an optimal solution of \ref{prob: convex}. Then $\prob \sum_{t: (u, t) \in E} x^*_{u, t} = 1$ if and only if for every optimal solution $\xbf^{\mathrm{OFF}}$ to $\offI(\inst)$ satisfies $\prob \sum_{t: (u,t) \in E} x^{\mathrm{OFF}}_{u,t} = 1$.
\end{lemma}

{Our third} 
lemma states that any optimal solution to $\offI(\inst)$, where $\inst$ admits a $\market$-imbalance pair, must have the supply constraints of nodes in $\undersup$ tight. This behavior is intuitive, {given that  increasing the right-hand side of all of these supply constraints results in an increase in the objective}. 

\begin{lemma}
    \label{lemma: kImbalancePairOffline}
    Suppose an instance $\inst$ admits a $\market$-imbalanced pair $(\undersup, \oversup)$ with $\market > 1$. Then, every optimal solution $\xbf^{\mathrm{OFF}}$ to $\offI(\inst)$ must satisfy $\prob \sum_{t: (u,t) \in E} x^{\mathrm{OFF}}_{u,t} = 1$, for all $u \in \undersup$.
\end{lemma}

{Applying the above results, we can now prove} \Cref{prop: k-imbalanced result} and \Cref{lemma: sol_does_not_use_edges}.

\noindent \textbf{Proof of \Cref{prop: k-imbalanced result}.} Suppose $\inst$ admits some $\market$-imbalanced pair $(\undersup, \oversup)$ with $\market > 1$. Denote an optimal solution of \ref{prob: convex} by $\xbf^*$. We will show that $\classU = \undersup$.
    \begin{itemize}
        \item $\classU \subseteq \undersup$: Let $u$ in $\classU$, i.e.  $\prob \sum_{t: (u,t) \in E} x^*_{u,t} = 1$. {Then, by \Cref{claim: convexProb1}, every optimal solution $\xbf^{\mathrm{OFF}}$ to $\offI(\inst)$ satisfies $\prob \sum_{t: (u,t) \in E} x^{\mathrm{OFF}}_{u,t} = 1$, so in particular, there is no optimal solution to $\offI(\inst)$ with $\prob \sum_{t: (u,t) \in E} x^{\mathrm{OFF}}_{u,t} \leq 1/\market<1$. As a result, $u \not \in \oversup$ and as $u\in\supply=\oversup\sqcup \undersup$, we conclude that $u \in \undersup$}.\\

        \item $ \undersup  \subseteq \classU$: Let $u \in \undersup$. {By \Cref{lemma: kImbalancePairOffline}, every optimal solution $\xbf^{\mathrm{OFF}}$ to $\offI(\inst)$ fulfills $\prob \sum_{t: (u,t) \in E} x^{\mathrm{OFF}}_{u,t} =~1$. By \Cref{claim: PconvexOptSol} this applies to $\xbf^*$, so we have that $\prob \sum_{t: (u,t) \in E} x^{*}_{u,t} = 1$, i.e., $u \in \classU$.}
    \end{itemize}
    Since this holds for any $\market$-imbalanced pair, the proof follows.
\hfill\Halmos

\noindent \textbf{Proof of \Cref{lemma: sol_does_not_use_edges}.} 
By \Cref{claim: PconvexOptSol}, any 
optimal solution $\xbf^*$ to \ref{prob: convex} is optimal for $\offI(\inst)$ and, by \Cref{cor: two-stage corollary} (i) and (iii),  $\xbf^*$ fulfills the necessary properties.
\hfill\Halmos

\subsection{Proof of \Cref{lemma: U and O are under/oversupplied}}
\label{proof: U and O are under/oversupplied}
To avoid repetition, we will define the following concept that will be useful in this proof;

\begin{definition}
    We say that we increase the capacity of a node $\hat{u} \in \undersup$ by $\varepsilon > 0$ in \ref{prob: offlineUndersupplied} if we change the right-hand side of the capacity constraint {of} 
    $\hat{u}$ {from} 
    $\market$ to $\market + \varepsilon$. Mathematically, {we change the set of constraints from $\sum_{t: (u,t) \in E } \prob \cdot  x_{u, t} \leq \market$ for all $u \in \undersup$ in \ref{prob: offlineUndersupplied} to} $\sum_{t: (u,t) \in E } \prob \cdot  x_{u, t} \leq \market$ for all $u \in \undersup \setminus \{ \hat{u}\}$ and $\sum_{t: (\hat{u},t) \in E } \prob \cdot  x_{\hat{u}, t} \leq \market + \varepsilon$.
\end{definition}

In order to prove {\Cref{lemma: U and O are under/oversupplied}} we need the following auxiliary result. We defer the proof to \Cref{ssec: auxiliary results}.

\begin{lemma}
    \label{claim: increasing_capacity}
    Suppose an instance $\inst$ admits a $\market$-imbalance pair $(\undersup, \oversup)$ with $\market > 1$. Then, for any $\market' \in [1, \market)$ and any $u \in \undersup$, increasing the capacity of $u$ in $\offI^{\undersup, \mathcal O}(\inst, \market')$ by $\varepsilon > 0$ increases the objective.
\end{lemma}

The proof of {\Cref{lemma: U and O are under/oversupplied}} is based on the following two claims:
\begin{itemize}
    \item \emph{Claim 1:} we will show that {any} 
optimal solution $\xbf^{1}$ to $\offI(\inst)$ that satisfies $\prob \sum_{t: (u,t)} x^{1}_{u,t} <1$ for all $u \in \mathcal O$ also satisfies the property of $x^{1}_{u,t} = 0$ for all $u \in \undersup$, $t \in \Toversup$.
    \item \emph{Claim 2}: we will show that {any} 
optimal solution $\xbf^{2}$ to \ref{prob: offlineUndersupplied} also satisfies the property of $x^{2}_{u,t} = 0$ for all $u \in \undersup$, $t \in \Toversup$.
\end{itemize}

With these two claims in place, we can show that the induced instances by $(\mathcal O, \Toversup)$ and $(\undersup, T^{\undersup})$ are $1/\market$-oversupplied and $\market$-undersupplied respectively. For simplicity, denote these instances as $\inst_{\oversup}$ and $\inst_{\undersup}$.

\begin{enumerate}
    \item $\inst_{\oversup}$ is $1/\market$-oversupplied: Consider the solution $\xbf^\oversup$ defined as the solution $\xbf^1$ to $\offI(\inst)$ that satisfies $\prob \sum_{t:(u,t)} x_{u,t}^1 \leq 1/\market<1$ (\Cref{def: new_over_undersupplied} (iii)), constrained to edges between $\oversup$ and $T^\oversup$. By \emph{Claim 1}, we know that $\xbf^1$ does not use any edge in between $\undersup$ and $T^\oversup$, implying that every demand node in $T^\oversup$ is fully matched in this induced instance (otherwise, there is one edge $(u,t) \in E$, $u \in \oversup$, $t \in T^\oversup$ that can be assigned more capacity without violating feasibility, contradicting optimality of $\xbf^1$). Since $\xbf^\oversup$ is fully matching all of the demand nodes, it achieves the optimal objective in $\offI(\inst_\oversup)$ and we have
    \begin{equation*}
        \offI(\inst_\oversup) = \prob |T^\oversup|=\offI(\inst_\oversup, 1/\market),
    \end{equation*}
    where the last equality follows by $\prob \sum_{t:(u,t)} x_{u,t}^\oversup \leq 1/\market$. Therefore, $\inst_\oversup$ is a $1/\market$-oversupplied instance.
    \item $\inst_{\undersup}$ is $\market$-undersupplied: Consider the solution $\xbf^\undersup$ as defined the solution $\xbf^2$
    constrained to edges between $\undersup$ and $T^{\undersup}$. By \Cref{def: new_over_undersupplied} (ii), we know that every node in $\undersup$ is fully matched in \ref{prob: offlineUndersupplied}, i.e., $\sum_{t:(u,t) \in E} \prob \cdot x_{u,t}^2 = \market$ for all $u \in \undersup$. Furthermore, by \emph{Claim 2}, we know that this can be achieved by only matching supply nodes in $\undersup$ with demand nodes in $T^{\undersup}$. Hence, $\offI(\inst_{\undersup}, \market) = \market |\undersup|$. Furthermore, the scaled solution $\frac{1}{\market} \xbf^\undersup$ is feasible in $\offI(\inst_{\undersup})$ as it is a re-scaling of $\xbf^\undersup$ that satisfies $\sum_{t:(u,t) \in E} \prob \cdot x_{u,t}^2 / \market = 1$ for all $u \in \undersup$. Since every supply node gets fully matched, we conclude that
    \begin{equation*}
        \market \cdot \offI(\inst_{\undersup}) = \market  |\undersup| = \offI(\inst_{\undersup}, \market),
    \end{equation*}
    which implies that $\inst_\undersup$ is a $\market$-undersupplied instance.
\end{enumerate}

{Thus, to conclude the proof of the lemma, we need to prove the first and second claim above.}

 \noindent\textit{Proof of Claim 1.} Let $\xbf$ be {an} 
 optimal solution to $\offI(\inst) = \offI^{\undersup, \mathcal O}(\inst, 1)$ that satisfies { $\prob \sum_{t: (u,t)} x_{u,t} <~1$ for all $u \in \mathcal O$}. 
 {We will prove that $x_{u',t'} =0$ for every $u' \in \undersup$, $t' \in \Toversup$. Suppose for the sake of contradiction that there exist $u' \in \undersup$, $t' \in \Toversup$ such that $x_{u',t'}>0$.} 
 {Note that any node $t \in \Toversup$ must be fully matched, i.e., 
 $\sum_{u:(u,t) \in E}x_{u,t}=1, \forall t \in\Toversup$ as nodes in $\Toversup$ have edges to nodes in $\oversup$, and nodes in $\oversup$ can accommodate additional demand, i.e., $\sum_{t:(u,t) \in E}x_{u,t} <1,  \forall u \in \oversup$. Hence, if we had $\sum_{u:(u,t) \in E}x_{u,t}<1$ for $t\in\Toversup$ we could increase $x_{u,t}$ for some $u\in\oversup$, which would contradict the optimality of $\xbf$.}  
 {Now, suppose we increase the capacity of the node $u'$ by $\varepsilon>0$. As every demand node $t$ in $T^\oversup$ is fully matched, and $\xbf$ is not optimal in the new problem (increasing the capacity of $u'$ increases the optimal objective by Lemma \ref{claim: increasing_capacity})}, there must exist an augmenting path\footnote{A path that starts with a node in $[T]$ and ends in $\supply$ in which each vertex has unused capacity. Note that a solution $\xbf$ is optimal if and only if it does not contains augmenting paths.} between a demand node $t_{u'}$ in $[T]$ and $u'$.

We denote this augmenting path as $P = (t_1, u_1, \ldots, t_n, u_n)$ where $t_1 = t_{u'}$ and $u_n = u'$ and extend it by adding the edges $(t', u_n)$ and $(t', o)$, where $o \in \oversup$ is a neighbor of $t'\in\Toversup$. Denote $t_{n+1} = t', u_{n+1} = o$, and
 \begin{equation*}
     \Tilde{\varepsilon} = \min \left \{ 1 - \sum_{u:(u,t_1)} x_{u, t_1},\; \min_{i \in [n]}\{ x_{u_i, t_{i+1}}\},\; \frac{1- \sum_{t:(u_{n+1},t) \in E} \prob \cdot x_{u_{n+1},t} }{\prob} \right \}.
 \end{equation*}
 
 Note that $\Tilde{\varepsilon} > 0$ since (i) $t_1$ has unassiged mass, (ii) there is an augmenting path between $t_{u'}$ and $u'$ (implying that $x_{u_i, t_{i+1}} > 0$ for $i \in [n]$) and (iii) {$\prob \sum_{t: (u,t)} x_{u,t} < 1$ for all $u \in \mathcal O$}. We now define a new feasible solution for $\offI(\inst)$:
 \begin{equation*}
     \Tilde{x}_{u,t} = \begin{cases}
         x_{u,t} + \Tilde{\varepsilon} & \text{if $(u,t) = (u_i,t_i)$ for all $i \in [n+1]$}\\
         x_{u,t} - \Tilde{\varepsilon} & \text{if $(u,t) = (u_i,t_{i+1})$ for all $i \in [n]$}\\
         x_{u,t}  & \text{else.}\\
     \end{cases}
 \end{equation*}
To prove feasibility of $\Tilde{\xbf}$, we first note that for demand node $t_1$ the definitions of $\Tilde{\xbf}$ and $\Tilde{\varepsilon}$ imply that
  \begin{equation*}
      \sum_{u:(u,t_1) \in E} \Tilde{x}_{u, t_1} = \sum_{u \not= u_1:(u,t_1) \in E} x_{u, t_1} + (x_{u_1, t_1} + \Tilde{\varepsilon}) = \sum_{u:(u,t_1) \in E} x_{u, t_1} + \Tilde{\varepsilon} \leq 1.
  \end{equation*}
  Next, for any demand node $t_i$ with $i \in \{2, \ldots, n+1\}$ we have
  \begin{equation*}
      \sum_{u:(u,t_i) \in E} \Tilde{x}_{u, t_i} = \sum_{u \not= u_i, u_{u+1}:(u,t_i) \in E} x_{u, t_i} + (x_{u_i, t_i} + \Tilde{\varepsilon}) + (x_{u_{i-1}, t_i} - \Tilde{\varepsilon}) = \sum_{u:(u,t_i) \in E} x_{u, t_i} \leq 1.
  \end{equation*}
  Next, we prove that $\Tilde{\xbf}$ satisfies the supply constraints. For $u \not \in \{u_1, \ldots, u_{n+1}\}$ the assignment of demand to $u$ under $\Tilde{\xbf}$ coincides with that under $\xbf$, so these supply constraints remain satisfied. For $u \in \{u_1, \ldots, u_n \}$  
  \begin{align}
      \sum_{t: (u_i, t) \in E} \prob \cdot \Tilde{x}_{u_i, t} = \sum_{t \not= t_i, t_{i+1}: (u_i, t) \in E} \prob \cdot x_{u_i, t} + \prob \cdot (x_{u_i,t_i} + \Tilde{\varepsilon}) + \prob \cdot (x_{u_i,t_{i+1}} - \Tilde{\varepsilon}) &= \sum_{t: (u_i, t) \in E} \prob \cdot x_{u_i, t} 
      \leq 1.\notag
  \end{align}
  Similarly, since { $\Tilde{\varepsilon}\leq (1- \sum_{t:(u_{n+1},t) \in E} \prob \cdot x_{u_{n+1},t})/\prob$, for $u_{n+1}$
  }
  \begin{align}
      \sum_{t: (u_{n+1}, t) \in E} \prob \cdot \Tilde{x}_{u_{n+1}, t} =  \sum_{t \not= t_{n+1}: (u_{n+1}, t) \in E} \prob \cdot x_{u_{n+1}, t} + \prob \cdot (x_{u_{n+1}, t_{n+1}} + \Tilde{\varepsilon}) = \sum_{t: (u_{n+1}, t) \in E} \prob \cdot x_{u_{n+1}, t} + \prob \Tilde{\varepsilon} 
       \leq 1. \notag
  \end{align}
  This proves that $\Tilde{\xbf}$ is feasible in $\offI(\inst)$. As we defined $\Tilde{\xbf}$ by starting from $\xbf$ and then adding $\Tilde{\varepsilon}$ to $n+1$ terms and subtracting $\Tilde{\varepsilon}$ from $n$ terms, $\Tilde{\xbf}$ has a greater objective than $\xbf$, which contradicts the optimality of $\xbf$ for $\offI(\inst)$; it follows that $\xbf$ fulfills that $x_{u',t'} =0$ for every $u' \in \undersup$, $t' \in \Toversup$. 
 
 \textit{Proof of Claim 2.} For the sake of contradiction, suppose there exists an optimal solution $\xbf^2$ to \ref{prob: offlineUndersupplied} with $x^2_{u',t'} > 0$ for some $u' \in \undersup$, $t' \in \Toversup$. Let
 \begin{equation*}
     x_{u,t} = \begin{cases}
            x^2_{u,t} \cdot \frac{1}{\market} & \text{if $u \in \undersup$},\\
            x^2_{u,t} & \text{else.}
    \end{cases}
 \end{equation*}
 Note that $\xbf$ is feasible in $\offI(\inst)$; it is a re-scaling of $\xbf^2$ that satisfies $\prob \sum_{t: (u,t)} x_{u,t} = \frac{1}{\market} \cdot \prob \sum_{t: (u,t)} x^2_{u,t}  \leq 1$ for $u \in \mathcal \undersup$, where the inequality follows from the feasibility of $\xbf^2$ in \ref{prob: offlineUndersupplied}. The remaining constraints are either unaffacted or scaled by $1/\market$.
  Also, $x_{u',t'} > 0$ follows from $x^2_{u',t'} > 0$. Without loss of generality,\footnote{Otherwise, take an optimal solution $\hat{x}$ to $\offI(\inst)$ that satisfies $\prob \sum_{t: (u,t)} \hat{x}_{u,t} \leq 1/\market$ for all $u \in \mathcal O$ (\Cref{def: new_over_undersupplied} (iii)) and define the solution $\xbf'=(\xbf + \hat{\xbf})/2$. Note that $\xbf'$ is an optimal solution to $\offI(\inst)$ with $x'_{u',t'} > 0$ and $\prob \sum_{t: (u,t)} x_{u,t}' < 1$ for all $u \in \mathcal O$.} we can assume that $\prob \sum_{t: (u,t)} x_{u,t} < 1$ for all $u \in \mathcal O$.
  But then, contradicting \emph{Claim 1}, there is  an optimal solution $\xbf$ to $\offI(\inst)$ with $\prob \sum_{t: (u,t)} x_{u,t} <1$ for $u \in \mathcal O$ and $x_{u',t'} > 0$ for some $u' \in \undersup$, $t' \in T^{\oversup}$. 
 \hfill \Halmos

\subsection{Proof of auxiliary results (\Cref{claim: PconvexOptSol}, \Cref{claim: convexProb1}, \Cref{lemma: kImbalancePairOffline}, \Cref{claim: increasing_capacity}, \Cref{cor: two-stage corollary})}
\label{ssec: auxiliary results}

We begin by proving \Cref{claim: PconvexOptSol}. The proof of this lemma employs techniques from convex programming. The key component of the argument is based on Lagrangian duality and the Karush-Kuhn-Tucker (KKT) conditions; see \citep{boyd2004convex}.  

\begin{definition}
    The Lagrangian dual of \ref{prob: convex} is defined as
    \begin{align*}
    \mathcal{L}(\xbf, \thetabf, \lambdabf, \gammabf) &= \sum_{u \in \supply}{ \exp\left( 1- \sum_{t: (u, t) \in E} \prob \cdot x_{u, t} \right)} + \sum_{u \in \supply} \theta_u \left( \sum_{t:(u,t) \in E} \prob\cdot x_{u,t} - 1\right) + \sum_{t \in [T]} \lambda_t \left( \sum_{u: (u,t) \in E} x_{u,t} - 1 \right) \\
    & \quad - \sum_{(u,t) \in E} \gamma_{u,t} x_{u,t}.
    \end{align*}
\end{definition}
\begin{proposition}[KKT Conditions, Proposition EC.1 \citep{feng2024two}]
    \label{prop_cite: KKT}
     Suppose $\xbf^*$ is {an} optimal solution of \ref{prob: convex}. Then, there exists dual variables $(\thetabf, \lambdabf, \gammabf)$, i.e., Lagrangian multipliers, such that 
    \begin{enumerate}[(i)]
        \item (Stationarity) $\forall u \in \supply$, $t \in [T]$, $(u,t) \in E$: $\frac{\partial \mathcal L}{\partial x_{u,t}}(\xbf^*, \thetabf, \lambdabf, \gammabf) = 0$,
        \item (Dual feasibility) $\forall u \in \supply$, $t \in [T]$, $(u,t) \in E$: $\thetabf, \lambdabf, \gammabf \geq \mathbf{0}$,
        \item (Complementary Slackness) $\forall u \in \supply$, $t \in [T]$, $(u,t) \in E$:
        \begin{enumerate}
            \item $\theta_u \cdot \left( \sum_{t:(u,t) \in E} \prob\cdot x_{u,t}^* - 1\right) = 0$.
            \item $\lambda_t \cdot \left( \sum_{u: (u,t) \in E} x_{u,t}^* - 1 \right) = 0$.
            \item $\gamma_{u,t} \cdot x_{u,t}^* = 0$.
        \end{enumerate}
    \end{enumerate}
\end{proposition}

\noindent \textbf{Proof of \Cref{claim: PconvexOptSol}.}
Let $\xbf^*$ be an optimal solution to \ref{prob: convex} and denote by $(\thetabf, \lambdabf, \gammabf)$ a dual solution of the convex problem  that satisfies the KKT conditions with $\xbf^*$. We will construct a feasible solution to the dual of the linear program $\offI(\inst)$ that satisfies complementary slackness with $\xbf^*$. In particular, since $\xbf^*$ is feasible in $\offI(\inst)$ this will show that it is also optimal for $\offI(\inst)$. We need to check that there exists $(\alphabf, \betabf)$ such that
    \begin{enumerate}
            \item (Complementary slackness) $\forall u \in \supply$, $t \in [T]$: \begin{equation*}
                \alpha_u \cdot \left( \sum_{t:(u,t) \in E} \prob\cdot x^*_{u,t} - 1\right) = 0 \quad \text{and} \quad \beta_t \cdot \left( \sum_{u: (u,t) \in E} x^*_{u,t} - 1 \right) = 0.
            \end{equation*}
            \item (Feasibility) $\forall u \in \supply, t \in [T], (u,t) \in E$: $\alpha_u + \beta_t \geq 1$ and $\alphabf$, $\betabf \geq \mathbf{0}$. 
    \end{enumerate}
    We define candidate solution as $\alphabf^* = \thetabf$ and $\betabf^* = \lambdabf/\prob$. We first check that complementary slackness holds for this solution. From \Cref{prop_cite: KKT} (iii, a) and (iii, b) we obtain 
    \begin{equation*}
                \alpha_u^* \cdot \left( \sum_{t:(u,t) \in E} \prob\cdot x^*_{u,t} - 1\right) = \theta_u \cdot \left( \sum_{t:(u,t) \in E} \prob\cdot x^*_{u,t} - 1\right) = 0
            \end{equation*}
     \begin{equation*}\text{and}\quad
        \beta_t^* \cdot \left( \sum_{u: (u,t) \in E} x^*_{u,t} - 1 \right) = \frac{\lambda}{\prob} \cdot \left( \sum_{u: (u,t) \in E} x^*_{u,t} - 1 \right) = 0.
    \end{equation*}
    We finish by showing that this solution is dual feasible. Clearly, $\alphabf^*, \betabf^* \geq \mathbf{0}$ by the dual feasibility of the Lagrangian multipliers.
    By the stationarity condition of \Cref{prop_cite: KKT} we have $\forall u \in \supply, t \in [T], (u,t) \in E$:
    \begin{align}
         \quad \frac{\partial \mathcal L}{\partial x_{u,t}}(\xbf^*, \thetabf, \lambdabf, \gammabf) &= - \prob \exp\left( 1- \sum_{t: (u, t) \in E} \prob \cdot x^*_{u, t} \right) + \prob \cdot \theta_u + \lambda_t - \gamma_{u,t} = 0, \notag\\
         \Longleftrightarrow & \quad \prob \theta_u + \lambda_t = \gamma_{u,t} + \prob \exp\left( 1- \sum_{t: (u, t) \in E} \prob \cdot x^*_{u, t} \right)\notag\\
         \Longrightarrow & \quad \prob \theta_u + \lambda_t \geq \prob \exp\left( 1- \sum_{t: (u, t) \in E} \prob \cdot x^*_{u, t} \right) \notag\\
         \Longrightarrow & \quad \prob \theta_u + \lambda_t \geq \prob \notag\\
         \Longrightarrow & \quad \alpha^*_u + \beta^*_t \geq 1 \notag,
    \end{align}
    where the first equivalence follows by re-arranging terms. The first implication follows from the non-negativity of the dual variable $\gamma_{u,t}$ for any $(u,t) \in E$. The second implication uses that $\xbf^*$ is feasible in \ref{prob: convex} and that the exponential function is increasing; mathematically this translates to $0 \leq 1 - \sum_{t: (u, t) \in E} \prob \cdot x^*_{u, t}$ and we lower bound the right-hand side term by $ \prob \exp(0) = \prob$. The last implication follows by substituting our definitions of $\alphabf^*$ and $\betabf^*$ and dividing through  by $\prob$. This proves the feasibility of our candidate solution and the proof of the lemma follows.
    \hfill
    \Halmos

\noindent \textbf{Proof of \Cref{claim: convexProb1}}. 
With $\xbf^*$ denoting an optimal solution of \ref{prob: convex}, we aim to show that every optimal solution $\xbf^{\mathrm{OFF}}$ to $\offI(\inst)$ satisfies $\prob \sum_{t: (u,t) \in E} x^{\mathrm{OFF}}_{u,t} = 1$ if and only if $\prob \sum_{t: (u, t) \in E} x^*_{u, t} = 1$.
Our proof relies on the decomposition of $\xbf^*$ described in \Cref{cor: two-stage corollary}. We prove the lemma by proving both directions of the if and only if statement. 

First, suppose that every optimal solution $\xbf^{\mathrm{OFF}}$ to $\offI(\inst)$ satisfies $\prob \sum_{t: (u,t) \in E} x^{\mathrm{OFF}}_{u,t} = 1$. By \Cref{claim: PconvexOptSol},  $\xbf^*$ is an optimal solution to $\offI(\inst)$ and thus $\prob \sum_{t: (u,t) \in E} x^{*}_{u,t} = 1$. 

    For the other direction, suppose $\prob \sum_{t: (u,t) \in E} x^{*}_{u,t} = 1$ for some $u\in\supply$ (in particular, $u \in \classU$) and there is a solution to $\xbf^{\mathrm{OFF}}$ to $\offI(\inst)$ that satisfies $\prob \sum_{t: (u,t) \in E} x^{\mathrm{OFF}}_{u,t} < 1$. We will prove that $x^{\mathrm{OFF}}$ is not an optimal solution to $\offI(\inst)$, yielding a contradiction. To do so, we first compute the exact value of $\offI(\inst)$ as a function of $|\classU|$ and $|\Toversup|$: 
    \begin{align*}
        \offI(\inst) = \prob \sum_{(u,t) \in E} x^*_{u,t} &= \prob \sum_{(u,t) \in E: u \in \classU} x^*_{u,t} + \prob \sum_{(u,t) \in E: u \in \classO} x^*_{u,t}\\
        &= \prob \sum_{u \in \classU} \sum_{t: (u,t) \in E} x^*_{u,t} + \prob \sum_{u \in \classO} \sum_{t: (u,t) \in E} x^*_{u,t}\\
        &= \prob \sum_{u \in \classU} \sum_{t: (u,t) \in E} x^*_{u,t} + \prob \sum_{t \in \Toversup} \sum_{u: (u,t) \in E} x^*_{u,t}\\
        &= \sum_{u \in \classU} 1 + \prob \sum_{t \in \Toversup} 1\\
        &= |\classU| + \prob |\Toversup|,
    \end{align*}
    where the first equality holds because $\xbf^*$ is an optimal solution to  $\offI(\inst)$ (\Cref{claim: PconvexOptSol}), the second separates the sum ($\classU$ and $\classO$ are a partition of $\supply$), the third rearranges the sums, and the fourth holds since $\xbf^*$ does not use edges between $T^{\oversup}$ and $\classU$ {(Proposition \ref{cor: two-stage corollary}, (iii))}, implying that the only demand nodes that have positive mass assigned to nodes in $\classO$ are the ones assigned by $T^{\oversup}$. The fifth equality follows because  all supply nodes in $\classU$ and demand nodes in $\Toversup$ are fully matched {(Proposition \ref{cor: two-stage corollary}, (ii))}.

    We now prove that the objective value of $\xbf^{\mathrm{OFF}}$ is less than $|\classU| + \prob |\Toversup|$, concluding the proof. Counting the assigned mass by the solution $\xbf^{\mathrm{OFF}}$ to supply nodes we have
    \begin{align*}
        \sum_{(u,t) \in E} \prob \cdot x^{\mathrm{OFF}}_{u,t} &= \sum_{(u,t) \in E : u \in \classU}  \prob \cdot x^{\mathrm{OFF}}_{u,t} + \sum_{(u,t) \in E : u \in \classO}  \prob \cdot x^{\mathrm{OFF}}_{u,t}\\
        &= \sum_{u \in \classU} \sum_{t: (u,t) \in E} \prob \cdot x^{\mathrm{OFF}}_{u,t} + \sum_{t \in \Toversup} \prob \cdot \sum_{u \in \classO, (u,t) \in E}  x^{\mathrm{OFF}}_{u,t}\\
        &< \sum_{u \in \classU} 1 + \sum_{t \in \Toversup} \prob \cdot \sum_{u \in \classO, (u,t) \in E}  x^{\mathrm{OFF}}_{u,t}\\
        &\leq \sum_{u \in \classU} 1 + \prob \sum_{t \in \Toversup} 1\\
        &= |\classU| + \prob |\Toversup|\\
        &= \offI(\inst).
    \end{align*}
    The first equality follows again by separating the sum since $\classU$ and $\classO$ are a partition of $\supply$. The second equality holds by re-arranging the sums. The strict inequality holds since by assumption there exists a supply node $u \in \classU$ such that $\prob \sum_{t: (u,t) \in E} x^{\mathrm{OFF}}_{u,t} < 1$. The inequality holds by using feasibility of $x^{\mathrm{OFF}}$; for any $t \in [T]$ we have $\sum_{u : (u,t) \in E} x^{\mathrm{OFF}}_{u,t} \leq 1$. The next two equalities hold by using the previous step in the proof in which we computed $\offI(\inst)$. The proof of the lemma follows.
    \hfill\Halmos

\noindent \textbf{Proof of \Cref{lemma: kImbalancePairOffline}.}
For the sake of contradiction, suppose there is some optimal solution $\xbf^{\mathrm{OFF}}$ to $\offI(\inst)$ and $\hat{u} \in \undersup$ such that $\prob \sum_{t: (u,t) \in E} x^{\mathrm{OFF}}_{\hat{u},t} < 1$. We will show that this enables the construction of a  solution to the dual of the problem of \ref{prob: offlineUndersupplied} with an objective strictly lower than $\offI(\inst) + (\market-1) |\undersup|$, yielding a contradiction with weak duality. The dual problem of $\offI(\inst)$ is 
    \begin{align*}
             \min_{\alphabf, \betabf \geq \mathbf{0}} \quad &  \sum_{u \in \supply} \alpha_u + \prob \sum_{t \in [T]} \beta_t\\
    \textrm{s.t.} \quad & \alpha_u + \beta_t \geq 1, \quad \forall (u,t) \in E.
\end{align*}
    By strong duality, there also exists $(\hat{\alphabf}, \hat{\betabf})$ solution to the dual problem of $\offI(\inst)$ such that
        \begin{itemize}
            \item $\sum_{u \in \supply} \hat{\alpha}_u + \prob \sum_{t \in [T]} \hat{\beta}_t = \offI(\inst)$.
            \item (Feasibility) For all $(u,t) \in E$: $\hat{\alpha}_u + \hat{\beta}_t \geq 1$ and $\hat{\alphabf}, \hat{\betabf} \geq \mathbf{0}$.
            \item (Complementary Slackness) For every $u \in \supply$, $(\prob \sum_{t: (u,t) \in E} x^{\mathrm{OFF}}_{u,t} -1) \cdot \hat{\alpha}_{u} = 0$. In particular, $\hat{\alpha}_{\hat{u}} = 0$.
        \end{itemize}
        Further, note that dual optimality implies that $\hat{\alpha}_u \leq 1$ for every $u$ because otherwise decreasing $\hat{\alpha}_u$ to $1$ would maintain feasibility and reduce the objective. Now, consider the  dual problem of \ref{prob: offlineUndersupplied}:
        \begin{align*}
             \min_{\alphabf, \betabf \geq \mathbf{0}} \quad & \market \sum_{u \in \undersup} \alpha_u + \sum_{u \in \oversup} \alpha_u + \prob \sum_{t \in [T]} \beta_t\\
    \textrm{s.t.} \quad & \alpha_u + \beta_t \geq 1. \quad && \forall (u,t) \in E.
\end{align*}
Note that $(\hat{\alphabf}, \hat{\betabf})$ is also feasible in the above problem. Furthermore, 
        since $\hat{\alpha}_{\hat{u}}$ is zero and $\alpha_u \leq 1$ for all $u \in \supply$, then by plugging in $(\hat{\alphabf}, \hat{\betabf})$ the dual problem achieves a value of $$\offI(\inst) + (\market - 1) \sum_{u \in \undersup \setminus \{ \hat{u} \}} \hat{\alpha}_u
            \leq \offI(\inst)+(\market - 1)( |\undersup|-1)$$
        which contradicts weak duality. The lemma follows.
\hfill\Halmos

 \noindent \textbf{Proof of \Cref{claim: increasing_capacity}.} The proof of this lemma uses that the supply constraints of nodes $u \in \mathcal \undersup$ of the problem \ref{prob: offlineUndersupplied} hold tight; {this result relies on and strictly generalizes \Cref{lemma: kImbalancePairOffline}.}
\begin{lemma}
    \label{claim: tightnessInU}
    Let $\xbf$ be an optimal solution of \ref{prob: offlineUndersupplied}. Then $\xbf$ satisfies $\prob \sum_{t: (u,t) \in E} x_{u,t} = \market$, for all $u \in \undersup$. Furthermore,
    \begin{equation*}
        \label{eq: tightnessInU}
        |\undersup| + \sum_{u \in \mathcal O} \sum_{(u,t) \in E} \prob \cdot x_{u,t} = \offI(\inst).
    \end{equation*}
\end{lemma}
We prove \Cref{claim: tightnessInU} immediately after proving \Cref{claim: increasing_capacity}.
Let $1 \leq \market' < \market$ and $\varepsilon \in (\market - \market', 0)$. We {first show that increasing the capacity of a node $\Tilde{u} \in \undersup$ by $\varepsilon>0$ in the problem $\offI^{\undersup, \mathcal O}(\inst, \market')$ increases the objective by $\varepsilon$}.
Take an optimal solution of $\xbf$ of \ref{prob: offlineUndersupplied} and define
    \begin{equation*}
        \hat{x}_{u,t} = \begin{cases}
            x_{u,t} \cdot \frac{\market'}{\market} & \text{if $u \in \undersup \setminus \{\Tilde{u}\}$}\\
            x_{u,t} \cdot \frac{\market' + \varepsilon}{\market} & \text{if $u  = \Tilde{u}$}\\
            x_{u,t} & \text{else.}\\
        \end{cases}
    \end{equation*}
    Since $\frac{\market'}{\market} \leq \frac{\market' + \varepsilon}{\market} \leq 1$ the above solution is a re-scaling of the original solution an hence it satisfies the constraints of $\offI^{\undersup, \mathcal O}(\inst, \market')$ with the capacity of the node $\Tilde{u}$ increased by $\varepsilon$. We compute the objective of the solution as follows
    \begin{align*}
        \sum_{(u, t) \in E}{ \prob \cdot \hat{x}_{u, t}} &= \sum_{u \in \undersup \setminus \{\Tilde{u}\}} \sum_{t: (u, t) \in E}{ \prob \cdot \hat{x}_{u, t}} + \sum_{t: (\Tilde{u}, t) \in E} \prob \cdot \hat{x}_{\Tilde{u}, t} + \sum_{u \in \mathcal O} \sum_{t: (u, t) \in E}{ \prob \cdot \hat{x}_{u, t}}\\
        &= \frac{\market'}{\market} \cdot  \sum_{u \in \undersup \setminus \{\Tilde{u}\}}\sum_{t: (u, t) \in E} { \prob \cdot x_{u, t}} + \frac{\market' + \varepsilon}{\market} \sum_{t:(\Tilde{u}, t) \in E} \prob \cdot x_{\Tilde{u}, t} + \sum_{u \in \mathcal O} \sum_{t: (u, t) \in E}{ \prob \cdot x_{u, t}}\\
        &= \market' \cdot (|\undersup|-1)   + (\market' + \varepsilon) + \sum_{u \in \mathcal O} \sum_{t: (u, t) \in E}{ \prob \cdot x_{u, t}}\\
        &= \market' \cdot |\undersup | + \varepsilon + \sum_{u \in \mathcal O} \sum_{t:  (u, t) \in E}{ \prob \cdot x_{u, t}}\\
        &= |\undersup| + \sum_{u \in \mathcal O} \sum_{t:  (u, t) \in E}{ \prob \cdot x_{u, t}} + (\market' - 1) |\undersup| + \varepsilon\\
        &= \offI(\inst) + (\market' - 1) |\undersup| + \varepsilon.
    \end{align*}
The first equality holds by separating the objective in $\undersup \setminus \{ \Tilde{u}\}$, $\Tilde{u}$ and $\oversup$. The second equality uses the definition of $\hat{\xbf}$. The third equality uses the tightness result of \Cref{claim: tightnessInU}. The fourth and fifth equalities only reorganize terms. The final equality holds by the equality result of \Cref{claim: tightnessInU}.
\hfill
\Halmos

\noindent \textbf{Proof of  \Cref{claim: tightnessInU}.}
    Suppose that there exists an optimal solution $\xbf$ to \ref{prob: offlineUndersupplied} that satisfies $\prob \sum_{t: (u',t)} x_{u',t} < \market$, for some $u' \in \undersup$. We will construct an optimal solution $\xbf^{\mathrm{OFF}}$ to $\offI(\inst)$ that satisfies $\prob \sum_{t: (u',t)} x^{\mathrm{OFF}}_{u',t} < 1$, {and thus find a contradiction to \Cref{lemma: kImbalancePairOffline}.}     
    Consider $\xbf^{\mathrm{OFF}}$ defined as
    \begin{equation*}
        x^{\mathrm{OFF}}_{u,t} = \begin{cases}
            x_{u,t} \cdot \frac{1}{\market} & \text{if $u \in \undersup$},\\
            x_{u,t} & \text{else.}
            \end{cases}
    \end{equation*}
    {First, $\xbf^{\mathrm{OFF}}$ satisfies $\prob \sum_{t: (u',t)} x^{\mathrm{OFF}}_{u',t} < 1$. Moreover, note} that $\xbf^{\mathrm{OFF}}$ must be feasible for $\offI(\inst)$: 
    it is a re-scaling of $\xbf$ that satisfies $\prob \sum_{t: (u,t)} x^{\mathrm{OFF}}_{u,t} = \frac{1}{\market} \cdot \prob \sum_{t: (u,t)} x_{u,t}  \leq 1$ for all $u \in \undersup$, where the inequality follows by the feasibility of $x$ in \ref{prob: offlineUndersupplied}. 
    {The remaining constraints are either unaffected or} 
    scaled by $1/\market \leq 1$ yielding the feasibility of $\xbf^{\mathrm{OFF}}$ in $\offI(\inst)$. 

    We now prove that $\xbf^{\mathrm{OFF}}$ is optimal in $\offI(\inst)$. We will use the following inequalities that follow by feasibility and optimality of $\xbf$ in \ref{prob: offlineUndersupplied}, respectively.
    \begin{equation}
        \label{eq: ineq_claim_appendix}
        \left(1 - 1/\market \right) \sum_{u \in \undersup }\sum_{t: (u,t) \in E} \prob \cdot x_{u,t} \leq \left(1 - 1/\market \right) \sum_{u \in \undersup } \market = (\market -1)|\undersup|.
    \end{equation}
    \begin{equation}
    \label{eq: ineq_claim_appendix2}
        \offI(\inst) + (\market-1)|\undersup| = \sum_{(u,t) \in E} \prob \cdot x_{u,t} = \sum_{u \in \undersup} \sum_{t: (u,t) \in E} \prob \cdot x_{u,t} + \sum_{u \in \mathcal O} \sum_{t: (u,t) \in E} \prob \cdot x_{u,t}
    \end{equation}
    {We now bound the objective of $\xbf^{\mathrm{OFF}}$ in terms of $\offI(\inst)$}: 
    \begin{align*}
        \sum_{(u,t) \in E} \prob \cdot x^{\mathrm{OFF}}_{u,t} &= \sum_{u \in \undersup} \sum_{t: (u,t) \in E} \prob \cdot x^{\mathrm{OFF}}_{u,t} + \sum_{u \in \mathcal O} \sum_{t: (u,t) \in E} \prob \cdot x^{\mathrm{OFF}}_{u,t}\\
        &= \frac{1}{\market} \cdot \sum_{u \in \undersup} \sum_{t: (u,t) \in E} \prob \cdot x_{u,t} + \sum_{u \in \mathcal O} \sum_{t: (u,t) \in E} \prob \cdot x_{u,t}\\
        &= \frac{1}{\market} \cdot \sum_{u \in \undersup} \sum_{t: (u,t) \in E} \prob \cdot x_{u,t} - \sum_{u \in \undersup} \sum_{t: (u,t) \in E} \prob \cdot x_{u,t} + \sum_{u \in \undersup} \sum_{t: (u,t) \in E} \prob \cdot x_{u,t} + \sum_{u \in \mathcal O} \sum_{t: (u,t) \in E} \prob \cdot x_{u,t}\\
        &= \left( \frac{1}{\market} - 1 \right) \sum_{u \in \undersup} \sum_{t: (u,t) \in E} \prob \cdot x_{u,t} + \offI(\inst) + (\market-1)|\undersup|\\
        &\geq -(\market-1)|\undersup| + \offI(\inst)+(\market-1)|\undersup|\\
        &= \offI(\inst).
    \end{align*}
    The first two equalities follow by separating the sum in the partition of the supply nodes $(\undersup, \mathcal O)$ and the definition of $\xbf^{\mathrm{OFF}}$. {In the third we add and subtract the term $\sum_{u \in \mathcal O} \sum_{t: (u,t) \in E} \prob \cdot x_{u,t}$, and the fourth  follows from \cref{eq: ineq_claim_appendix2} and grouping terms}. The inequality follows {from \cref{eq: ineq_claim_appendix}}. Since $\xbf^{\mathrm{OFF}}$ {is feasible for $\offI(\inst)$, this proves its optimality and contradicts \Cref{lemma: kImbalancePairOffline}}.

    Finally, the equality in \Cref{eq: tightnessInU} follows from the optimality of $x^{\mathrm{OFF}}$ in $\offI(\inst)$ and that for every $u \in \undersup$ we have $\prob \sum_{t: (u,t) \in E} x_{u,t} =~\market$:
\begin{align*}
        \offI(\inst) = \sum_{(u,t) \in E} \prob \cdot x^{\mathrm{OFF}}_{u,t} 
        &= \frac{1}{\market} \cdot \sum_{u \in \undersup} \sum_{t: (u,t) \in E} \prob \cdot x_{u,t} + \sum_{u \in \mathcal O} \sum_{t: (u,t) \in E} \prob \cdot x_{u,t}\\
        &= \frac{1}{\market} \sum_{u \in \undersup} \market + \sum_{u \in \mathcal O} \sum_{t: (u,t) \in E} \prob \cdot x_{u,t}\\
        &= |\undersup| + \sum_{u \in \mathcal O} \sum_{t: (u,t) \in E} \prob \cdot x_{u,t}.
    \end{align*}
    \hfill \Halmos

\subsection{Comparing Algorithm \ref{alg: new_greedyD} and Algorithm \ref{alg: adv_greedy_alg}}
\label{ssec: example_newgreedy}
We now show that the adaptation from Algorithm \ref{alg: adv_greedy_alg} to Algorithm \ref{alg: new_greedyD}, which relies on knowing ex ante the $\market$-imbalanced pair $(\undersup,\oversup)$, is necessary to obtain the CR guarantees of \Cref{theo: CRguarantee_extended_def}. 
Specifically, we show that, in contrast to Algorithm \ref{alg: new_greedyD}, Algorithm \ref{alg: adv_greedy_alg} may fail to achieve a CR of $\market/(\market+1)$ on instances that admit a $\market$-imbalanced pair $(\undersup,\oversup)$. To do so, we rely on the instance in Figure \ref{fig:instance_generalized_imbalance}.
\begin{figure}[h]
    \centering

\tikzset{every picture/.style={line width=0.75pt}} 

\begin{tikzpicture}[x=0.75pt,y=0.75pt,yscale=-.55,xscale=.9]

\draw   (257.14,91.73) .. controls (257.17,100.56) and (250.04,107.76) .. (241.2,107.79) .. controls (232.37,107.83) and (225.17,100.7) .. (225.14,91.86) .. controls (225.1,83.02) and (232.23,75.83) .. (241.07,75.79) .. controls (249.91,75.76) and (257.1,82.89) .. (257.14,91.73) -- cycle ;
\draw   (256.86,55.73) .. controls (256.9,64.56) and (249.76,71.76) .. (240.93,71.79) .. controls (232.09,71.83) and (224.9,64.7) .. (224.86,55.86) .. controls (224.82,47.03) and (231.96,39.83) .. (240.79,39.8) .. controls (249.63,39.76) and (256.82,46.89) .. (256.86,55.73) -- cycle ;
\draw   (126.14,90.73) .. controls (126.17,99.56) and (119.04,106.76) .. (110.2,106.79) .. controls (101.37,106.83) and (94.17,99.7) .. (94.14,90.86) .. controls (94.1,82.02) and (101.23,74.83) .. (110.07,74.79) .. controls (118.91,74.76) and (126.1,81.89) .. (126.14,90.73) -- cycle ;
\draw   (257.86,129.73) .. controls (257.9,138.56) and (250.76,145.76) .. (241.93,145.79) .. controls (233.09,145.83) and (225.9,138.7) .. (225.86,129.86) .. controls (225.82,121.03) and (232.96,113.83) .. (241.79,113.8) .. controls (250.63,113.76) and (257.82,120.89) .. (257.86,129.73) -- cycle ;
\draw   (123.14,207.73) .. controls (123.17,216.56) and (116.04,223.76) .. (107.2,223.79) .. controls (98.37,223.83) and (91.17,216.7) .. (91.14,207.86) .. controls (91.1,199.02) and (98.23,191.83) .. (107.07,191.79) .. controls (115.91,191.76) and (123.1,198.89) .. (123.14,207.73) -- cycle ;
\draw   (258,169.73) .. controls (258.03,178.56) and (250.9,185.76) .. (242.06,185.79) .. controls (233.23,185.83) and (226.03,178.7) .. (226,169.86) .. controls (225.96,161.03) and (233.09,153.83) .. (241.93,153.79) .. controls (250.76,153.76) and (257.96,160.89) .. (258,169.73) -- cycle ;
\draw   (259,206.73) .. controls (259.03,215.56) and (251.9,222.76) .. (243.06,222.79) .. controls (234.23,222.83) and (227.03,215.7) .. (227,206.86) .. controls (226.96,198.03) and (234.09,190.83) .. (242.93,190.79) .. controls (251.76,190.76) and (258.96,197.89) .. (259,206.73) -- cycle ;
\draw    (123.14,207.73) -- (227,206.86) ;
\draw    (126.14,90.73) -- (226,169.86) ;
\draw    (126.14,90.73) -- (225.86,129.86) ;
\draw    (126.14,90.73) -- (225.14,91.86) ;
\draw    (126.14,90.73) -- (224.86,55.86) ;
\draw    (126.14,90.73) -- (227,206.86) ;

\draw (234,80) node [anchor=north west][inner sep=0.75pt]   [align=left] {{$t_4$}};
\draw (234,45) node [anchor=north west][inner sep=0.75pt]   [align=left] {{$t_5$}};
\draw (102,82) node [anchor=north west][inner sep=0.75pt]   [align=left] {{$u_2$}};
\draw (100,200) node [anchor=north west][inner sep=0.75pt]   [align=left] {{$u_1$}};
\draw (234,118) node [anchor=north west][inner sep=0.75pt]   [align=left] {{$t_3$}};
\draw (235,158) node [anchor=north west][inner sep=0.75pt]   [align=left] {{$t_2$}};
\draw (235,195) node [anchor=north west][inner sep=0.75pt]   [align=left] {{$t_1$}};
\end{tikzpicture}
\caption{The described instance $\inst$ with $\prob=1/2$ and $\sigmabf = (t_1, \ldots, t_5)$.}\label{fig:instance_generalized_imbalance}
\end{figure}

 First off, observe that  a possible realization of $\greedy$ results in every arrival being matched to $u_2$, so $\greedy(\inst) = 1 - (1/2)^5$.  Next, notice that $\offI(\inst) \geq 3/2$ by matching $t_1$ to $u_1$ and $t_2$ and $t_3$ can be matched fully to $u_2$. Finally, notice that the instance admits a $2$-imbalanced pair $(\undersup, \oversup) = (\{ u_2 \}, \{ u_1\})$ because reducing the capacity of $u_1$ to $1/2$ does not affect the optimal solution whereas increasing the capacity of $u_2$ to $2$ increases the objective by $(\market-1)=1$, as it allows to additionally match $t_4$ and $t_5$ to $u_2$. Combining all these we find that
\begin{equation*}
    \frac{\mathbb E[\greedy(\inst, \sigmabf)]}{\offI(\inst)} \leq \frac{1 - (1/2)^5}{3/2} = \frac{2}{3} \cdot (1 - (1/2)^5) < \frac{2}{3} \leq \frac{\mathbb E[\altgreedy(\inst, \sigmabf, (\undersup, \oversup))]}{\offI(\inst)}.
\end{equation*}
\section{The tradeoff between tighter benchmarks}

{Recent work in online bipartite matching has introduced tighter benchmarks that allow for tighter CR guarantees. Notable examples include the Configuration LP proposed by \citet{huang2020online}, the Path-Based Program by \citet{goyal2023online} for adversarial arrivals, and the benchmark introduced by \citet{brubach2020online} for stochastic arrivals.} Despite their ability to provide sharper performance guarantees, tighter benchmarks come with a substantial computational cost. In this section, we illustrate, with a focus on the Path-Based Program by \citet{goyal2023online}, how the exponential complexity inherent in these benchmarks makes them difficult to use in practice, even for small instances.

We begin by defining the set $\Omega$ of \emph{sample paths}, where each $\omega \in \Omega$ indicates which edges in $E$ realize (success of consumption) and which do not. Since each edge may independently succeed or fail, the size of this sample space grows exponentially with $|E|$, in particular, $|\Omega| = 2^{|E|}$. In this section, we focus on the \emph{Path-Based Program} (PBP) as described in \citet{goyal2023online}, a tighter benchmark that exploits this sample space. Although PBP has an exponential number of constraints, it is still possible -- at least in principle -- to simulate it. As we will see, however, this simulation fails to accurately describe the value of PBP even on small instances.

 We now introduce the path-based formulation, whose decision variables  $(\xbf^\omega)_{\omega \in \Omega}$ represent if an offline algorithm (that knows the input graph in advance), matches $u$ to $t$ in the sample path $\omega$.
\begin{align}
\mathrm{PBP}(\inst) = \max_{\mathbf{x}} \quad &  \mathbb E_{\omega} \left[ \sum_{(u,t) \in E} x_{u,t}^\omega  \cdot \bm{1}^\omega_{(u,t) \in E} \right] \label{pbp: obj}\\
\text{s.t.} \quad & \sum_{t: (u,t) \in E} x_{u,t}^\omega  \cdot \bm{1}^\omega_{(u,t) \in E} \leq 1, & \forall u \in \supply, \omega \in \Omega \label{pbp: supply}\\
& \sum_{u: (u,t) \in E} x_{u,t}^\omega  \leq 1, & \forall t \in [T], \omega \in \Omega \label{pbp: demand}\\
& x^\omega_{u,t} = x^{\omega'}_{u,t} &\forall (u,t) \in E, \omega, \omega' \in \Omega: \omega_{-ut} = \omega'_{-ut} \label{pbp: exp}\\
& 0 \leq x^\omega_{u,t} \leq 1, & \forall (u,t) \in E, \omega \in \Omega. \label{pbp: relax}
\end{align}
\noindent
Constraint $ (\ref{pbp: supply}) $ enforces that every supply node can be consumed at most once whenever the edge $ (u,t) $ is present in $ E $ and realizes in the sample path $ \omega $.
Similarly, constraint $ (\ref{pbp: demand}) $ ensures that $ t $ is matched to at most one supply node in every sample path.
Of particular interest is constraint $ (\ref{pbp: exp}) $, which captures the idea that an algorithm must commit to (or refrain from) matching $ (u,t) $ without knowing whether that specific edge will ultimately succeed.
Formally, the choice of matching $ (u,t) $ depends only on the realization of \emph{other} edges, and not on the realization of $ (u,t) $ itself.
Lastly, constraint $ (\ref{pbp: relax}) $ allows fractional decision variables. The objective $ (\ref{pbp: obj}) $ takes the weighted average (expectation) of these matchings across all realizations.

Since every algorithm must satisfy constraints $ (\ref{pbp: supply}) $, $ (\ref{pbp: demand}) $, $ (\ref{pbp: exp}) $, and $ (\ref{pbp: relax}) $, the value of $ \pbp $ provides a valid upper bound for any algorithm.
Note the one-to-one correspondence between the supply and demand constraints $ (\ref{pbp: supply}) $, $ (\ref{pbp: demand}) $ of $ \pbp $ and the supply and demand constraints $ (\ref{adversarial_lp: kappa-constraint}) $, $ (\ref{adversarial_lp: demand_constraint}) $ of $ \offI(\inst) $.
Since the constraints of $ \pbp $ hold for every sample path---whereas those of $ \offI(\inst) $ hold only in expectation---the path-based formulation provides a strictly tighter upper bound.

\subsection{The infeasibility of computing PBP for large instances}
\label{ssec: pbp_infeasible}

The tighter upper bound of PBP comes with a significant disadvantage: to solve the relaxation exactly, one needs to (i) enumerate all sample paths in $\Omega$, and (ii) enumerate,  for each edge $(u,t)$, \emph{all} pairs of sample paths 
whose realizations differ solely in the realization of $(u,t)$. 
Concretely, we have $\vert \Omega \vert = 2^{|E|}$, and the number of subsets of the form
$\{\,(\omega,\omega') \in \Omega^2 : \omega_{-ut} = \omega'_{-ut}\}$
has size $2^{|E|+1}$. 
This holds because each of the $|E|-1$ other edges has two possible outcomes (success/failure), which 
leads to $2^{|E|-1}$ assignments, 
and for each such assignment, $(u,t)$ can be realized or not in both $\omega$ and $\omega'$; thus, there are four distinct ways to define $(\omega,\omega')$ for that fixed assignment. As a result, computing $\pbp$ even for small instances requires significant computational effort and becomes intractable rapidly.

A natural approach to circumvent this complexity is by simulating only a subset of sample paths. Concretely, fix a sample size $n_s$ and draw $n_s$ realizations of $E$ where each edge succeeds w.p. $\prob$. This generates a set of sample paths $\Omega_{n_s} \subseteq \Omega$ and one can then replace $\Omega$ by $\Omega_{n_s}$ in the definition of $\pbp$. However, as we next discuss, this approach may fail to yield an upper bound on the true value of $\pbp$. 


The main reason why sampling-based $\pbp$ may yield a worse bound than $ \offI(\inst) $ relates to constraint \eqref{pbp: exp}, which acts as a linking constraint between sample paths. 
In the absence of that constraint, $\pbp$ can solve a maximum matching on the graph of realized edges. For example, if $\prob=1/2$ and a node $t$ has 6 neighbors, then with probability $63/64$ at least one of the incident edges realizes, meaning that on about 63 out of 64 sample paths, $\pbp$ without \eqref{pbp: exp} can successfully match that node in a maximum matching. In contrast, both $\offI(\inst)$ and any adaptive algorithm have a chance of at most $1/2$ to assign that node $t$. Now, \eqref{pbp: exp} ensures that this issue does not occur for $\pbp$, but we argue that for sample-based $\pbp$ to include \emph{any} constraints of the form \eqref{pbp: exp}, the number of samples needs to be exponential in $|E|$. Specifically, this holds true because, for any edge $(u,t)$ and sample path $\omega$, the probability that a different sampled sample path $\omega'$ coincides with $\omega$ on all but one value is $(1/2)^{|E|-1}$ (for $\prob=1/2$). Therefore, unless one samples an exorbitant number of edges, even for medium-sized graphs, the sampled version of $\pbp$ is likely to yield a worse bound than $\offI(\inst)$ would.


We now numerically illustrate this finding about sampled-PBP. We rely on the graph structure in \Cref{fig:instance_pbp} and vary $\prob$ at 40 equidistant values from $10^{-2}$ to $1$; for each $\prob$, we simulate $\pbp$, with either 100, 500, or 1000 sample paths, 30 times and average the result. 
\Cref{fig: pbpbad1} illustrates the poor convergence of these sampled versions of $\pbp$. Though $ \pbp $ is a tighter benchmark than $ \offI(\inst) $, and thus lies below it, the curves corresponding to the sampling approach often lie above $ \offI(\inst) $, even for $n_s=1000$ and in an instance with just 8 edges. 
\Cref{fig: pbpbad1hist} also shows that the computational effort of this sampling approach rapidly increases with $n_s$. As a result, we view these state-of-the-art benchmarks as mostly useful for the theoretical design of algorithms, but not to numerically evaluate algorithms. 

\begin{figure}[h]
    \centering
        \begin{minipage}{0.29
        \textwidth}
        \centering
        \includegraphics[width=\textwidth]{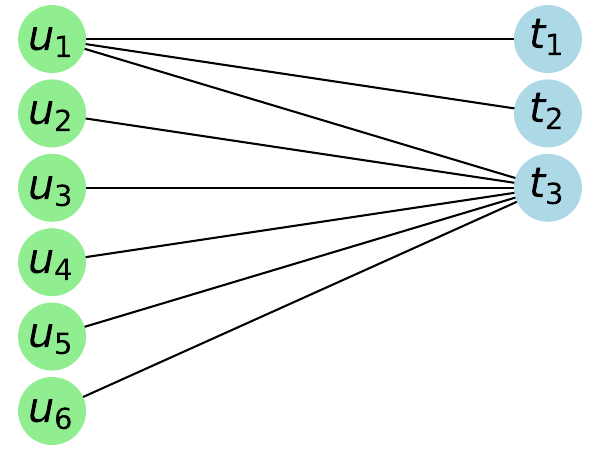}
        \caption{\small{Graph structure considered for the tests.}}
        \label{fig:instance_pbp}
    \end{minipage}\hfill
    \begin{minipage}{0.33\textwidth}
        \centering
        \includegraphics[width=\textwidth]{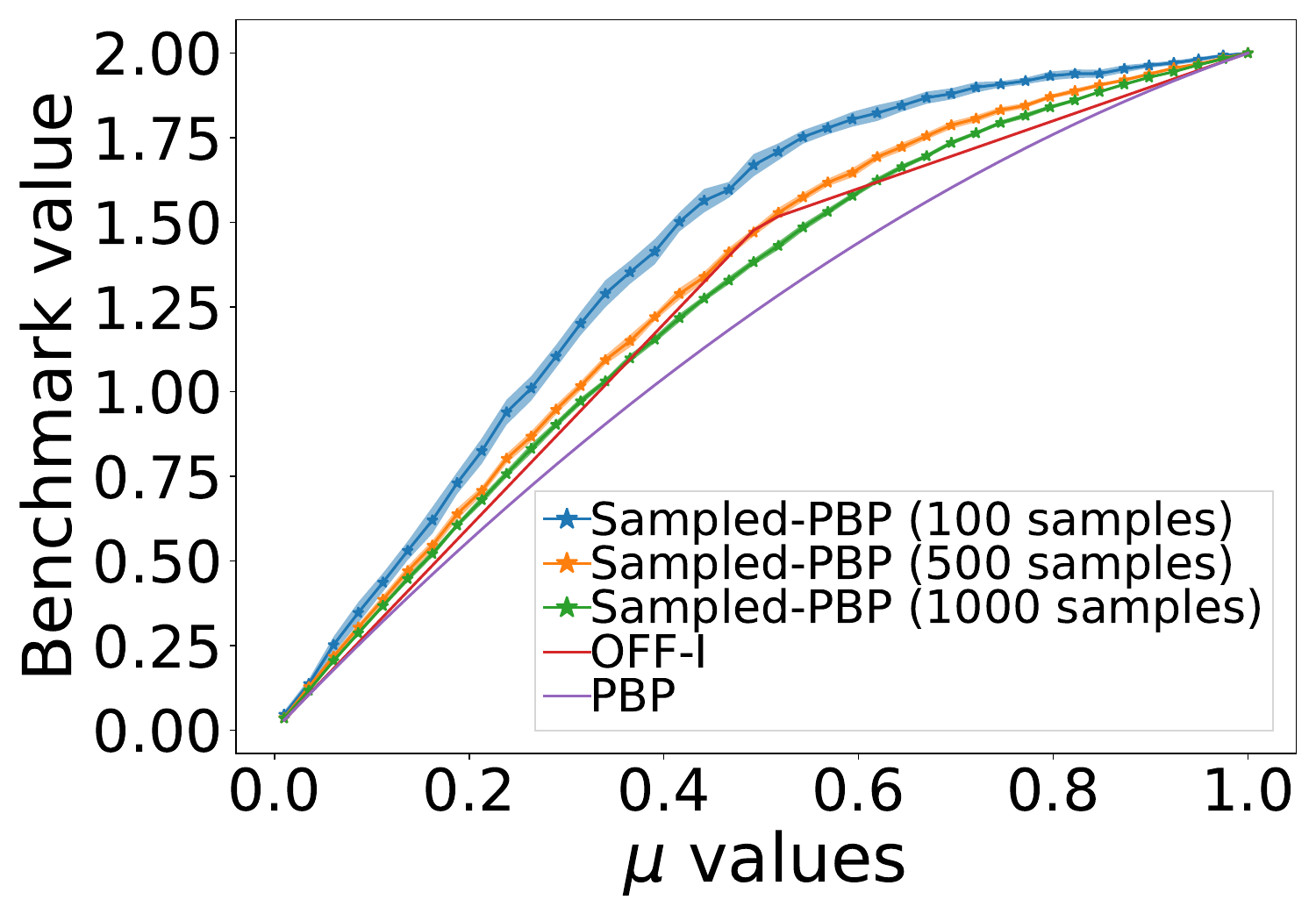}
        \caption{\small{Comparison of sampling PBP (number of samples).}}
        \label{fig: pbpbad1}
    \end{minipage}\hfill
    \begin{minipage}{0.33\textwidth}
        \centering
        \includegraphics[width=\textwidth]{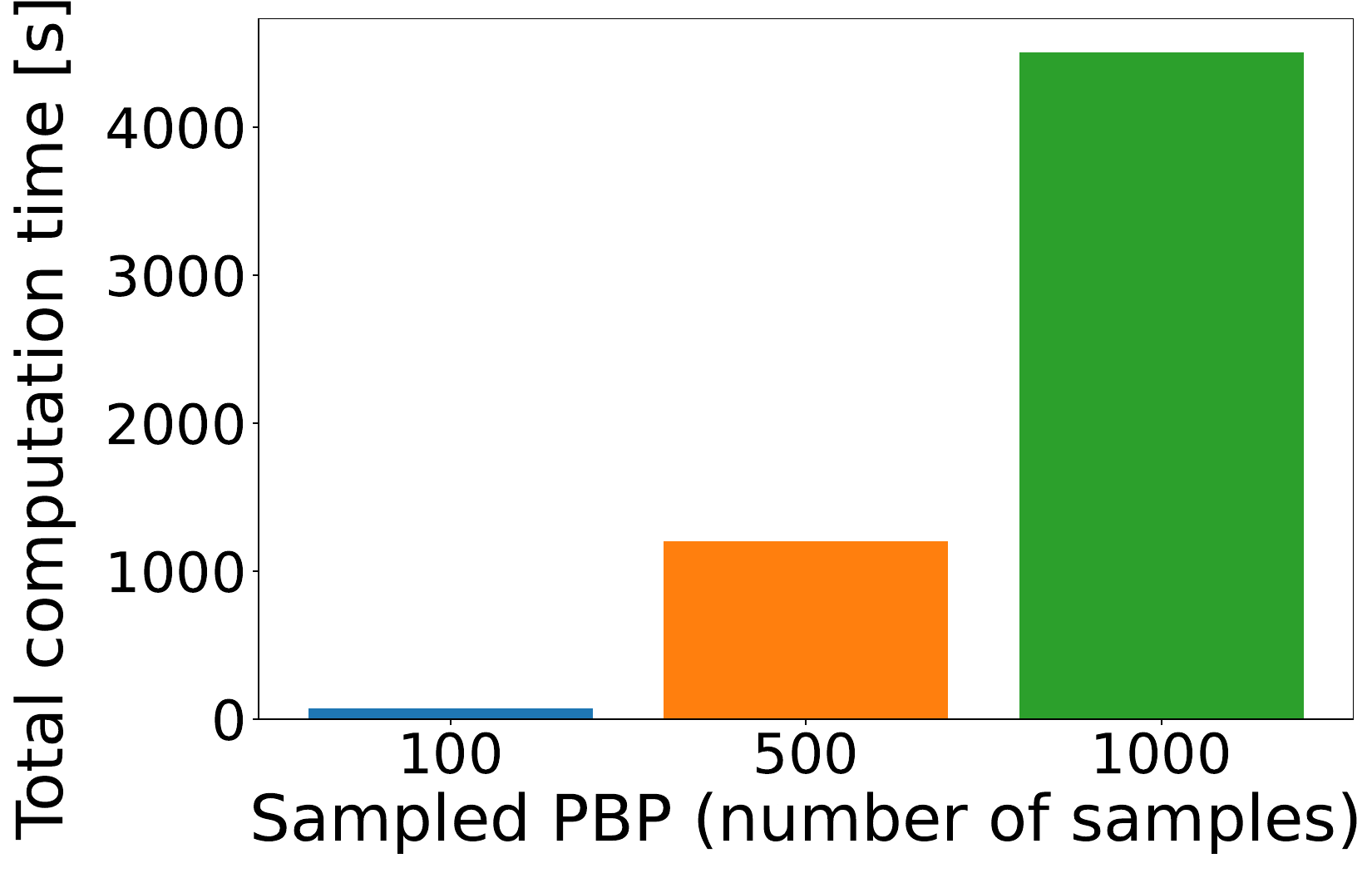}
        \caption{\small{Histogram representing the total time taken for each sampled PBP method.}}
        \label{fig: pbpbad1hist}
        \vfill \hfill
    \end{minipage}
\end{figure}


\subsection{Numerical Evaluation of instantaneous algorithms under Imbalanced Instances}

\label{ssec:imbalanced}

The main goal of this subsection is to illustrate that in highly imbalanced settings, both delayed and instantaneous algorithms produce better CRs. 
In \Cref{section: impossibility}, we introduced a class of instances $\inst_s[\market]$, parameterized by $s$ and $\market$, to prove the tightness result of \Cref{prop: adversarial CR UB}. Here, we focus on a simplified version of those instances and numerically compare the performance of $\greedy$ and the generalized fully-adaptive algorithm of \citet{goyal2023online}.

We set $L = 10$ and $\prob = 1/100$. We use an upper triangular instance as our test instance; it consists of 10 demand types and 10 supply nodes, structured so that demand nodes of type 1 connect to every supply type 1, demand nodes of type 2 connect to all supply nodes except for node 1, demand nodes of type 3 connect to all except for nodes 1 and 2, and so forth. Each type of demand consists of $\lceil \market / \prob \rceil$ nodes (with demand type 1 arriving first, then 2, and so forth), where $\market$ takes 40 logarithmically spaced values between $10^{-1.2}$ and $10^{1.2}$. For each instance considered, we compute $\greedy(\inst)$, $\offI(\inst)$, and the generalized fully-adaptive algorithm a total of {1,500} times, taking their mean. We plot the results in \Cref{fig: goyal_greedy_comparison}.
\begin{figure}[h]
    \centering
    \includegraphics[width=0.5\linewidth]{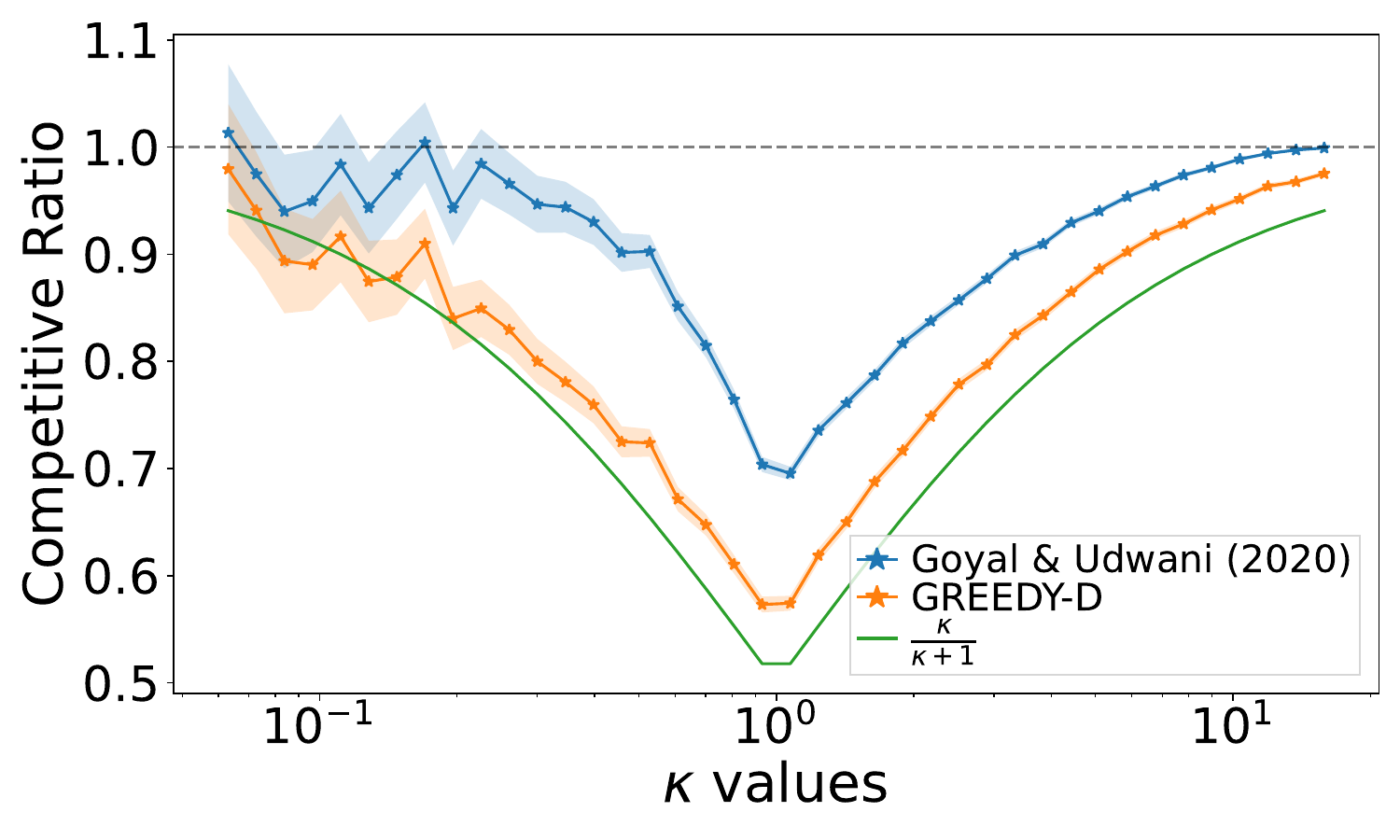}
    \caption{Comparative analysis of CRs between $\greedy$  and the generalized fully-adaptive algorithm from \citet{goyal2023online} across the described class of instances. Each data point represents the mean CR obtained from 1,500 simulation runs; the shaded regions denote {95\% confidence intervals}. }
    \label{fig: goyal_greedy_comparison}
\end{figure}

As \Cref{fig: goyal_greedy_comparison} shows, unsurprisingly and in line with \Cref{prop: adversarial CR UB}, the CR of $\greedy$ is close to its theoretical lower bound. We also observe the difference of instantaneous feedback in these instances: the generalized fully-adaptive algorithm consistently obtains better CRs than $\greedy$ for all of the considered instances. As the imbalance $\market$ becomes large or small, both algorithms approach a CR of 1 at a similar rate, indicating that imbalance yields better CR guarantees for instantaneous algorithms also, not just for $\greedy$ which our theoretical analysis revealed. 

{Lastly, we remark that the wider confidence intervals for small $\market$ arise from the scale of $\offI(\inst)$ (proportional to $\market L$) and the structure of the instance. For lower values of $\market$, the variance increases since there are fewer demand nodes per type, which makes it more likely for algorithms to assign demand nodes to supply nodes that will no longer be available for future demand arrivals.}


\subsection{Computation of $\market$ on a given instance}
\label{ssec: computation_kappa_instance}

Given $\xbf^*$ an optimal solution to \ref{prob: convex}, we compute the imbalance of an imbalanced pair $(\undersup, \oversup)$ via the following optimization problem:
\begin{align}
\max_{\market,\, z,\, \mathbf{x}} \quad 
& \market  \notag\\
\text{s.t.}\quad 
& \market \cdot z \;=\; 1, \label{cons: kappa_z}\\
& \sum_{t : (u,t) \in E} \prob \cdot x_{u,t} \leq \market,
  \quad \forall\, u \in \undersup, \label{cons: kappa_undersup}\\
& \sum_{t : (u,t) \in E} \prob \cdot x_{u,t} \leq z,
  \quad \forall\, u \in \oversup, \label{cons: kappa_oversup}\\
& \sum_{u : (u,t) \in E} x_{u,t} \leq 1,
  \quad \forall\, t \in [T], \label{cons: kappa_demand}\\
& \sum_{(u,t) \in E: u \in \undersup}
    \prob \cdot x_{u,t} 
  = \market  |\undersup|, \label{cons: kappa_undersup_guarantee}\\
& \sum_{(u,t) \in E: u \in \oversup}
    \prob \cdot x_{u,t} =
  \sum_{(u,t) \in E : u \in \oversup}
      \prob \cdot x^*_{u,t}, \label{cons: kappa_oversup_guarantee}\\
& \mathbf{x} \;\ge\; \mathbf{0}, 
  \quad \market \;\ge\; 1, 
  \quad z \;\ge\; 0 \label{cons: kappa_feasibility}.
\end{align}

As observed, this optimization problem is \textit{non-convex}, primarily due to constraint \eqref{cons: kappa_z}. This constraint, together with \eqref{cons: kappa_oversup}, enforces that $\sum_{t : (u,t) \in E} \prob \cdot x_{u,t} \leq 1/\market, \forall u \in \oversup$, ensuring that condition (iii) of \Cref{def: new_over_undersupplied} holds. Given that solving non-convex problems is NP-hard and may require exponential time, it is practical to consider an alternative approach for computing $\market$. Specifically, by first computing $\offI(\inst)$ and with knowledge of $\undersup$ and $\oversup$, one can employ a binary search on $\market$ in the right-hand side constraint of \ref{prob: offlineUndersupplied}. In this process, conditions (ii) and (iii) in \Cref{def: new_over_undersupplied} are ensured at each step. While the binary search itself is logarithmic in the range size, the complexity of each step depends on solving a LP, resulting in an overall polynomial time complexity.

We now state and show that the previous non-convex problem objective is $\market$, for some $\market$-imbalance pair.

\noindent\textit{Claim.}
    Given $(\undersup, \oversup)$ a $\market$-imbalance pair and $x^*$ an optimal solution to \ref{prob: convex}, then the above optimization problem returns $\market$.

\noindent \textbf{Proof of Claim.} Start off by noting that the set of constraints is feasible: the solution $(\market, z, \xbf) = (1,1,\xbf^*)$ satisfies all the constraints. Since $\xbf^*$ is an optimal solution to $\offI(\inst)$, then the set of constraints \eqref{cons: kappa_undersup}, \eqref{cons: kappa_oversup} are exactly the set of constraints \eqref{adversarial_lp: kappa-constraint} with $\market = 1$. Constraint \eqref{cons: kappa_demand} is exactly constraint \eqref{adversarial_lp: demand_constraint}. Constraint \eqref{cons: kappa_undersup_guarantee} follows by definition of $\classU = \undersup$, the set of consumed nodes by $\xbf^*$ (where the equality follows by \Cref{prop: k-imbalanced result}). The rest of the constraints follows directly.

Now, let $(\market, z, \xbf)$ an optimal solution to the above problem. We show that $\xbf$ is optimal for \ref{prob: offlineUndersupplied}:
\begin{align*}
    \sum_{(u,t) \in E : u \in \undersup} \prob \cdot x_{u,t} + \sum_{(u,t) \in E : u \in \oversup} \prob \cdot x_{u,t} &= \market \cdot |\undersup|  + \sum_{(u,t) \in E : u \in \oversup} \prob \cdot x_{u,t}\\
    &= \market \cdot |\undersup|  + \sum_{(u,t) \in E : u \in \oversup} \prob \cdot x^*_{u,t}\\
    &= (\market -1)  |\undersup| + \offI(\inst).
\end{align*}
The first equality uses constraint \eqref{cons: kappa_undersup_guarantee}. The second equality uses constraint \eqref{cons: kappa_oversup_guarantee} and the last equality uses $\sum_{(u,t) \in E : u \in \undersup} \prob \cdot x^*_{u,t} = |\undersup|$ and that $x^*$ is optimal for $\offI(\inst)$.
Since constraints \eqref{cons: kappa_undersup}, \eqref{cons: kappa_oversup}, \eqref{cons: kappa_demand} and \eqref{cons: kappa_feasibility} impose feasibility of $\xbf$, we have that $\xbf$ is an optimal solution to \ref{prob: offlineUndersupplied}.
Finally, the objective  ensures that we find the maximum $\market$ for which constraints \eqref{cons: kappa_undersup}, \eqref{cons: kappa_oversup} ensure that $\xbf$ is optimal in \ref{prob: offlineUndersupplied}. 
\clearpage

\end{document}